\documentclass[hidelinks,preprintnumbers,%
 reprint,
nofootinbib,
amsmath,amssymb,
 aps,onecolumn,
]{revtex4-2}
\usepackage[justification=justified,singlelinecheck=false]{caption}
\usepackage{ragged2e}
\usepackage{appendix,float,comment}
\usepackage{graphicx,setspace}
\usepackage{dcolumn}
\usepackage{bm}
\usepackage{xcolor,subcaption,caption}
\usepackage{hyperref}
\usepackage[normalem]{ulem}
\usepackage{slashed}

\def \hat{\widehat}
\def \<{\langle}
\def \>{\rangle}
\def \+{\dagger}

\def \hA{\hat{A}}
\def \tA{\widetilde{A}}
\def \hmu{\hat{\mu}}
\renewcommand\({\left(}
\renewcommand\){\right)}
\renewcommand\<{\left<}
\renewcommand\>{\right>}
\renewcommand\[{\left[}
\renewcommand\]{\right]}
\def \bes {\begin{subequations} }
\def \ees {\end{subequations}}
\def \be{\begin{eqnarray}}
\def \ee{\end{eqnarray}}

\def \e{\epsilon}

\begin{document}

\title{Quantifying fluctuation signatures of the QCD critical point using maximum entropy freeze-out}

\author{Jamie M. Karthein }
 \email{jmkar@mit.edu}
\author{Krishna Rajagopal}%
 \email{krishna@mit.edu}
\affiliation{%
Center for Theoretical Physics -- a Leinweber Institute, Massachusetts Institute of Technology, Cambridge, MA 02139,
USA
}%


\author{Maneesha Pradeep}
 \email{mpradeep@umd.edu}
\affiliation{
 Department of Physics and Maryland Center for Fundamental Physics, University of Maryland, College Park, MD 20742 USA
}%

\author{Mikhail Stephanov}
\email{misha@uic.edu}
\affiliation{%
Department of Physics and Laboratory for Quantum Theory at the Extremes, University of Illinois, Chicago, Illinois 60607, USA
}%

\author{Yi Yin}
\email{yiyin@cuhk.edu.cn}
\affiliation{%
School of Science and Engineering,
The Chinese University of Hong Kong,
Shenzhen, Guangdong 518172, China
}%


\date{\today}

\begin{abstract}
A key question about the QCD phase diagram is whether there is a critical point somewhere on the boundary between the hadronic and quark-gluon plasma phases, and if so where. Heavy-ion collisions offer a unique opportunity to search for signatures of such a critical point by analyzing event-by-event fluctuations in particle multiplicities. To draw meaningful conclusions from experimental data, a theoretical framework is needed to link QCD thermodynamics with the particle spectra and correlations observed in detectors. 
The Equation of State (EoS) of QCD near a critical point can be related to the universal Gibbs free energy of the 3D Ising model using four currently unknown non-universal mapping parameters whose values are determined by the microscopic details of QCD. We utilize the maximum entropy approach to freeze-out the fluctuations in order to make estimates for factorial cumulants of proton multiplicities, assuming thermal equilibrium, for a family of EoS with a 3D Ising-like critical point, varying the microscopic inputs that determine the strength and structure of the critical features. We quantify the effect of the non-universal mapping parameters, and the distance between the critical point and the freeze-out curve, on the factorial cumulants of proton multiplicities.
\end{abstract}

\preprint{MIT-CTP/5906}

\maketitle

\tableofcontents
\section{Introduction}
\label{Sec:Introduction}

The QCD Equation of State (EoS) at large densities remains poorly understood and therefore our knowledge of the phase diagram is largely schematic~\cite{Bzdak:2019pkr}. 
At zero baryon chemical potential, $\mu_B=0$, we have a reliable quantitative understanding of the EoS from first-principles via lattice QCD calculations at higher temperatures~\cite{HotQCD:2014kol,Borsanyi:2010cj} and with empirical input via the Hadron Resonance Gas model at lower temperatures.  These agree well with each other over an intermediate range of temperatures below the pseudo-critical chiral crossover temperature 
$T_{pc} \sim 158$ MeV~\cite{Borsanyi:2010cj,Goswami:2020yez}. 
The quantitative understanding of hot QCD thermodynamics in the vicinity of the crossover temperature and at higher temperatures where QCD matter is a strongly coupled 
quark-gluon liquid (conventionally called quark-gluon plasma) relies upon nonperturbative computations done via lattice methods, methods which are 
hindered at non-zero values of $\mu_B$ by the infamous sign problem~\cite{Gattringer:2016kco,Pawlowski:2022rdn,Aarts:2023vsf}.  
As noted above, lattice QCD calculations have confirmed that the phase  transition at $\mu_B=0$ is a continuous crossover~\cite{Aoki:2006we}. 
One of the most important long-standing questions in the field  is whether the transition between the hadronic and quark-gluon matter becomes first-order above some non-vanishing baryon chemical  potential at which a second order critical point is located. 
Determining whether such a critical point exists and, if so, where on the $(\mu_B,T)$ phase diagram it is located is to date unresolved and has garnered considerable theoretical and experimental interest. Although we do not yet know whether such a critical point exists, we do know that if present it is in the same universality class as the 3D Ising model~\cite{Rajagopal:1992qz}, which means that universal features of the critical fluctuations in its vicinity are well understood. 

Over the years, there have been many model calculations that have yielded indications for the presence of a critical point at different locations on the phase diagram; 
for reviews see Refs.~\cite{Stephanov:2004wx, Luo:2017faz, Busza:2018rrf,Bzdak:2019pkr,Luo:2022mtp,Aarts:2023vsf}. 
In recent years, functional renormalization group studies \cite{Fu:2019hdw}, holographic calculations tuned to agree with lattice QCD results at $\mu_B=0$ \cite{Cai:2022omk,Hippert:2023bel,Jokela:2024xgz,Chen:2024mmd}, Dyson-Schwinger based approaches~\cite{Gunkel:2021oya,Gao:2020fbl}, and several different extrapolations of lattice QCD calculations~\cite{Basar:2023nkp, Clarke:2024ugt, Shah:2024img} all point towards a critical point in the range of baryon chemical potentials $500$~MeV $< \mu_B < 700$~MeV, and all find no critical point at lower values of $\mu_B$. 
Moreover, inferences drawn from lattice QCD calculations of expansions of the EoS in powers of $\mu_B$ also place constraints on the location of a possible critical point,
in particular excluding its presence in the region of the phase diagram with 
$\mu_B/T \lesssim 2-2.5$~\cite{Borsanyi:2021sxv,HotQCD:2018pds}. More recent refinements of this approach have been able to exclude the presence of a critical point anywhere in the range of chemical potentials $\mu_B < 450$ MeV~\cite{Borsanyi:2025dyp}.

On the experimental front, the Beam Energy Scan (BES) program at the Relativistic Heavy Ion Collider (RHIC) 
which has now successfully completed its data taking has, as one of its major objectives, determining or constraining the location of the critical point by scanning through the phase diagram with collisions of heavy-ions at different center-of-mass energies. The STAR collaboration has recently reported measurements 
from the high-statistics second phase of this program,
in particular measurements
of non-Gaussian higher cumulants of the (net-) proton multiplicity fluctuations from Au+Au collisions with center-of-mass energies as low as $\sqrt{s_{\rm NN}}=7.7$~GeV ~\cite{STAR:2025zdq}, which freeze out at $\mu_B\leq 420$~MeV. The moments of the proton distribution are sensitive to the presence of a critical point in the phase diagram, with higher orders diverging more strongly because they are proportional to higher powers of the correlation length $\xi$, a quantity which diverges at the critical point. (For a review, see Ref.~\cite{Bzdak:2019pkr}.)
These results show no evidence of a critical point located in the region of the phase diagram with $\mu_B\leq 420$~MeV. 
However, theoretical models of the noncritical baseline expected for these fluctuations do not appear to describe the data at all energies. 

It is crucial to understand the critical point contribution to proton fluctuations in order to help interpret the recent results from STAR and further experimental measurements coming soon. We shall take significant steps toward this goal in this paper, although our calculations cannot yield quantitative predictions as our analysis 
focuses only on quantifying the proton fluctuations expected assuming that the fluctuations in heavy ion collisions that freeze out not far from a critical point are in thermal equilibrium when they freeze out.

We look forward with particular anticipation to measurements coming soon from STAR using data taken in fixed-target collisions during the BES program that extend its reach to lower $\sqrt{s_{\rm NN}}$, higher $\mu_B$, and lower temperatures. As noted above, a variety of different theoretical approaches, none of which are under quantitative control but which have quite different systematic limitations, all point to the possibility of a critical point not far beyond the region of the phase diagram that STAR has explored in collider mode. 
There are also further planned experimental programs, such as the CBM detector at GSI-FAIR, which also aim to study the QCD phase diagram at larger values of baryon chemical potential.

The analysis of heavy-ion collisions has been extremely successful in describing various aspects of the EoS. 
The mid-rapidity yields of various hadronic species have been shown to be in agreement with a hadron resonance gas in equilibrium just prior to chemical freeze-out \cite{Andronic:2017pug}.
The transport coefficients such as bulk viscosity and shear viscosity in QCD have now been constrained using the experimental data on particle multiplicity, transverse momentum and collective flow data using sophisticated Bayesian techniques \cite{Bernhard:2016tnd,Bernhard:2019bmu,JETSCAPE:2020mzn,Nijs:2020ors,Nijs:2020roc}.
It has long been a goal to constrain the EoS of QCD at non-zero chemical potential from experimental data such as the cumulants of particle multiplicities.  The
recent STAR measurements~\cite{STAR:2025zdq} are a big success in this regard. It is also pleasing that their results and the recent lattice-QCD-based and model calculations are fully consistent, as all find no critical point with $\mu_B<420$~MeV.

A crucial feature that distinguishes the EoS of QCD from that of a hadron resonance gas is its non-trivial susceptibilities, i.e.~its correlations of thermodynamic densities. At $\mu_B=0$, the lowest few Taylor coefficients of the expansion of the pressure in powers of $\mu_B^2$ 
are well described by the hadron resonance gas only at low temperatures, below and near the crossover temperature.
Their non-trivial behavior at higher temperatures has been reliably determined via controlled lattice QCD calculations at temperatures comparable to the crossover temperature and above~\cite{Bazavov:2017dus,Guenther:2017hnx,Borsanyi:2021sxv}. 
The goal of exploring the QCD phase diagram is to extend these successes to $\mu_B>0$.
The recently developed maximum entropy freeze-out procedure~\cite{Pradeep:2022eil} proposes a method to connect the cumulants of particle multiplicities measured in heavy-ion collisions to the correlation functions of the fluctuating thermodynamic densities at freeze-out with $\mu_B>0$, thus enabling the use of experimental measurements in the exploration of the phase diagram.

It has been long argued that the non-monotonic collision energy dependence of cumulants of particle multiplicities, in particular those of proton multiplicities, can serve as a crucial signature of the presence of a critical point somewhere in the phase diagram with $\mu_B>0$~\cite{Stephanov:1998dy,Stephanov:1999zu,Hatta:2003wn,Stephanov:2008qz,Athanasiou:2010kw,Stephanov:2011pb,Fu:2021oaw,Fu:2023lcm}.
These early estimates assumed  fluctuations that are in equilibrium at freezeout and employed models that include a critical $\sigma$ field to describe the universal long-wavelength fluctuations of the chiral order parameter near a QCD critical point.  
Fluctuations in $\sigma$ couple to hadrons, for example coupling to protons via an effective $\sigma \bar p p$ coupling. This means that $\sigma$ fluctuations induce fluctuation in the masses, and consequently the equilibrium multiplicities, of protons and antiprotons~\cite{Stephanov:1999zu,Hatta:2003wn,Stephanov:2008qz,Athanasiou:2010kw,Stephanov:2011pb}.
However, in these early studies the magnitude and form of the couplings between the critical $\sigma$ field and hadrons were based on phenomenological considerations. This was sufficient to determine the universal power law 
scalings of the particle multiplicities but does not allow for a quantitative determination of their absolute magnitudes.
Further progress can be made upon noting that fluctuations in the multiplicities of protons, antiprotons, and other hadrons at freezeout constitute fluctuations in thermodynamic densities such as energy and baryon densities, which must therefore be enhanced near a critical point. This can also be determined directly from thermodynamic considerations, as
the derivatives of pressure, which determine the correlations of the fluctuating hydrodynamic variables, are known to diverge at the critical point with the same universal (3D Ising) power law exponents~\cite{Stephanov:2011pb} that govern the scaling of the cumulants of proton (hadronic) multiplicities~\cite{Stephanov:2008qz,Athanasiou:2010kw,Stephanov:2011pb}.
However, if the cumulants of the particle multiplicities are determined via universality arguments applied to the fluctuating $\sigma$ field without a quantitative understanding of the coupling between these fluctuations and the hadrons, it is impossible to determine the quantitative relationship between the cumulants of particle multiplicities and those of thermodynamic densities.
This is where the maximum entropy procedure for describing how fluctuations of thermodynamic densities freeze out into fluctuations of particle multiplicities comes in~\cite{Pradeep:2022eil}. This procedure is the least biased way of translating the thermodynamic correlation functions to the cumulants of particle multiplicities at freeze-out while respecting all the local conservation laws for energy, momentum and conserved charge densities. 
The role of conservation laws in the determination of cumulants and factorial cumulants~\cite{Ling:2015yau,Bzdak:2016sxg} of particle multiplicities as a fluctuating fluid particlizes, as well as the sensitivity of these observable quantities to experimental acceptance cuts, have been investigated and highlighted previously in Refs.~\cite{Bzdak:2012an,Bzdak:2012ab,Luo:2013bmi,Ling:2015yau,Bzdak:2016sxg,He:2017zpg,Bzdak:2017ltv,Brewer:2018abr,Oliinychenko:2019zfk,Oliinychenko:2020cmr,Vovchenko:2021yen,Vovchenko:2020tsr,Vovchenko:2020gne,Vovchenko:2020kwg,Vovchenko:2021kxx,Pihan:2022xcl,Aasen:2022cid,Kuznietsov:2024xyn,Bzdak:2025rhp,Wang:2025fve}.

In this work, we apply maximum entropy freeze-out to obtain the factorial cumulants of proton multiplicities in an equilibrium setting. 
That is, we make the assumption that the fluctuations in thermodynamic densities just before freezeout near a critical point are in
thermal equilibrium, and determine the resulting factorial cumulants of the proton multiplicity just after freezeout.
We leave the inclusion of dynamical effects~\cite{Berdnikov:1999ph}, in particular the out-of-equilibrium dynamics of the critical fluctuations before freezeout incorporating critical slowing down and the effects of conservation laws -- as described via Hydro+~\cite{Stephanov:2017ghc,Rajagopal:2019xwg,Du:2020bxp,Pradeep:2022mkf} --
to future work.
This is the first of two central reasons why we cannot make {\it ab initio} predictions for heavy-ion collisions.  
The second is that in
order to make any estimates for the cumulants of particle multiplicities even upon assuming thermal equilibrium at freezeout we would need to know the EoS of QCD 
at the large nonzero values of $\mu_B$ where a critical point may be found.  And, of course, this is exactly what we do not know from first principles, and are seeking to learn about via experimental measurements.
Our work is not at all on a path toward some future determination of the QCD EoS at large $\mu_B$ (including perhaps a critical point) from theoretical calculations alone.  Not at all. Rather, it represents a significant advance in the effort to build a framework that (if nature is kind) will allow us to determine the QCD EoS at large $\mu_B$ near a critical point using experimental measurements of the cumulants of particle multiplicities in heavy-ion collisions that freeze out near said critical point.

We shall focus in this paper on how to 
connect experimentally measured cumulants of particle multiplicities to the equation of state of QCD in the vicinity of a possible critical point. 
In this study, we do so upon assuming that the  fluctuations are in thermal equilibrium at freezeout, which cannot actually be the case for heavy ion collisions that freeze out near a critical point on account of critical slowing down, and given the fact that the matter produced in such collisions can only spend a short period of time in the vicinity of a critical point.
By assuming thermal equilibrium at freezeout, we are taking a step backwards relative to previous Hydro+ treatments of the dynamics of fluctuations before freezeout~\cite{Stephanov:2017ghc,Rajagopal:2019xwg,Du:2020bxp,Pradeep:2022mkf}. A direct consequence of this simplifying assumption is that sensitivity to initial hydrodynamic fluctuations are completely washed out from the final particle multiplicity fluctuations.
We shall, however, take a substantial step forward in the analysis of freezeout itself.  The maximum entropy method that we shall employ here allows us to eliminate the ad hoc assumptions used previously. The treatment of freezeout in Ref.~\cite{Pradeep:2022mkf} and much prior literature is based upon assuming that the long wavelength critical fluctuations in a scalar field $\sigma$ couple solely to the masses of hadrons, and furthermore required the introduction of a coupling parameter (for example $g_p$ in $g_p\sigma \bar{p} p$) between $\sigma$ and each hadron species. The maximum entropy method, instead, allows us to relate hadron multiplicity fluctuations immediately after freezeout directly to thermodynamic fluctuations immediately before freezeout.
That said, because we are assuming equilibrium fluctuations at freezeout, and also because we shall not make any quantitative estimates of the contribution of decay particles to cumulants of multiplicities of observed particles, we will not present quantitative predictions for experimental measurements here. 
Our goal is that this work already sets a quantitative framework, to which additional elements such as pre-freezeout Hydro+ dynamics can be incorporated, so as to determine the QCD EoS near the critical point from future experimental measurements. 

The paper is organized as follows.
We review the family of EoSs for QCD with a critical point 
obtained via mapping the universal features of the 3D Ising EoS onto the QCD phase diagram that we employ in our analysis in Section~\ref{Sec:EoS}. In Section~\ref{Sec:ME} we introduce
the maximum entropy freeze-out procedure that we use 
to determine the fluctuations of particle multiplicities, assuming them to be in thermal equilibrium, and present our calculation of the factorial cumulants of particle multiplicities. 
We demonstrate the effect of the nonuniversal EoS mapping parameters in Section~\ref{Sec:main} on both the hydrodynamic correlators and the cumulants of particle multiplicity. 
We provide quantitative estimates for the equilibrium cumulants of proton multiplicity for our parameter choices in EoSs in Section~\ref{Sec:CumulantsPlots}. In Section~\ref{Sec:Conclusion}, we summarize and conclude with future prospects, while the appendices provide details on the scaling equation of state (Appendix \ref{app:ScalingEoS}), the contribution of feed-down protons from resonance decays to the factorial cumulants of the proton multiplicity distribution (Appendix \ref{app:resonances}), quantifying the effect of subleading contributions by comparing our results to what we would have obtained had we kept only the most divergent contributions to the flucutuations (Appendix \ref{app:subleading}), as well as a glossary of the notations used in this work (Appendix \ref{app:notations}).

\section{Equation of State of QCD Near a Critical Point}

The conjectured critical point of QCD, if it exists, would belong to the 3D Ising universality class~\cite{Rajagopal:1992qz}. By the universality of critical phenomena, the EoS near this critical point can be mapped to the Gibbs free energy of the 3D Ising model. A general parametrization for the mapping near the QCD critical point was introduced in Ref.~\cite{Parotto:2018pwx} and later extended in Refs.~\cite{Karthein:2021nxe,Kahangirwe:2024cny}. This approach yields a parametrized family of Equations of State, each with a critical point at a location specified via two parameters, $\mu_c$ and $T_c$, with the orientation and shape of the critical region on the phase diagram as well as the strength of the critical fluctuation determined via two further parameters. For
each EoS (i.e.~for each choice of the parameters that specify the mapping from the Ising phase diagram to the QCD phase diagram) the correlation functions of hydrodynamic densities in thermal equilibrium and the corresponding equilibrium fluctuations in the hydrodynamic densities
are determined, and in all cases they have the universal features expected in the 3D Ising universality class.
The microscopic, nonuniversal, information unique to QCD is encoded in the values of the mapping parameters. In this Section we shall review the mapping from the Ising model to QCD introduced in Refs.~\cite{Parotto:2018pwx,Kahangirwe:2024cny}, describe the mapping of contours of constant correlation length to illustrate the size and shape of the critical region and set up the calculation of hydrodynamic correlations in equilibrium.

In Section~\ref{Sec:ME}, we shall use the maximum entropy method to translate the correlations of hydrodynamic densities into momentum correlations between observed particles, and thereby the cumulants. We shall discuss quantitatively how different features of the QCD EoS (including its universal features and the nonuniversal mapping parameters that we introduce in this Section) modify the equilibrium estimates for the cumulants of particle multiplicities along the freeze-out curve.
We shall discuss, qualitatively and semi-quantitatively, the dependence of various properties such as the hierarchy of the magnitude of cumulants, and the scaling dependence of the cumulants on the separation between the freeze-out point and the cross-over trajectories on the various mapping parameters in the family of EoS.

\label{Sec:EoS}
\subsection{Mapping from the Ising Model to QCD}
\label{Subsec:Mapping}

The critical point characterizing the end-point of the conjectured first-order phase transition curve demarcating the QGP phase from the hadron resonance gas phase is expected to fall in the 3D Ising universality class.
This universality of critical phenomena follows from renormalization group theory~\cite{Rajagopal:1992qz}.
Thus, the EoS near the critical point can be mapped to the Gibbs free energy of the 3D Ising Model EoS, as follows. 
We write the QCD EoS as
\be
\label{Eq:EoS}
P(\mu,T)=P^{\text{reg}}(\mu,T)+P^{\text{sing}}(\mu,T)\,,
\ee
where 
here and henceforth we shall denote $\mu_B$ just by $\mu$, where
$P^{\text{reg}}$ is an analytic 
(more precisely, less singular than $P^{\rm sing}$) 
function of $\mu$ and $T$ near the critical point $(\mu_c,T_c)$,
and where we can equate the most singular part of the logarithm of QCD partition function to that of 3D Ising model yielding
\be
\label{Eq:Psing}
 P^{\text{sing}}(\mu,T)=-T^{4}_c\, G_{\rm{Ising}}(r(\mu,T)\, h(\mu,T))
\ee
 where $G_{\rm Ising}(r,h)$, the negative of the logarithm of the partition function, is the Gibbs free energy as  a function of the reduced temperature $r$ and the magnetic field $h$ in the three-dimensional Ising model.
We have taken the coefficient of proportionality between $P^{\text{sing}}$ and $-G_{\text{Ising}}$ to be $T^4_c$, as in Ref.\cite{Parotto:2018pwx}. A different choice of normalization would correspond to an overall rescaling of both the $h$ and $r$ axes on the $T-\mu$ plane, and hence can be compensated by an overall rescaling of the two mapping parameters $\rho$ and $w$, introduced below.  
Because the explicit form of $G_{\rm Ising}$ is rather cumbersome, we present the details of the scaling Equation of State \eqref{Eq:Psing} in Appendix~\ref{app:ScalingEoS}. 

The framework for mapping the universal physics of a critical point from the Ising model onto the QCD phase  diagram was first developed in Ref.~\cite{Parotto:2018pwx} and was later updated in Refs.~\cite{Karthein:2021nxe,Kahangirwe:2024cny}.
The method of mapping has been studied in further detail in Refs.~\cite{Pradeep:2019ccv,Mroczek:2022oga}.
This framework has also been utilized for various studies both in and out of equilibrium~\cite{Dore:2020jye,
Monnai:2021kgu,Dore:2022qyz},  
and in studies of a first-order phase transition~\cite{Pradeep:2024cca,Karthein:2024zvs}.
Another approach to implementing a critical point from universality can be found in Ref.~\cite{Kapusta:2022pny}.
The Ising model phase diagram has axes $r$ (Ising-temperature direction) and $h$ (Ising-magnetic-field direction) with a critical point at $r=h=0$.  Upon mapping to the QCD phase diagram whose axes are $\mu$  and $T$, the
Ising
variables $r$ and $h$ are analytic functions of $\mu$ and $T$. By definition, $r=h=0$ is mapped onto 
the QCD critical point at $T=T_c$ and $\mu=\mu_c$. 
In the close vicinity of the critical point, the Ising $r$ and $h$ axes will be mapped onto straight lines on the QCD phase diagram and in this region a linear mapping between $(r,h)$ and $\Delta T\equiv T-T_c$ and $\Delta \mu \equiv \mu-\mu_c$ suffices to describe the leading singular behavior.
Following the notation introduced by Parotto et al \cite{Parotto:2018pwx}, a linear mapping between the variables can be parametrized as:
\bes
\label{Eq:linmap}
\be \label{Eq:linmap1}
h(\mu,T)&=&-\frac{\Delta T \cos \alpha_1 +\Delta \mu \sin \alpha_1}{w T_c \sin (\alpha_{1}-\alpha_2)}\\ \label{Eq:linmap2}
r(\mu,T)&=&\frac{\Delta T \cos \alpha_2 +\Delta \mu \sin \alpha_2}{\rho w\, T_c \sin (\alpha_{1}-\alpha_2)}\ .
\ee
\ees
The parameters $\alpha_1$ and $\alpha_2$ are the angles between the axes of the QCD phase diagram and those of the Ising model, and $w$ and $\rho$ are the scaling parameters between  Ising and QCD coordinates: $w$ determines the overall scale of both $r$ and $h$, while $\rho$ determines the relative scale between them. We shall visualize the 
consequence of changes in the values of $w$ and $\rho$ in the next Subsection.

Note that if one always chooses the critical point to lie on the chiral crossover line as in Refs.~\cite{Parotto:2018pwx,Karthein:2021nxe,Kahangirwe:2024cny}, this fixes $T_c$ once $\mu_c$ has been chosen. Furthermore, requiring that the Ising crossover line $h=0$; $r>0$ maps onto a line on the QCD phase diagram that is tangent to the chiral crossover line fixes $\alpha_1$ once $\mu_c$ has been chosen. This means that the linear mapping from the Ising phase diagram to the QCD phase diagram \eqref{Eq:linmap} is specified by the four parameters $\mu_c$, $\alpha_2$, $w$ and $\rho$.

The linear mapping \eqref{Eq:linmap} is not sufficient for our purposes precisely because it is linear: it maps the Ising $r$-axis ($h=0$ with $r>0$) which corresponds to the crossover onto a straight line on the QCD phase diagram, whereas in QCD the crossover follows a curve in the $(\mu_B,T)$ phase diagram. The authors of Ref.~\cite{Kahangirwe:2024cny} have taken advantage of the charge conjugation symmetry of the QCD Lagrangian to extend the linear mapping in $(\mu-\mu_c,T-T_c)$ to a mapping in $(\mu^2-\mu^2_c, T-T_c)$.
Employing a quadratic mapping in $\mu$ makes it possible to map the Ising crossover line ($h=0$, $r>0$) onto a curve in the QCD phase diagram that follows the chiral crossover. Ref.~\cite{Kahangirwe:2024cny} further improved the
extrapolation of lattice data from small $\mu$ to large $\mu$  by using the $T^{'}$ expansion introduced in Ref.~\cite{Borsanyi:2021sxv}. We shall employ the resulting mapping since even though it is more complicated than \eqref{Eq:linmap} it represents a significant improvement. 
The parametrization of the mapping from the Ising model variables to QCD variables introduced in Ref.~\cite{Kahangirwe:2024cny} is given by:  
 \be
 \label{Eq:quadmap}
h(\mu,T)=-\frac{\Delta T'\cos \alpha_1}{T_c w \, \sin(\alpha_1-\alpha_2)}\, , \quad r(\mu,T)=-\frac{\mu^2-\mu^2_{c}}{2\mu_c T_c \, \rho w\, \cos\alpha_1}+\frac{\Delta T'\cos \alpha_2}{T_c\, \rho w \,  \sin(\alpha_{1}-\alpha_2)}\,
 \ee
where we have defined 
\begin{equation}
  \label{eq:DT'-def}
   \Delta T'(\mu,T)\equiv \frac{T'(\mu,T)-T_0}{(\partial T'/\partial T)_{\rm cp}},
 \end{equation}
 with $T_0=158$~MeV being the (known) crossover temperature at $\mu=0$ and the subscript ``cp" referring to the evaluation of the derivative at the critical point at $(\mu_c,T_c)$. 
Here, $T^{'}-T$ can be specified order by order in $(\mu/T)^2$, and following Ref.~\cite{Kahangirwe:2024cny} we truncate this expansion at quadratic order, yielding
\be
\label{Eq:TprimeExp}
T^{'}(\mu,T)=T\left[1+\kappa^{BB}_2(T)\(\frac{\mu}{T}\)^2+{\cal O}\(\(\frac{\mu}{T}\)^4\)\],
\ee
where $T'=T_0$ is the crossover curve, $\kappa_2^{BB}(T_0)$ is the curvature of the 
chiral crossover curve at $\mu=0$, and the full expression for $\kappa^{BB}_2(T)$ can be found in Ref.~\cite{Kahangirwe:2024cny}. 
Note that 
at $\mu=0$ the variable $T'$ becomes simply $T$. This means that when the Ising $h=0$ axis is mapped onto the QCD phase diagram it maps onto a curve that crosses $\mu=0$ at $T=T_0$, ensuring that it maps onto a quadratic curve connecting $(\mu=0,T=T_0)$ to $(\mu=\mu_c,T=T_c)$,   following the chiral crossover curve $\Delta T'=0$. 
Note also that, as for the linear map \eqref{Eq:linmap}, the quadratic mapping from the Ising phase diagram to the QCD phase diagram defined by Eqs.~(\ref{Eq:quadmap})-(\ref{Eq:TprimeExp}) is specified by the four parameters $\mu_c$, $\alpha_2$, $w$ and $\rho$ as long as the critical point is indeed chosen to lie along the chiral crossover curve.

We want to focus on the EoS close to the critical point and the effect of it on the hydrodynamic fluctuations and experimental observables, the cumulants of proton multiplicity. 
We shall compute the cumulants of proton multiplicity along a freezeout curve that we shall take to be parallel to the crossover curve
$T'=T_0$, shifted downward in temperature relative to it by the same amount $\Delta T$ at all $\mu$. We shall quote results for $\Delta T=4$, 6 and 9 MeV, which is to say that we shall  consider scenarios where the freezeout curve passes within only 4, 6 or 9 MeV below the critical point. We shall present more details with the cumulant results in Sec.~\ref{Sec:CumulantsPlots}. (We are assuming that the fluctuations are in equilibrium at freezeout; were we to analyze out-of-equilibrium fluctuations including ``memory effects'', the enhanced critical contributions to the cumulants of proton multiplicity would persist even if the time between when the fluid cools past the critical point and when it freezes out is longer.) 
Focusing on the critical fluctuations also allows us to ignore the subtleties associated with the descriptions of the regular part $P^{\rm reg}$ in this paper. That is, the only deviations from the  hadron resonance gas EoS that we shall take into account in this paper are those coming from the $P^{\text{sing}}$ part of the QCD EoS.

\subsection{Correlation Length Contours Mapped from Ising Model to QCD}
\label{Sec:CorrelationLength}

In this Subsection, we build our intuition for the mapping from the Ising phase diagram to the QCD phase diagram by illustrating where contours of constant values of the critical magnitude of the equilibrium correlation length $\xi$ land on the QCD phase diagram. 
Since $\xi$ diverges at the critical point, the constant-$\xi$ contours encircle the critical point, and it is conventional to refer to a ``critical region'' as the region on the phase diagram around the critical point where $\xi$ is greater than some specified value. The diverging correlation length is responsible for the divergence of thermodynamic susceptibilities and, more generally, for the enhancement of fluctuations in the critical region, with higher order fluctuations diverging more strongly as higher powers of $\xi$~\cite{Stephanov:2008qz}.
In this Subsection, we shall illustrate how the shape of the critical region in the QCD phase diagram depends on the nonuniversal parameters that specify the mapping from the Ising model to the QCD phase diagram.
This intuition will be very helpful in interpreting our results for the cumulants of particle multiplicity in Section~\ref{Sec:ME}, but it is important to note that we shall use the maximum entropy freeze-out procedure~\cite{Pradeep:2022eil} to calculate these cumulants directly from the critical equation of state, and will not need to 
evaluate $\xi$ per se. 
This represents an advance over previous 
studies~\cite{Stephanov:1999zu,Stephanov:2008qz,Athanasiou:2010kw} that have relied upon phenomenologically-motivated couplings between the scalar $\sigma$ field whose mass is the inverse of the correlation length to the
masses of protons and other hadrons to calculate such fluctuations.

We shall employ universality to calculate the contours of constant $\xi$ near the critical point in the Ising model, and then employ
Eq. \eqref{Eq:quadmap} concretely to determine the shape of these contours in the QCD phase diagram. 
The intuition that we shall gain by visualizing the effect of the non-universal  parameters in the mapping \eqref{Eq:quadmap} on the shape of contours of constant correlation length will enable us to understand the effects of these parameters on the hydrodynamic fluctuations that we shall compute in Section~\ref{Sec:ME}, starting from the equation of state.  
(The scaling form of the correlation length $\xi$
that we compute here is consistent with the scaling form of the equation of state.)
The scaling form of the correlation length in the Ising model can be written in terms of the parametric variables $R$ and $\theta$~\cite{ZinnJustin}, where $r=R(1-\theta^2)$ and where $h$ is given by $R^{\beta\delta}$ multiplied by an odd function of $\theta$~\cite{ZinnJustin,Guida:1996ep}, meaning that the critical point where $\xi$ diverges is at $R=0$. In terms of these variables, $\theta=\pm 1$ corresponds to the Ising $h$-axis, where $r=0$, and $\theta=0$ corresponds to the Ising $r$-axis, where $h=0$ and $r>0$.
The scaling form of $\xi$ is 
\begin{equation}
\label{ZinnJustin}
    \xi^2(r,h) = R^{-2 \nu} g_\xi(\theta), \quad
\end{equation}
where $\nu=0.63$ is the 3D Ising critical exponent for the correlation length and the scaling function $g_\xi(\theta)$,
which determines the $\theta$-dependence of the correlation length, is given to first order in the $\epsilon$-expansion by~\cite{ZinnJustin}
\begin{equation}
    g_\xi(\theta) = 1-\frac{5}{18}\epsilon \theta^2 + {\cal O}(\epsilon^2).
\end{equation}
Along the Ising $r$-axis where $\theta=0$, the scaling form \eqref{ZinnJustin} becomes $\xi^2(r,0)=r^{-2\nu}$. In the next Section, we shall employ the Ising equation of state calculated to ${\cal O}(\epsilon^2)$. For consistency, we need 
the 
correlation length and hence $g_\xi$ to the same order, which we have determined by extending the $\mathcal O(\epsilon)$ result in Ref.~\cite{ZinnJustin}, yielding
\begin{equation}
\label{xiOeps2}
    {g}_\xi(\theta) = 1-\frac{5}{18}\epsilon \theta^2 + \epsilon^2\Big[\frac{1}{972}( 24 I-25)\theta^2 + \frac{1}{324} ( 4 I + 41) \theta^4\Big]  +{\cal O}(\epsilon^3) ,
\end{equation}
where $I \equiv \int_0^1 \frac{\ln [x (1-x)]}{1 - x(1-x)}dx \approx -2.3439$.  We shall employ \eqref{xiOeps2} with $\epsilon$ set to $1$.

When mapping the correlation length to the QCD phase diagram, we introduce a parameter, $f_{\xi}$, such that the critical magnitude of the equilibrium correlation length in QCD is given by 
\begin{equation}\label{eq:fxi}
    \xi_{\mathrm{QCD}}= f_{\xi} \, \xi.
\end{equation}
Thus, $f_{\xi}$ sets the scale between the Ising correlation length $\xi$ and the one in QCD, $\xi_{\rm QCD}$.
Without a determination of the correlation length from first-principles, for example via lattice QCD methods, $f_{\xi}$ remains a free parameter.
This is an additional non-universal parameter (in addition to the non-universal mapping parameters introduced in Eq. \eqref{Eq:quadmap}) that is only relevant for the correlation length.
Our aim here is to visualize the effect of these non-universal parameters on the contours of constant $\xi_{\rm QCD}$ in the QCD phase diagram.

\begin{figure}[h!]
    \centering
    \includegraphics[width=0.98\linewidth]{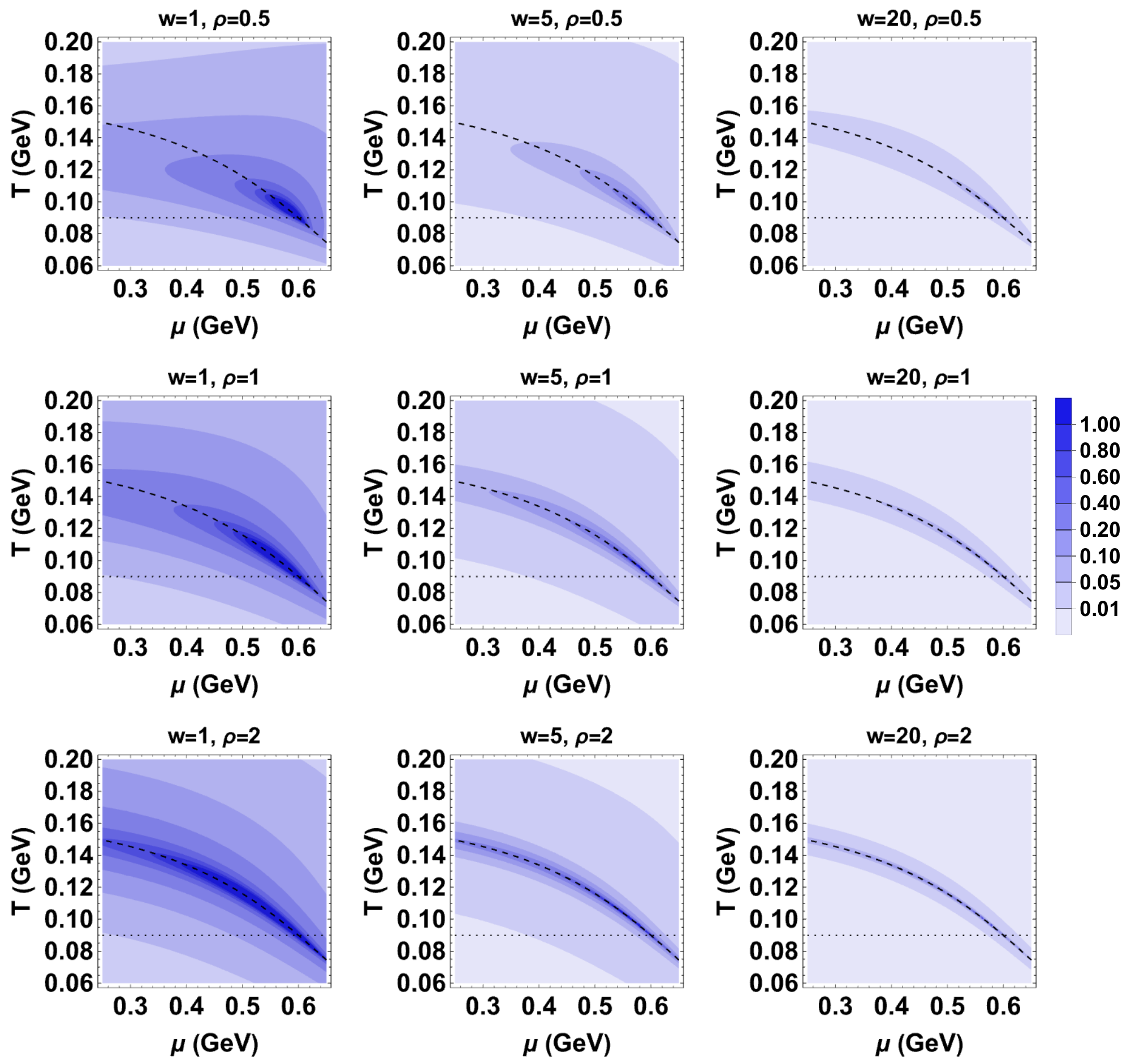}
    \caption{
    \justifying
    Contours of the critical magnitude of the squared correlation length scaled by $w^2$, $\xi_{\rm{QCD}}^2/w^2$ in units of $f_\xi$, for several values of the equation of state parameters $w$ and $\rho$. We have placed the critical point at $\mu_c=600$~MeV with a temperature $T_c=90$~MeV chosen so as to put the critical point on the chiral crossover curve estimated from lattice QCD calculations~\cite{Borsanyi:2020fev}. We have chosen the second mapping angle to be given by $\alpha_2=0^{\circ}$. The dotted line shows the Ising $h$-axis, while the dashed line shows the Ising $r$-axis which follows the chiral crossover line by construction.  
        }    \label{Fig:CorrLengthFixxi0}
\end{figure}

Based upon the prior work in Refs.~\cite{Stephanov:2008qz,Athanasiou:2010kw}, we expect the second moment of the proton number multiplicity distribution to be proportional to $g_p^2 (\xi_{\rm QCD})^{2-\eta}$ ($\eta\approx 0.036$ in the 3-dimensional Ising model, and as in Refs.~\cite{Stephanov:2008qz,Athanasiou:2010kw} we shall at times set it to zero for simplicity), where $g_p$ is a non-universal dimensionless parametrization of the coupling $\sigma \bar p p$ between the scalar order parameter $\sigma$ whose fluctuations diverge with correlation length $\xi_{\rm QCD}$ at the critical point and proton mass. The estimates in Ref.~\cite{Athanasiou:2010kw} relied upon making a rough estimate of the value of $g_p$ on phenomenological grounds. More recently, the authors of Ref.~\cite{Pradeep:2022eil} showed that the maximum entropy freezeout prescription predicts that $g_p$ depends on the non-universal parameter $w$ in the mapping from the Ising phase diagram to the QCD phase diagram that is related to the overall size of the critical region in the QCD phase diagram according to $g_p\propto 1/w$.  This observation is of qualitative importance: without noticing this, one would conclude from either the mapping \eqref{Eq:linmap} or \eqref{Eq:quadmap} that increasing $w$ would yield a larger critical region on the QCD phase diagram; the fact that $g_p \propto 1/w$ suggests the opposite.  To inform our expectations quantitatively, in
Fig.~\ref{Fig:CorrLengthFixxi0} we plot the behavior of the contours of $\xi_{\rm QCD}^2/w^2$ in units of $f_\xi^2$ in the QCD phase diagram for three different choices of the mapping parameter $w=\{1,5,20\}$ and three different choices of the mapping parameter $\rho=\{0.5,1,2\}$.
We have placed the critical point at $\mu_c=600$~MeV, and placing it on the chiral crossover line at this chemical potential corresponds to placing it at $T_c=90$~MeV with $\alpha_1=16.6^{\rm{o}}$. In this Figure, we have made the (arbitrary, but simplifying) choice $\alpha_2=0$. From Eqs.~\eqref{Eq:quadmap}, \eqref{eq:DT'-def} and \eqref{Eq:TprimeExp}, we see that with this choice when the Ising $h$-axis is mapped onto the QCD phase diagram it maps onto the horizontal line $T=T_c$  (and at the QCD critical point it is not perpendicular to the Ising-$r$ axis which follows the chiral crossover curve $\Delta T'=0$).
To get a better sense of the scale of the contours we have plotted in Fig.~\ref{Fig:CorrLengthFixxi0}, let us pick an arbitrary point, say $(\mu,T)=(250,90)$~MeV, far from the critical point (on the Ising $h$-axis, for simplicity). The critical contribution to the correlation length at that point is $\xi_{\rm QCD}=
    \{0.22,0.44,0.79\}$ in units of $f_\xi$ for $w=\{1,5,20\}$. If the value of $f_\xi$ is somewhere around $0.5$~fm (which is to say somewhere around the inverse $\sigma$-mass in vacuum) then, for all the values of $w$ we have chosen to plot in Fig.~\ref{Fig:CorrLengthFixxi0}, any critical contributions will in fact be small at the chosen point far away from the critical point, which is reasonable at a point where the physics is noncritical.

Our goal in Fig. \ref{Fig:CorrLengthFixxi0} is to train our intuition for how varying $w$ and $\rho$ changes the shape of the critical region.  
The contours of constant $\xi_{\rm QCD}^2/w^2$ are specified once these mapping parameters have been chosen.
Let us first focus on the shape of the contours.
We see that by increasing $\rho$ the contours become stretched along the transition line given by $h=0$.
On the other hand, increasing $w$ shrinks the contours around the transition line
as well as reducing the overall magnitude for these contours, indicating the importance of plotting  $\xi_{\rm QCD}^2/w^2$ rather than just $\xi_{\rm QCD}^2$. In Sect.~\ref{Sec:main} we shall compare our results for the second moment of the proton multiplicity distribution to $\xi_{\rm QCD}^2/w^2$ plotted in Fig.~\ref{Fig:CorrLengthFixxi0}.
From Refs.~\cite{Stephanov:2008qz,Athanasiou:2010kw},
we expect that the $k^{th}$ moment of the proton multiplicity distribution should scale like $g_p^k(\xi_{\rm QCD})^{-3+k(5-\eta)/2}$; so, plotting $\xi_{\rm QCD}^{9/2}/w^3$ and $\xi_{\rm QCD}^7/w^4$ would serve to shape our expectations for the third and fourth factorial moments of the proton multiplicity distribution. At a qualitative level, however, doing so does not add to the intuition 
that we have gained from Fig.~\ref{Fig:CorrLengthFixxi0} for how $w$ and $\rho$ control the shape and size of the critical region.

To further understand the relationship between the correlation length and the mapping parameters that we have visualized in Fig.~\ref{Fig:CorrLengthFixxi0}, we shall start from
the parametric expression \eqref{ZinnJustin} for the correlation length that determines the leading scaling behavior of $\xi$, and thus, by \eqref{eq:fxi} $\xi_{\rm QCD}$. 
Let us first consider the scaling behavior along the Ising $h$-axis, where $r=0$. With $\alpha_2=0$ as in Fig.~\ref{Fig:CorrLengthFixxi0}, this corresponds to
the horizontal dotted line depicted on the QCD phase diagrams there.
We can use the expression for $r=0$ obtained from \eqref{Eq:quadmap} to write the Ising variable $h$ along $r=0$ (the $h$-axis) as 
\begin{equation}\label{eq:h-dmu2-w}
h=-\cos\alpha_1\,\frac{\Delta T'}{T_c w \sin(\alpha_1-\alpha_2)}=\frac{\mu_c^2-\mu^2}{2\mu_c T_c  w \cos\alpha_2}
\end{equation}
from which we learn the scaling behavior of the correlation length $\xi_{\rm QCD}$ along this $\theta=1$ direction, 
which yields 
\begin{eqnarray}\label{eq:correlation-scaling-h-axis}
\frac{(\xi_{\rm QCD})^{2-\eta}}{w^2} &=& \frac{f_\xi^{2-\eta} g_\xi(1)^{1-\eta/2}}{w^2 R^{(2-\eta)\nu}} = \frac{f_\xi^{(2-\eta)} g_\xi(1)^{(1-\eta/2)}}{w^2}\left(\frac{h_0\tilde{h}(1)}{h} \right)^{(2-\eta)\nu/\beta\delta} \nonumber\\
&=& \frac{f_\xi^{2-\eta} g_\xi(1)^{(1-\eta/2)}}{w^2}\left(\frac{2\mu_c T_c w  h_0\tilde{h}(1)\cos\alpha_2}{\mu_c^2-\mu^2}\right)^{(2-\eta)\nu/\beta\delta}
\nonumber\\
&\propto & 
\frac{w^{-2+(2-\eta)\nu/\beta\delta}}{\left(\mu_c^2-\mu^2\right)^{(2-\eta)\nu/\beta\delta}} 
= \frac{1}{w^{1+1/\delta} \left(\mu_c^2-\mu^2\right)^{1-1/\delta}}\ ,
\end{eqnarray}
where we have used Eq.~\eqref{IsingEoS} from Appendix A and the scaling relations $\nu d = \beta(\delta+1)$ and $2-\eta=d(\delta-1)/(\delta+1)$.  If we set $\eta=0$ and $\nu=2/3$, which is a good approximation for the 3-dimensional Ising model, this yields $\delta=5$ and
\be
\frac{\xi_{\rm QCD}^2}{w^2} \propto \frac{1}{w^{6/5}(\mu_c^2-\mu^2)^{4/5}}\ ,
\ee
which is a good description of how the contours in Fig.~\ref{Fig:CorrLengthFixxi0} crossing the dotted $r=0$ axis compress toward the critical point as $w$ is decreased
and also confirms that -- along the dotted line -- $\xi_{\rm QCD}^2/w^2$ does not depend on $\rho$ as can be seen in Fig.~\ref{Fig:CorrLengthFixxi0}.

Next, we turn to the scaling behavior along the Ising $r$-axis, where $h=0$, which corresponds to the crossover curve on the QCD phase diagram. There, $\theta=0$ and $R=r$ and from Eqs.~\eqref{ZinnJustin} and \eqref{Eq:quadmap} we obtain
\begin{eqnarray}
\label{eq:xiscaling}
\frac{(\xi_{\rm QCD})^{2-\eta}}{w^2}  &=& 
\frac{f_{\xi}^{2-\eta}  g_\xi(0)^{1-\eta/2}}{w^2r^{(2-\eta)\nu}}
=\frac{f_{\xi}^{2-\eta}  g_\xi(0)^{1-\eta/2}}{w^2} \Big(\frac{2 \mu_c T_c \, \rho w \, \cos\alpha_1}{\mu_c^2-\mu^2}\Big)^{(2-\eta)\nu} \nonumber\\
&\propto&  \frac{1}{w^{2-(2-\eta)\nu}}\Bigg(\frac{\rho }{\mu_c^2-\mu^2} \Bigg)^{(2-\eta)\nu} ,
\end{eqnarray}
which becomes
\be
\frac{\xi_{\rm QCD}^{2}}{w^2} \propto \frac{1}{w^{2/3}}\Bigg(\frac{\rho }{\mu_c^2-\mu^2} \Bigg)^{4/3}
\ee
upon setting $\eta=0$ and $\nu=2/3$.
This confirms that the contours of constant $(\xi_{\rm QCD})^2/w^2$ in Fig.~\ref{Fig:CorrLengthFixxi0} should stretch out along the chiral crossover curve as $\rho$ is increased and should compress along the crossover curve as $w$ is increased, as is indeed manifest in Fig.~\ref{Fig:CorrLengthFixxi0}.

In Section~\ref{Sec:main} we shall utilize the 
perspectives from this Section on how changing the EoS mapping parameters changes the contours of the correlation length to help us to interpret the behavior of thermodynamic fluctuations just before freezeout and (via maximum entropy freezeout) the particle multiplicity fluctuations just after freezeout.
To complete our preparations, in the next Subsection we shall review the universal behavior expected for various different hydrodynamic fluctuations.

\subsection{Hydrodynamic Correlations in Equilibrium}
\label{Subsec:Universal}
Fluctuations of the hydrodynamic quantities such as the energy density and
the baryon number density  are characterized by correlation functions. In equilibrium, these correlators are local on the hydrodynamic scales (longer than $\xi$) and their magnitudes are completely determined by the equation of state, i.e., pressure as a function of temperature and chemical potential $P(T,\mu)$. Taking derivatives of $\beta P V$ (which is the logarithm of the partition function, where $V$ is the hydrodynamic cell volume) with respect to $\beta\equiv1/T$ and $\hmu\equiv\mu/T$ treated as independent variables, one obtains
\be
\label{eq:ExplicitCorrelationsAll}
\langle (\delta\e)^{j}(\delta n)^{k-j} \rangle_{\rm connected} &=&
\frac{(-1)^{j}}{V^{k-1}}\frac{\partial^{k}(\beta P)}{\partial\beta^{j}\partial\hmu^{k-j}}\,.
\label{Eq:en}
\ee
For example, with $j=0$ and $k=4$ what we are computing is ``connected'' in the sense that the RHS is in that case an expression for $\langle(\delta n)^4\rangle_{\rm connected}=\langle(\delta n)^4\rangle-3\langle(\delta n)^2\rangle^2$.  The factor of $(-1)^j$ on the RHS of Eq.~\eqref{Eq:en} arises because we are differentiating $j$ times with respect to $-\beta$. 
The $1/V^{k}$ in Eq.~\eqref{Eq:en} makes the hydrodynamic correlators that we compute intensive quantities, as can be seen if we rewrite Eq.~\eqref{Eq:en} as
\be
\label{eq:ExplicitCorrelationsV}
\langle (\delta\e V)^{j}(\delta n V)^{k-j} \rangle_{\rm connected} &=&
{(-1)^{j}}\frac{\partial^{k}(\beta PV)}{\partial\beta^{j}\partial\hmu^{k-j}}\,.
\ee
Here the LHS is clearly extensive and, since $\e V$ and $nV$ are extensive and a cumulant of an extensive quantity is extensive, so is the RHS. 
Finally, we note that in all subsequent expressions, we shall drop the subscript ``connected"; we shall always be computing the connected correlations (or, later, combinations thereof).

Similar to the decomposition of regular and singular contributions to $P$ in Eq.~(\ref{Eq:EoS})
, we can separate the regular parts (that we shall evaluate using the hadron resonance gas equation of state) and the singular parts (that we shall evaluate via mapping the Ising EoS onto the QCD phase diagram) of the connected hydrodynamic correlations at the critical point in the equation.
The most singular part of the connected hydrodynamic correlations is given by
\be
\label{eq:ExplicitCorrelations}
\langle (\delta\e)^{j}(\delta n)^{k-j} \rangle^{\rm sing} &=&
\frac{(-1)^{j}}{V^{k-1}}\frac{\partial^{k}(\beta_c P^{\text{sing}})}{\partial\beta^{j}\partial\hmu^{k-j}}\,,
\ee
where $\beta_c\equiv 1/T_c$. (In the calculation of the most singular contribution of the $k^{th}$ order hydrodynamic correlations via mapping QCD to the 3D Ising model, only the terms in which all of the derivatives $\partial^k$ act on the Gibbs free energy of the 3D Ising model contribute.) 
The identification of the {\em singular} part of a quantity such as pressure or a correlator in Eq.~\eqref{eq:ExplicitCorrelations} is inherently ambiguous. The ambiguity is the contribution of a regular part or a subleading singularity. In this paper we focus on {\em estimating} the magnitude of the singular contributions,
and therefore will choose the definition of the singular part of the hydrodynamic (and later the particle multiplicity) correlations which simplifies our calculations. For example, the appearance of $\beta_c P^{\rm sing}$ instead of $\beta P^{\rm sing}$ in Eq.~\eqref{eq:ExplicitCorrelations} reflects such a choice. At the critical point, the corresponding ambiguity contributes at a lower order in $\xi$ as $\xi$ diverges (lower by a factor $\xi^{-\beta\delta/\nu}\approx\xi^{-2.5}$). 
Similarly, when the derivatives in Eq.~\eqref{eq:ExplicitCorrelations} act on $P^{\rm sing}$ we shall not include contributions that arise from the nonlinearities in the mapping functions $h(\mu,T)$ and $r(\mu,T)$ defined in Eq.~\eqref{Eq:quadmap}. In this case also, deciding whether or not to include such contributions represents an ambiguity at a lower order in $\xi$, again lower by $\approx\xi^{-2.5}$.
It is important to keep in mind that such ambiguities are not an inherent limitation of the maximum entropy approach, which can handle the full equation of state (and even non-equilibrium effects on the correlators) without separating  singular and regular parts. We shall defer such a more precise, complete, and difficult calculation to future work and focus here on estimating the magnitude of the effects originating from critical fluctuations.

In this paper, we will use the notation $H_{a_1\dots a_k}$ to represent the $k$-point connected correlation functions of some list of $k$ hydrodynamic variables as follows:

\be \label{Eq:Hdef}
H_{a_1\dots a_k}\equiv 
\langle\delta \psi_{a_1}\dots \delta \psi_{a_k}\rangle 
=
\langle\delta \psi_{a_1}\dots \delta \psi_{a_k}\rangle_{\rm reg}
+
\langle\delta \psi_{a_1}\dots \delta \psi_{a_k}\rangle_{\rm sing}
\ .
\ee
Here, each $\delta\psi_a$ is a hydrodynamical field that is a function of position ---
$\delta\epsilon({\bf x})$ when $a=\e$ and $\delta n({\bf x})$ when $a=n$. We shall take the labels $a$ to be collective labels that incorporate the position variable that specifies the location of the fluid cell in the fireball as well as the index that identifies whether the field is $\delta\e$ or $\delta n$. 
We have also noted that
each correlation function $H_{a_1 \dots a_k}$ has a regular part and a singular part. 
In our simplified description of the regular part of QCD EoS, we shall assume that 
\be\label{Eq:barHreg}
\langle\delta \psi_{a_1}\dots \delta \psi_{a_k}\rangle_{\rm reg}=
\bar{H}_{a_1\dots a_k} \ ,
\ee
where $\bar{H}$ is the relevant thermodynamic correlation function in the hadron resonance gas, as we describe in more detail in Section~\ref{Sec:ME}. The derivatives of $P^{\text{sing}}$ that serve to define the singular contributions to each correlation function divergence at a critical point with power law exponents dictated by the 3D Ising universality class~\cite{Stephanov:2008qz}. Near the critical point, the hydrodynamic correlations scale (i.e.~diverge) as power laws in the correlation length as follows: $\<(\delta \e)^{j}(\delta n)^{k-j}\>^\text{sing}\sim \xi^{-3+k(5-\eta)/2}$ \cite{Stephanov:2008qz}. 
In this paper, we will study how these characteristic divergences show up in the cumulants of particle multiplicities, and shall do so quantiatively upon assuming that the freezeout from hydrodynamic fluctuations to particle multiplicity fluctuations occurs in thermal equilibrium.

\section{Cumulants of Particle Multiplicity Using the Maximum Entropy Procedure}
\label{Sec:ME}

The freeze-out hypersurface is the space-time surface 
at which the expanding cooling droplet of matter produced in a heavy ion collision has become so dilute that beyond this hypersurface no multiplicity-changing collisions occur and the particle multiplicity distributions
are ``frozen''.
Throughout this paper, we shall assume that
we can use hydrodynamics to describe the system all the way up to the freeze-out hypersurface. Beyond the freeze-out hypersurface, we shall describe the system in terms of particles.  That is, for simplicity we shall assume throughout that the particlization hypersurface is the same as the freeze-out hypersurface.  
Furthermore, as we have noted at many points already,
we shall assume that the system is in thermal equilibrium at the freeze-out/particlization hypersurface.
This means that the {\it mean} mid-rapidity yields of the particle multiplicities are specified according to the Cooper Frye freezeout prescription~\cite{Cooper:1974mv} by the values of the temperature and the baryon chemical potential at freezeout, which can then be obtained from data via using the statistical hadronization model \cite{Andronic:2017pug,Vovchenko:2019pjl,Alba:2020jir}. 
We assume that the cumulants of the particle multiplicities also get frozen on the same freeze-out hypersurface. After freeze-out, there are no further multiplicity-changing processes in the hadron resonance gas. 
(It does include heavier mass resonances that subsequently decay to stable or long-lived particles.) 
The hadron gas that is ``born" when the hydrodynamic fluid particlizes and freezes out near a critical point has multiplicity fluctuations in it that originate from the hydrodynamic fluctuations just before freezeout. When we wish
to distinguish it from the conventional hadron resonance gas, we shall refer to this hadron resonance gas that includes non-trivial multiplicity fluctuations and correlations imprinted on it due to the prior hydrodynamic evolution as the \textit{correlated} hadron resonance gas. 
We shall reserve the abbreviation HRG for the conventional uncorrelated hadron resonance gas that is usually employed in the literature, as appropriate when freezeout occurs in equilibrium far from any critical point.  
In this Section, we shall develop and employ the maximum entropy freeze-out procedure to fix the factorial cumulants of the fluctuating particle multiplicities just after freeze-out.  We shall often quote results for the difference between these non-trivial cumulants
and the cumulants in the uncorrelated HRG. 

We shall develop a formalism that could be employed to describe freezeout at a realistically shaped freezeout hypersurface based upon a realistic hydrodynamic simulation of a heavy ion collision. In all the calculations that we do in this paper, however, we shall instead employ the (standard) simplified hydrodynamical solution, known as Bjorken expansion, that describes a fluid with boost invariant expansion in the $z$ (``beam'') direction that is uniform in
the $x$ and $y$ (transverse) directions.  In this case, freezeout occurs at a  hypersurface with a constant value of the proper time $\tau\equiv\sqrt{t^2-z^2}$ that we shall denote by $\tau_f$ at which the fluid has cooled to a temperature $T_f$. Boost invariance in the beam direction and uniformity in the transverse directions mean that freezeout occurs at the same $\tau=\tau_f$ at all values of $x$, $y$ and spacetime rapidity $\eta \equiv \frac{1}{2} \ln[(t+z)/(t-z)]$.
Note also that, although for calculational simplicity we shall not do so in this paper, it is conceptually straightforward to introduce acceptance cuts on the momenta of the particles produced at freezeout~\cite{Ling:2015yau}.

\subsection{Hydrodynamic and Particle Multiplicity Correlators}

At freeze-out (particlization), there are 
two descriptions of strongly-interacting matter --- one is the hydrodynamic description in terms of the conserved densities
and the fluid velocity 
and the other is the description in terms of particle distribution functions in phase space (denoted by $f_A\equiv f_{\hA}(x_A,p_A)$, where $\hA$ refers to the hadron species) of the hadron resonance gas. 
There is much more information in the particle description, which includes distribution functions for every hadron species. 
At minimum, the two descriptions must agree with one another to the extent that they agree on the values of the conserved densities.  That is, freeze-out must respect conservation laws. 
The idea behind the maximum entropy freezeout procedure is that the additional information associated with fully specifying the hadron resonance gas distribution functions for every hadron species (additional in the sense that it goes beyond what is known via conservation of conserved densities across the freezeout) 
should be minimized -- which is to say the relative entropy (the entropy of the fluctuations relative to the entropy of a hadron resonance gas) should be maximized.\footnote{
   Here the term `relative entropy' refers to a negative quantity -- the reduction of the entropy due to the constraints on the fluctuations. The {\em negative} of this quantity, which is thus minimized in the maximum entropy approach, is known as information for discrimination~\cite{Kullback1951OnIA}, information gain, or $I$-divergence~\cite{Csiszar1975}. Unfortunately, and confusingly, the $I$-divergence is also often called relative entropy in the literature.}
Applying this logic only to the mean values of the conserved densities (neglecting their fluctuations) yields the Cooper-Frye freezeout prescription~\cite{Cooper:1974mv},
in which the hadron distributions just after freezeout have boosted Bose-Einstein or Fermi-Dirac distributions 
specified via the $T$ and $\mu$ and fluid velocity just before freezeout. The maximum entropy freeze-out procedure introduced in Ref.~\cite{Pradeep:2022eil} extends this logic to circumstances where we are interested in the
{\em fluctuations} of conserved densities, and want to ensure that information about these fluctuations is preserved
through freezeout. That is, the maximum entropy procedure ensures that freezeout respects conservation laws in each individual event, not just on average.  
By utilizing the maximum entropy method in this context, we can extract the information about the particle multiplicity fluctuations after freezeout, taking the equation of state in the hydrodynamic description as input~\cite{Pradeep:2022eil}. 
We briefly review the maximum entropy freeze-out method in this Section.

In the local rest-frame of the fluid, the matching of the energy and baryon densities between the hydrodynamic description in terms of the hydrodynamic fields $\epsilon$ and $n$ (which we shall denote by $\psi_\e\equiv\epsilon$ and $\psi_n\equiv n$ or, generically, by $\psi_a$ where $a\in\{\e,n\}$)
and the particlized hadron resonance gas description on an event-by-event basis (which is to say including fluctuations) can be expressed via the following integral of the occupation number $f_{\hA}(\bm x_A,\bm p_A)$ over all single particle states: 
    \be\label{Eq:conEN}
\psi_{a}({\bm x})=\int_A f_{\hA}({\bm x}_A,{\bm p}_A) P^A_a({\bm x})\ ,
\ee
where $\int_{A}$ denotes integration over the phase space variables $x_A$ and $p_A$ of the particles and summation over all hadronic species in the particle description, namely
\begin{equation}\label{Eq:intA}
    \int_A\equiv
    \int d^3\bm x_A
    \int d^3\left(\frac{\bm p_A}{2\pi}\right)
\sum_{\hA}
\end{equation}
where the single particle states labeled by a composite index $A$ are characterized by the position $\bm x_A$, momentum $\bm p_A$ as well as discrete particle characteristics (quantum numbers) such as mass $m_A$,  (baryon)
charge $q_A$, spin $s_A$, etc. collectively denoted by $\hA$ in Eqs.~\eqref{Eq:conEN}, and \eqref{Eq:intA}. The phase space integral in Eq.~\eqref{Eq:intA} represents the Lorentz invariant measure for integration over the freezeout hypersurface written in terms of the local tangent space variables, $\bm x$ and $\bm p$, namely 
\be
\label{eq:d3xd3p}
d^3\bm x\, d^3\bm p = 
p^\mu
d^3\Sigma_\mu
\, d^4 p \, \delta(p^2-m^2)\,2\theta(p\cdot n)
= \tau_{\rm f} \, d\eta \, dx_\perp^2\, m_\perp\cosh(\eta-\eta_p)\,d\eta_p\,dp_\perp^2
\ ,
\ee
where $ d^3\Sigma_\mu\equiv \varepsilon_{\mu\alpha\beta\gamma}(\partial x^\alpha/\partial\theta_1) (\partial x^\beta/\partial\theta_2)(\partial x^\gamma/\partial\theta_3)d\theta_1d\theta_2d\theta_3$ is the hypervolume for a freezeout hypersurface defined parametrically as $x=x(\theta_1,\theta_2,\theta_3)$ and 
$p\cdot n$ is the projection of the particle 4-momentum onto the 4-vector $n_\mu$ normal to the freezeout surface, i.e., $n_\mu\sim d^3\Sigma_\mu$. 
The second equality applies when we consider freezeout at Bjorken coordinate time $\tau=\tau_{\rm f}$ on a hypersurface parameterized by Bjorken coordinates $\{\theta_i\}=\{\eta, x, y\}$, from a fluid undergoing boost invariant expansion,
where $m_\perp^2\equiv m^2+p_\perp^2$, $\eta_p\equiv  \frac{1}{2} \ln[(p_0+p_z)/(p_0-p_z)]$ is the momentum-space rapidity,
and where we have omitted the particle index $A$ throughout for simplicity.
In Eq.~\eqref{Eq:conEN}, we have also introduced the collective notation $P^{A}_{a}$, as before with $a\in\{\e,n\}$,  to denote the contribution of each occupied particle state to the corresponding conserved quantity in the hydrodynamic cell at position $\bm x$:
\be \label{Eq:P}
P_{\e}^{A}({\bm x})\equiv E_A\, \delta^3({\bm x}-{\bm x}_A)\, , \qquad P_{n}^{A}({\bm x})\equiv q_A \,\delta^3({\bm x}-{\bm x}_A)\,
\ee
When $a=\epsilon$, Eq.~(\ref{Eq:conEN}) is the matching condition for the energy densities between the hydrodynamic and particlized descriptions, where $E_A$ is the energy of the particle state $A$ in the rest frame of the hydrodynamic cell. The three-dimensional delta function is defined as a density normalized to unity when integrated over the freezeout hypersurface. When $a=n$, Eq.~(\ref{Eq:conEN}) represents the matching condition for the baryon densities. In Eqs.~\eqref{Eq:conEN} and \eqref{Eq:P}, ${\bm x}$ is an arbitrary spacetime point on the freezeout surface. 
Note that if we were to boost from the local fluid rest-frame back to the lab frame, we would obtain conditions matching from the momentum, energy, and baryon densities (conserved quantities which, in the standard Cooper-Frye freezeout, are related to local fluid velocity, temperature, and chemical potential) to appropriately boosted integrals over the distribution functions.

Let $\<\psi_a\>$  (here again, the subscript $a$ denotes the identity of the hydrodynamic variable as well as the coordinates of the fluid cell) and $\<f_A\>$ 
be the event-by-event average of the hydrodynamic variables $\psi_a$ and the particle distribution functions $f_A\equiv f_{\hA}(\bm x_A,\bm p_A)$, respectively. And, let $\delta \psi_a\equiv \psi_a-\langle \psi_a\rangle$ and $\delta f_A\equiv f_A-\langle f_A\rangle$ denote the fluctuations from their mean values. 
The correlation functions of the particle multiplicities in phase space 
for the particlized 
description of the system just after freezeout, including particle multiplicity correlations coming from the critical hydrodynamic fluctuations just before freezeout as well as non-critical multiplicity correlations,
are denoted by $G$, with subscripts denoting collectively the species index and the phase space variables, i.e.,
\be\label{Eq:G}
G_{A_1\dots A_k}\equiv\langle\delta f_{A_1}\delta f_{A_2}\dots \delta f_{A_k}\rangle\,.
\ee
The matching of conserved densities on an event-by-event basis can then be represented by equating their means, as in traditional Cooper-Frye freezeout~\cite{Cooper:1974mv}, and correlation functions, via the maximum entropy procedure that we shall describe. The matching conditions between the hydrodynamic correlation functions defined in Eq.~\eqref{Eq:Hdef} and the particle multiplicity correlation functions defined in Eq.~\eqref{Eq:G} can be written explicitly as follows:
\be\label{Eq:H}
H_{a_1\dots a_k}&=&
\int_{A_1\dots A_{k}}G_{A_1\dots A_k}P_{a
_1}^{A_1}\dots P_{a_k}^{A_k}\,,
\ee
where the subscripts in lower-case letters correspond to the hydrodynamic variables/indices, e.g. $\epsilon$ and $n$, as well as the spatial coordinates of the fluid cell, and where
the subscripts and superscripts in upper-case letters correspond to hadron gas variables/indices, e.g. proton with position ${\bm x}_p$ and momentum ${\bm p}_p$.
Generalizing Eq.~\eqref{Eq:intA}, the condensed integral notation subsumes all position ${\bm x}$ and momentum ${\bm p}$ integrals and a sum over the hadronic species $A$:
\be \label{IntA}
\int_{A_1 \dots A_k} \equiv \int_{{\bm x}_{A_1}} \dots \int_{{\bm x}_{A_k}} \int_{{\bm p}_{A_1}} \dots \int_{{\bm p}_{A_k}} \sum_{\hA_1 \dots \hA_k}.
\ee

We denote the correlation functions in the uncorrelated HRG with a bar over the quantity.
For example, the $G_{A_1\dots A_{k}}$ in the HRG will be denoted by $\bar{G}_{A_1\dots A_{k}}$. 
Similarly, the matching conditions shown in Eq. \eqref{Eq:H} become:
   \be \label{Eq: BarH}
\bar{H}_{a_1\dots a_k}&=&
\int_{A_1\dots A_{k}}\bar{G}_{A_1\dots A_k}P_{a_1}^{A_1}\dots P_{a_k}^{A_k}\ .
\ee
In the absence of any critical fluctuations, the correlations in the hydrodynamic densities obtained using Eqs.~(\ref{Eq:Hdef}),
would be equivalent to the correlations of the same order in the HRG, i.e $H=\bar{H}$ at all orders.  This simplification corresponds to our assumption that the regular part of the EoS is simply that of the HRG.
The correlations that we are interested in, namely correlations in the correlated hadron resonance gas that originate from critical fluctuations of thermodynamic quantities, are denoted by $\Delta H$, with 
   \be\label{Eq:DelH}
\Delta H_{a_1\dots a_{k}}\equiv H_{a_1\dots a_{k}}-\bar{H}_{a_1\dots a_{k}}\ ,
\ee
where $k$ refers to the order of the correlation function. 
Since the regular contribution to the equation of state is simply that of the HRG, the contributions $\Delta H$ to the correlation functions originate entirely from the singular contribution to the equation of state. Upon going beyond our simplifying approximation, when the regular part of the pressure deviates from that of the hadron resonance gas, $\Delta H$ (which is still given by Eq.~(\ref{Eq:DelH})) will have nonzero contributions which are subleading relative to the critical contribution at the critical point.
Similarly, let us define
   \be\label{Eq:DelG}
   \Delta G_{A_1\dots A_{k}} \equiv G_{A_1\dots A_{k}}-\bar{G}_{A_1\dots A_{k}}
   \ee
where $\bar G$ are the correlations in the HRG.

We shall follow Ref.~\cite{Pradeep:2022eil} 
and introduce correlation measures which are combinations of
the $\Delta G$'s (or $\Delta H$'s) that serve to quantify the contribution to the correlations in a hadron gas (or hydrodynamic fluid) at a given order that are not composed from lower order correlations. These are 
referred to as the irreducible relative cumulants (IRCs)~\cite{Pradeep:2022eil}, and are denoted by  $\widehat{\Delta}G$'s (or $\widehat{\Delta}H$'s).  The second, third and fourth irreducible relative cumulants  of the particle multiplicities are defined as follows:
 \bes \label{Eq:IRCG} 
\be \label{Eq:IRCG2}
\widehat\Delta G_{AB}&=&\Delta G_{AB}\\  
\label{Eq:IRCG3}
\widehat{ \Delta}  G_{ABC} &\equiv & \[\Delta G_{ABC}-3\widehat\Delta G_{AD}\bar{G}^D_{BC}\]_{\overline{ABC}}
\\ 
\label{Eq:IRCG4}
\widehat{\Delta}  G_{ABCD} & \equiv &
\[\Delta G_{ABCD}-6\widehat{\Delta} G_{ABF}\bar{G}^F_{CD}-4\widehat{\Delta} G_{AF}\bar{G}^F_{BCD}-3\widehat{\Delta}G_{EF}\bar{G}^E_{AB}\bar{G}^F_{CD}\]_{\overline{ABCD}}
\ee
\ees  

Here the indices are ``raised" by the inverse of the two-point correlator in the uncorrelated HRG, $(\bar G^{-1})^{AB}$, as in $\bar G^D_{BC}\equiv  (\bar G^{-1})^{DK}\bar G_{KBC}$
and $
\bar G^F_{BCD}
\equiv\(\bar{G}^{-1}\)^{FK}\bar{G}_{KBCD}$.
Each pair of matching upper and lower indices implies summation over particle quantum numbers and integration over the particle's phase space variables.  For example, a repeated $K$ index denotes $\int_K$, defined as in Eq.~\eqref{Eq:intA}.
Finally, as in Ref.~\cite{Pradeep:2022eil}, we use the notation
$\[\dots\]_{\overline{ABCD}}$ for the average over the permutations of indices. 
Note that the IRCs are constructed iteratively: each line in Eq.~\eqref{Eq:IRCG} serves to define $\widehat{\Delta}G$ at a specified order in terms of $\Delta G$ at that order and $\widehat{\Delta}G$ at all lower orders.

One can similarly define the hydrodynamic irreducible relative cumulants, as follows:
 \bes\label{eq:hatH}
\be\label{eq:hatH2}
   \widehat\Delta H_{ab} &\equiv & \Delta H_{ab}\,;\\ \label{eq:hatH3}
    \widehat\Delta H_{abc}&\equiv &\Big[\Delta H_{abc}
    - 3\widehat\Delta H_{ad} \bar H^d_{bc}\Big]_{\overline{abc}}
    \\   
   \widehat\Delta H_{abcd}  &\equiv&\Big[\Delta H_{abcd}
    - 6 \widehat\Delta H_{abf} \bar H^{f}_{cd}\,
      -  4\widehat\Delta H_{af}\bar H^{f}_{bcd}    - 3\widehat\Delta H_{ef}\bar H^{e}_{ab}
   \bar H^{f}_{cd} \Big]_{\overline{abcd}}\, , \label{eq:hatH4}
\ee
\ees
employing notation that is analogous to that introduced in Eq.~\eqref{Eq:IRCG}. Here,
each pair of matching sub- and superscript lower-case
indices implies summation over the hydrodynamic 
density labels $\e$ and $n$ and integration over the fluid cells at the freezeout hypersurface with the integration
measure $\tau_f \, d\eta \,dx_\perp^2$. 
Note that here the indices are raised by the inverse of the two point thermodynamic correlator in the uncorrelated HRG, $(\bar{H}^{-1})^{ab}$, as in $\bar{H}^{d}_{bc}\equiv \(\bar{H}^{-1}\)^{de}\bar{H}_{ebc}$  and $\bar{H}^{f}_{bcd}\equiv \(\bar{H}^{-1}\)^{fe}\bar{H}_{ebcd}$. 
(We note parenthetically that the integration over the freezeout hypersurface is trivial in both Eqs.~\eqref{Eq:IRCG} and \eqref{eq:hatH}
due to the locality of the hydrodynamic and kinetic correlators in equilibrium, but the equations are more general and apply to correlators out of equilibrium which are not local.) 

At second order, the irreducible relative cumulants are the same as their ordinary connected counterparts. 
As is evident from Eqs.~\eqref{Eq:IRCG3} and~\eqref{Eq:IRCG4} or Eqs.~\eqref{eq:hatH3} and~\eqref{eq:hatH4}, 
this is not the case for higher orders. The physical interpretation of $\widehat{\Delta}G$ and $\widehat{\Delta}H$ begins from the observation that, according to their definitions \eqref{Eq:DelG} and \eqref{Eq:DelH}, these quantities vanish in the uncorrelated HRG. 
Nonzero values of $\widehat{\Delta}G$ imply the presence of non-trivial correlations in the hadron gas, 
arising from deviations of the EoS from that of the HRG, or critical fluctuations, or non-equilibrium dynamics in the fluid just before freezeout -- any of which would be encoded 
in nonzero values of $\widehat{\Delta}H$.

Upon making the assumptions that we are making in this paper, 
namely that just before freezeout the hydrodynamic fluid is in thermal equilibrium, with the equilibrium critical fluctuations being the only source of deviations of the EoS from that of the HRG, we
can now use the maximum entropy procedure to translate the hydrodynamic IRCs $\widehat \Delta H$ into IRCs of particle multiplicity distributions $\widehat \Delta G$. 
Eq.~\eqref{Eq:H} specifies necessary conditions that the matching 
from the hydrodynamic $\widehat \Delta H$ to the particle multiplicity $\widehat \Delta G$ must satisfy at freeze-out. However, 
although Eq.~\eqref{Eq:H} specifies the $H$ that corresponds to any given phase space distribution $G$, for any specified $H$ 
there are infinitely many solutions $G$ to these matching conditions.
That is, Eq.~\eqref{Eq:H} by itself does not suffice for our purposes. 
The authors
of Ref.~\cite{Pradeep:2022eil} have used the prescription that the
kinetic theory distribution functions for the hadrons immediately after 
freezeout should be chosen from among the space of possibilities 
that satisfy Eq.~\eqref{Eq:H} in such a way as to maximize the entropy of the correlated hadron gas to derive the matching condition

\be\label{Eq:IRCGAsToIRCH}
	\hat{\Delta} G_{A_1\dots A_k}=\hat{\Delta}H_{a_1a_2\dots a_n} P^{a_1}_{A_1} P^{a_2}_{A_2}\dots P^{a_n}_{A_k}\ ,
	\ee	
    where, consistent with the notation that we have introduced above, we have defined
\begin{equation} \label{eq:P-lowerA-uppera-defn}   
    P^{a}_{A}\equiv (\bar{H}^{-1})^{ab}P^B_b\bar{G}_{BA}\,.
\end{equation}
 The central result \eqref{Eq:IRCGAsToIRCH} from Ref.~\cite{Pradeep:2022eil} is the generalization
of Cooper-Frye freezeout~\cite{Cooper:1974mv} that 
maps (critical) fluctuations in the hydrodynamic fluid just before freezeout to correlations among the hadrons just after freezeout, in a way that faithfully respects all conservation laws and minimizes additional information. 
In Eq.~\eqref{Eq:IRCGAsToIRCH}, as before, $\hat{\Delta}$ refers to irreducible relative cumulants of either the hydrodynamic fluctuations (H) or the hadron gas phase space densities (G). The sub(super)scripts denoted by lower-case letters refer to the hydrodynamic variables 
$(\e(x),n(x))$ 
at a space-time coordinate $x$. 
The subscripts denoted by upper-case letters collectively refer to the quantum numbers and phase space coordinates of the particles in the hadron resonance gas, as already discussed at the beginning of this Section where we introduced Eq.~\eqref{Eq:H}. 
And, the repeated upper-case indices inside
each $P_A^a$ imply summation over hadron species and spins as well as integrations over hadron momenta and over the freezeout hypersurface.
Also as above, in Eq.~(\ref{Eq:IRCGAsToIRCH}) 
the repeated lower-case indices imply summation over hydrodynamic fields as well as integrations over the freezeout 
hypersurface. 
The quantities with overbars refer to the correlations that arise in the HRG with appropriate quantum statistics. 
We wish to emphasize that although in this
first numerical demonstration of the maximum entropy freeze-out prescription that we present in this paper we have made simplifying assumptions concerning $P^{\text{reg}}$ and have assumed
that the critical fluctuations are in thermal equilibrium at freezeout, the maximum entropy freeze-out prescription itself, as
derived in Ref.~\cite{Pradeep:2022eil} and reviewed in Section.~\ref{Sec:ME}, is general enough to be applicable for any form of the regular EoS and also in situations where non-equilibrium effects are important.

Note that the IRCs $\widehat\Delta G_{A_1\dots A_k}$ 
are correlations in $6k$-dimensional phase space.  In the next Subsection, we shall integrate over all of the phase space variables, so as to obtain irreducible relative cumulants of particle multiplicities.  We shall also see how these quantities are related to the factorial cumulants of particle multiplicities, quantities that (for protons in particular) have recently been measured by STAR~\cite{STAR:2025zdq}.

\subsection{Factorial Cumulants of Particle Multiplicities}
\label{Sec:FactorialCumulants}

Let us denote the total multiplicity of particle species $\hA$ in a specified acceptance window in an event by $N_{\hA}$. (To make contact with any particular analysis of experimental data, the acceptance window in the definition of $N_{\hA}$ -- and consequently of all the quantities that follow below -- should be chosen as in the 
experimental analysis.)
Let $\<N_{\hA}\>$ be the event-averaged, or mean, multiplicity of $\hA$.  Event-by event fluctuations of the particle multiplicity, are then denoted by $\delta N_{\hA}\equiv N_{\hA}-\<N_{\hA}\>$. In the literature, $\omega_{k\hA}$ has been used to denote the $k^{th}$ cumulant (ordinary cumulant; not factorial cumulant) of the multiplicity of species $\hA$, namely $\<(\delta N_{\hA})^{k}\>_{\rm connected}$, divided by its mean multiplicity, $\<N_{\hA}\>$. As in Ref.~\cite{Athanasiou:2010kw}, we generalize this notation, defining $\omega_{\hA_{1}\dots \hA_{k}}$ so as to describe mixed cumulants of the multiplicities of one or more particle species, normalized by suitable powers of mean-multiplicities, $\<N_{\hA_i}\>$. That is, we define
\be\label{Eq:Omegak}
\omega_{\hA_{1}\dots \hA_{k}}\equiv \frac{\<\prod^{k}_{i=1}\delta N_{\hA_i}\>_{\!\rm connected}}{\prod^{k}_{i=1}\<N_{\hA_i}\>^{1/k}}
\ee
where 
\be\label{Eq:DelNA}
\<\prod^{k}_{i=1}\delta N_{\hA_i}\>_{\rm connected}=
\int_{\tA_1\dots \tA_k}G_{A_1\dots A_k}
\ee
and where the integration is performed over the phase space coordinates of particles whose multiplicity is actually being counted, which in the case of protons will include the sum over spins, namely:
\begin{equation}
    \int_{\widetilde A}\equiv \int_{\bm x_A}\int_{\bm p_A} \sum_{s_A}
    \label{eq:tA}
\end{equation}
The tilde on $\widetilde A_i$  serves to distinguish the notation from Eq.~(\ref{Eq:intA}), as here the discrete summation is only over the spins of hadrons of a single species, whereas in Eq.~\eqref{Eq:intA} the summation over 
the hadronic species index $\hA$ is also included.  Here, we are only integrating over phase space variables (and summing over spins), not summing over species. 
Next, we can define 
\be\label{eq:Delta-omega-Ahats}
\Delta\omega_{\hA_1\dots \hA_{k}} \equiv \omega_{\hA_1\dots \hA_{k}}-\bar\omega_{\hA_1\dots \hA_{k}}\,,
\ee
where the $\bar\omega$'s are the same normalized cumulants
as the $\omega$'s evaluated in the HRG,
defined so as to focus on the the critical contribution as we have done for the hydrodynamic and hadron gas correlators in Eqs.~\eqref{Eq:DelH} and \eqref{Eq:DelG}, respectively.

The reader may guess the next step. We have defined $\Delta\omega_{\hA_{1}\dots \hA_{k}}$, which is the suitably normalized ordinary cumulant of $k$ particle multiplicities, via integrating $\Delta G_{A_{1}\dots A_{k}}$ over $6k$ phase space variables. Next we do the same for $\widehat\Delta G_{A_{1}\dots A_{k}}$ which we defined in Eq.~\eqref{Eq:IRCG} and which according to the Maximum Entropy procedure is determined at freezeout from the hydrodynamic correlations by Eq.~\eqref{Eq:IRCGAsToIRCH}. That is, we define 
\be\label{Eq:IRCNa}
\hat{\Delta}\<\prod^{k}_{i=1}\delta N_{\hA_i}\>_{\rm connected}\equiv \int_{\tilde{A}_1\dots\tilde{A}_k} 
\hat{\Delta} G_{A_1\dots A_k}\ ,
\ee
via an analogous phase space integration. This in turn allows us to define
\be\label{Eq:IRCOmega}
\hat{\Delta}\omega_{\hA_{1}\dots \hA_{k}}\equiv \frac{\hat{\Delta}\<\prod^{k}_{i=1}\delta N_{\hA_i}\>_{\rm connected}}{\prod^{k}_{i=1}\<N_{\hA_i}\>^{1/k}}
\ee
which we shall refer to as the integrated IRCs, or the integrated IRCs of particle multiplicities. 

The integrated IRCs are generalizations of the normalized factorial cumulants of particle multiplicities. This is most easily seen starting from the example where all of the species in the integrated IRCs are the same, for example protons.  
In this case, we can simplify our notation for the $\omega$'s defined in Eq.~\eqref{Eq:Omegak} by defining $\omega_{2p}\equiv \omega_{pp}$, $\omega_{3p}\equiv\omega_{ppp}$ and $\omega_{4p}\equiv\omega_{pppp}$, and it is also convenient to define $\omega_{1p}\equiv \langle N_p \rangle/\langle N_p \rangle =1$.
The standard definition of the first few factorial cumulants of the proton multiplicity in terms of its first few ordinary cumulants then takes the form
\begin{eqnarray}\label{eq:factorial-vs-ordinary-cumulants}
\omega^F_{2p} &\equiv& \omega_{2p} - \omega_{1p}\nonumber\\
\omega^F_{3p} &\equiv& \omega_{3p} - 3 \omega_{2p} +2\omega_{1p}\nonumber\\
\omega^F_{4p} &\equiv& \omega_{4p} - 6 \omega_{3p} +11\omega_{2p}-6\omega_{1p}\ .
\end{eqnarray}
We shall now argue that, up to the effects of quantum statistics in the HRG, these factorial cumulants of the proton multiplicity are the same as the integrated IRCs of the proton multiplicities: 
\be
\omega^F_{kp}\approx \hat\Delta\omega_{kp}.
\ee

To make this argument, let us consider the case where the HRG is an HRG in which every hadron species has a classical Boltzmann distribution, rather than a Bose-Einstein or Fermi-Dirac distribution.  
In this case, in the HRG we have $\bar \omega_{4p}=\bar\omega_{3p}=\bar\omega_{2p}=\bar\omega_{1p}= 1$, and all the $k>1$ 
factorial cumulants vanish. 
Furthermore, upon noting that $\omega_{1p}=\bar\omega_{1p}$ it is straightforward to 
see that $ \omega^F_{2p} = \omega_{2p} - \bar \omega_{2p}$  meaning that with this classical HRG the full non-trivial normalized second factorial cumulant  $\omega^F_{2p}$
is precisely the phase space integral of  $\widehat \Delta G_{pp}$, defined in Eq.~\eqref{Eq:IRCG2}, normalized by dividing by $\langle N_p\rangle$. 
Turning to third order, we can see that $\omega^F_{3p}=(\omega_{3p}-\bar\omega_{3p})-3(\omega_{2p}-\bar\omega_{2p})$ 
which means that the full non-trivial normalized third factorial cumulant $\omega^F_{3p}$ is the normalized phase space integral of $\widehat{\Delta} G_{ppp}$, defined 
in Eq.~\eqref{Eq:IRCG3}.  And, upon using this result and the fourth order result
$\omega^F_{4p}=(\omega_{4p}-\bar\omega_{4p})-6(\omega_{3p}-\bar\omega_{3p})+11(\omega_{2p}-\bar\omega_{2p})$ we see
that $\omega^F_{4p}$ is the integral of $\widehat{\Delta} G_{pppp}$ from Eq.~\eqref{Eq:IRCG4}, normalized.
So, we have shown explicitly that with this Boltzmann HRG, $\omega^F_{kp}=\hat\Delta\omega_{kp}$ for $k\leq 4$.

The factorial cumulants of the proton multiplicity, for example as measured by STAR~\cite{STAR:2025zdq}, are useful correlation measures precisely because they have been defined so as to subtract the trivial contributions to the ordinary cumulants that come just from the statistics of a classical Boltzmann HRG.
The integrated IRCs $\hat{\Delta}\omega_{\hA_1,\dots, \hA_{k}}$ 
that we have defined in Eq.~\eqref{Eq:IRCOmega}
and that, via Eqs.~\eqref{Eq:IRCNa} and \eqref{Eq:IRCGAsToIRCH}, we shall calculate using the Maximum Entropy freezeout procedure,
are generalizations of the ordinary factorial cumulants.
They, too, measure the nontrivial multiplicity correlations relative to those that come just from the statistics of a (not necessarily Boltzmann) HRG.

For protons, whose mass $m_p$ is much larger than the freeze-out temperature $T_f$, the $k^{th}$ integrated IRCs for proton multiplicity, $\hat{\Delta}\omega_{kp}$ 
is in fact very close to the conventional $k^{th}$ factorial cumulant of the proton multiplicity because in the HRG the Fermi-Dirac distribution for protons is very well approximated by a Boltzmann distribution.  In this sense, $\omega^F_{kp}\approx\hat\Delta\omega_{kp}$.
Therefore, in this paper when we set $\hA_1=\ldots = \hA_k=p$ we
shall refer to the integrated IRCs
$\hat{\Delta}\omega_{\hA_1\dots \hA_{k}}$ as the factorial cumulants of the proton multiplicity. We have generalized the definition of the integrated IRCs to encompass other (mixed) cumulants also, but this generalization will not be important to us in this paper, as we shall focus on proton multiplicity fluctuations throughout.   

The important reason why we have  introduced the integrated 
IRCs is that these are the proton multiplicity fluctuation measures 
that are specified immediately after freezeout in terms of corresponding measures of the hydrodynamic fluctuations immediately before freezeout via the maximum entropy procedure. Specifically, what we need to do is to integrate Eq.~\eqref{Eq:IRCGAsToIRCH} over
$6k$ phase space variables, thus obtaining the integrated IRCs from its left-hand-side, and see what we obtain from its right-hand-side. To this we now turn.

In calculating $\widehat{\Delta}\omega_{kp}$ at freeze-out by integrating Eq.~\eqref{Eq:IRCGAsToIRCH} over phase space, the integration of Eq.~\eqref{Eq:IRCGAsToIRCH} over $3k$ position variables means integrating the position coordinates in each $A_i$ over the freezeout hypersurface in spacetime, see Eq.~\eqref{eq:d3xd3p}.  
We are assuming throughout that the fluid four-velocity at freezeout is normal to the freezeout hypersurface, 
And, we are making the further simplifying assumption of Bjorken flow:  the fluid
is expanding longitudinally in a boost invariant fashion
and is spatially uniform in the transverse directions, with no transverse expansion, see Eq.~\eqref{eq:d3xd3p}.
In this case, the freeze-out happens on a constant $\tau=\tau_{\rm f}$ hypersurface where the temperature and chemical potential equal take on the values at a point on the freezeout curve in the $(T,\mu)$ plane.
In performing the integrations over the $3k$ momentum variables using Eq.~\eqref{eq:d3xd3p}, we shall not impose any kinematic acceptance cuts.  Note that if the fluid is homogeneous and if the hydrodynamic correlations are local, then as long as we do not impose any kinematic cuts the boost invariance of the particle multiplicities after freezeout means that the fluid velocity does not actually influence the results. With these assumptions, doing the integrals over the hypersurface $\tau=\tau_{\rm f}$ using Eq.~\eqref{eq:d3xd3p} yields the same result that would have been obtained by ignoring the fluid velocity, assuming freezeout at constant time, and doing the integral using $d^3\bm x\, d^3\bm p$. 
In either case, the hydrodynamic correlation functions are given by their equilibrium values as specified just by the freeze-out temperature and chemical potential. 
Also in either case, the $\delta$-functions in Eq.~\eqref{Eq:P} mean that each  integral over the freezeout hypersurface gives a factor of the volume $V$.   
In this scenario, we can substitute the expression 
Eq.~(\ref{Eq:IRCGAsToIRCH}) for $\widehat \Delta G$ obtained from the maximum entropy freezeout prescription into the definition~(\ref{Eq:IRCOmega}) of the integrated IRCs of the particle multiplicities,  $\hat{\Delta}\omega^k_{\hA_1\dots \hA_{k}}$, to obtain the following expression:
\be\label{Eq:IRCOmegaSimple}
\hat{\Delta}\omega_{\hA_1\dots \hA_{k}}=\hat{\Delta}H_{a_1\dots a_k}P^{a_1}_{\hA_1}\dots P^{a_k}_{\hA_k}\(\prod_{i=1}^{k} \<N_{\hA_i}\>\)^{-1/k}
\ee
where the $P^{a}_{\hA}$'s are given by 
\be
\label{Eq:Xa}
P^{a}_{\hA}\equiv \int_{\widetilde A}P^a_A
=\int_{\tA}\, (\bar{H}^{-1})^{ab}\,P_{b}^{B} \,\bar{G}_{BA} 
\ee
Recall that the repeated index $b$ corresponds to integrating over fluid cells on the freezeout hypersurface and summing over $b\in\{\e,n\}$
and the repeated index $B$ corresponds to integrating over phase space variables according to Eq.~\eqref{eq:d3xd3p} and summing over all hadronic species in the particle description of the system just after freezeout. The $\int_{\tA}$ is only an integration over phase space variables and summation over spin as defined in Eq.\eqref{eq:tA}.)
The singular contribution to the IRCs of the particle multiplicities near the critical point can now be evaluated directly from the EoS using the expression~\eqref{Eq:IRCOmegaSimple}  because $\hat{\Delta}H$ is specified directly by the EoS according to Eqs.~\eqref{Eq:Hdef}, \eqref{Eq:barHreg}, \eqref{Eq: BarH}, \eqref{Eq:DelH}, and \eqref{eq:hatH}. In order to evaluate the critical contribution, it suffices to consider the $\mu$- and $T$-dependence of $\Delta H$ only. 
Note that the normalization by mean multiplicity ensures that the $\widehat{\Delta}\omega_{k}$'s are volume-independent, intensive, quantities.

Without loss of generality, we restrict the analysis in this work to the study of factorial cumulants of proton multiplicity. The extension to cumulants of other particles such as pions, or even mixed particle cumulants is straightforward. We start from Eq.~(\ref{Eq:IRCOmegaSimple}) and specialize to the factorial cumulant of proton multiplicity, $\hat{\Delta}\omega^{k}_{p_1 \dots p_k}$, which we denote by $\hat{\Delta}\omega_{kp}$: 
\be \label{Eq:omegaP}
\hat{\Delta}\omega_{kp}=\frac{\widehat{\Delta}H_{a_1\dots a_k}P^{a_1}_{p}\dots P^{a_k}_{p}}{ \<N_{p}\>}
\ee
where the $P^{a}_{p}$'s are defined in Eq.~\eqref{Eq:Xa} with $\hA=p$ here, denoting protons.

Note that in any experimental measurement of the factorial cumulants of the proton multiplicity distribution, 
only some of the detected protons are produced directly 
at freezeout, as described via the particlization procedure that we are treating.  Other protons in the final state measured by the detector are the daughters coming from  
strange baryons and excited baryons that were produced at freezeout via the particlization procedure that we describe,
and that subsequently decayed.
The total contribution from both direct as well as daughter protons to $\hat{\Delta}\omega_{kp}$ can be determined using the following freeze-out formula:   
   \be\label{Eq:IRCOmegaGen}
	\widehat{\Delta} \omega_{kp}=\frac{ \hat{\Delta}H_{a_1\dots a_k} \(\sum_{B_1}\Gamma_{B_1\rightarrow p}P^{a_1}_{B_1}\)\dots\(\sum_{B_k}\Gamma_{B_2\rightarrow p}P^{a_k}_{B_k}\) }{\sum_{B}\Gamma_{B\rightarrow p}\<N_{B}\>}
	\ee	
where the summations go over all hadrons that decay into a proton, including the proton itself, and where $\Gamma_{B\rightarrow p}$ represents the probability that a baryon $B$ decays into a proton. In the main part of this paper, we restrict ourselves to computing the contribution of direct protons only, which is to say we compute, $\widehat{\Delta}\omega_{kp}$ using Eq.~(\ref{Eq:omegaP}). The calculation of factorial cumulants of total proton multiplicity including both direct and feed-down protons, using Eq.~(\ref{Eq:IRCOmegaGen}), is reported in Appendix~\ref{app:resonances}. The plots in Appendix~\ref{app:resonances} show that including the daughter protons makes only a quantitative change to our results; the qualitative form of all of our results are unchanged.

Since the maximum entropy freeze-out procedure yields multiplicity fluctuations for all hadron species present at freeze-out, it can be straightforwardly extended to compute other experimentally relevant observables, such as, cumulants of the event-by-event distribution of net electric charge or net strangeness or net proton number. Let $\tilde{q}_A$ denote the amount of the quantity of interest
carried by a hadron of species $A$. 
For instance, in the case of net-proton fluctuations, $\tilde{q}_A=+1$ for protons, $-1$ for anti-protons and $0$ for all other hadrons. The $k^{th}$ cumulant of the corresponding net (proton) number distribution is then given by: 
    \be\label{eq:net-number-case}
    \<\delta N^{k}_{\tilde{q}}\>_{\text{connected}} = \int_{A_1\dots A_{k}}\tilde{q}_{A_1}\tilde{q}_{A_2}\dots\tilde{q}_{A_k}\, G_{A_1A_2\dots A_k}\ .
    \ee
Note that if $\tilde q_A$ is chosen to be 1 for the hadrons of species $\hat A$ (e.g., protons) and 0 otherwise, then Eq.~\eqref{eq:net-number-case} reduces to Eq.~\eqref{Eq:DelNA}.

\section{Results in the QCD Phase Diagram}
\label{Sec:main}

With the maximum entropy freeze-out prescription, we are able to determine the critical contribution to the fluctuations using only the equation of state as an input.
This is a significant improvement over earlier approaches that relied upon making a parametrized ansatz for how the correlation length $\xi$ varies with $T$ and $\mu_B$~\cite{Athanasiou:2010kw}. We shall see, though, that our results for the 
fluctuations of thermodynamic quantities (in particular, the baryon number density) and, consequently via the maximum entropy freeze-out prescription, for the factorial cumulants of the proton multiplicity are fully consistent with 
our results from 
Section \ref{Sec:EoS} for how the correlation length 
depends on $T$ and $\mu_B$.
In that Section,
we presented an update on $\xi_{\rm QCD}(\mu_B,T)$ utilizing what is known from universality \cite{ZinnJustin} and the mapping \eqref{Eq:quadmap} between the universal Ising variables and the ones for QCD.  In Fig.~\ref{Fig:CorrLengthFixxi0} we 
illustrated the effects of the nonuniversal mapping parameters $w$ and $\rho$ on the contours of  $\xi_{\rm QCD}^2/w^2$ obtained via mapping the universal features of an Ising critical point onto the QCD phase diagram. 
In the first part of this Section, Sect.~\ref{Sec:HydroCorrelations}, we use the full 3D Ising EoS to study the behavior of contour plots of the critical contribution to ordinary cumulants of the baryon number density,  i.e $\Delta H_{kn}\equiv \<\delta n^k\>$, (where 
$k$ indicates the order of the correlation function and $n$ stands for baryon number density) while varying the mapping parameters $w$ and $\rho$
in the same way that we did in Sect.~\ref{Sec:CorrelationLength},
choosing $w= \{1,5,20\}$ and $\rho=\{0.5,1,2\}$.
These choices were motivated by previous studies on the causality and stability of the critical 
EoS~\cite{Parotto:2018pwx,Mroczek:2022oga,Kahangirwe:2024cny}.
We shall compare our results for $\Delta H_{2n}=\langle \delta n^2\rangle$ directly 
to $\xi_{\rm QCD}^2/w^2$ plotted in Fig.~\ref{Fig:CorrLengthFixxi0}.
We also fix the remaining parameters in the mapping between QCD and the Ising model as in Sect.~\ref{Sec:CorrelationLength}, choosing:  $\mu_c=600$~MeV (leading to $T_c=90$~MeV and $\alpha_1=16.6^{\rm{o}}$) and $\alpha_2=0$.
The choice $\alpha_2=0$ is somewhat special for baryon density cumulants, since this reduces derivatives with respect to $\mu$ to derivatives with respect to $h$ at constant $r$. We do know from the work of Ref.~\cite{Pradeep:2019ccv}, though, that $\alpha_2-\alpha_1$ vanishes in the chiral limit and so is reasonably small in the real world. Since $\alpha_1=16.6^\circ$ at $\mu_c=600$~MeV, this
means that $\alpha_2$ cannot be far from zero.

\begin{figure}[t]
    \centering
\includegraphics[width=0.5\textwidth]{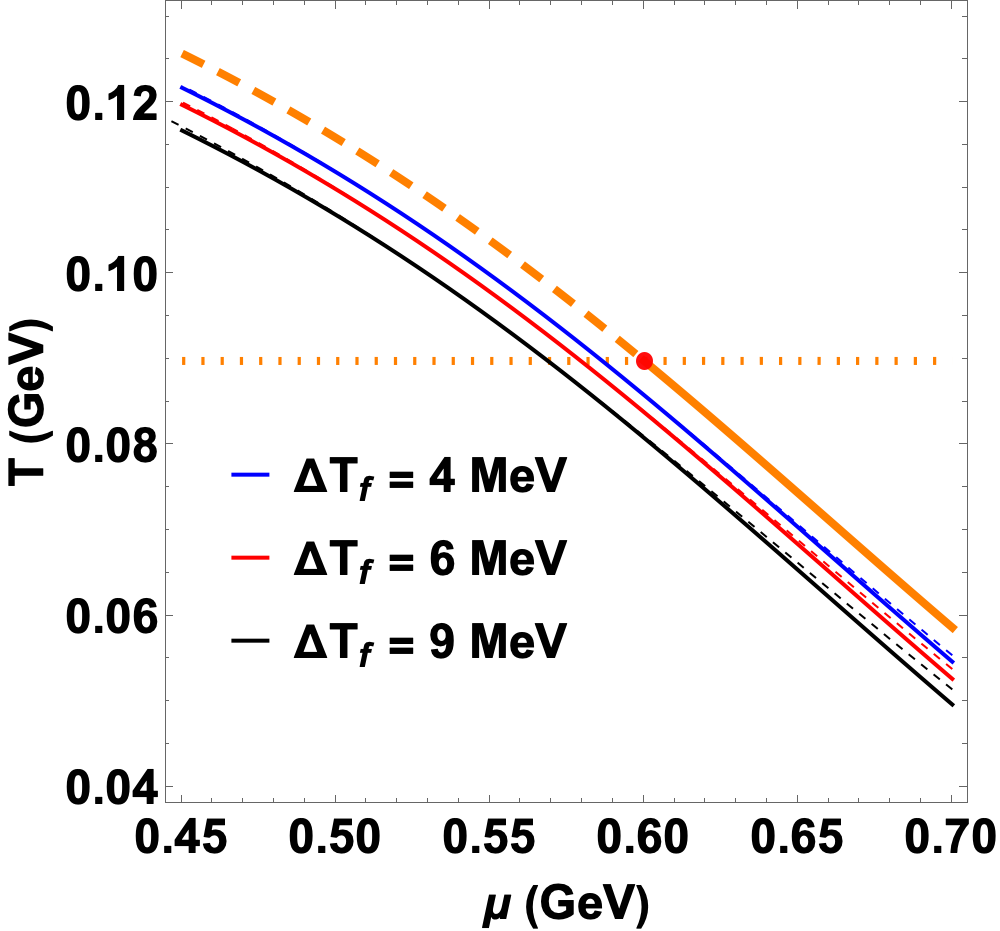}
    \caption{\justifying Three freezeout curves displaced downward relative to the crossover curve $\Delta T'=0$ (orange) by $\Delta T_f=4$, 6 and 9 MeV (solid blue, red and black curves, respectively).  By construction, the Ising-$r$ axis maps onto the crossover curve. Because we have chosen $\alpha_2=0$, the Ising-$h$ axis maps onto the horizontal orange dotted line.
    The dashed blue, red and black curves are curves of constant Ising-$h$ that are coincident with the three freezeout curves where the freezeout curves each cross the Ising-$h$ axis.  We see that the freezeout curves are close to being curves of constant $h$.
    }
    \label{Fig:FreeezeOutCurves}
\end{figure}

We vary the critical contribution to the QCD equation of state via varying the non-universal mapping parameters $w$ and $\rho$ in order to demonstrate the range of possible results not only for the order of magnitude of the moments of the fluctuations 
of $n$ as a function of position on the QCD phase diagram  (Sect.~\ref{Sec:HydroCorrelations})
but also to see how they influence
the shape of the features 
in the factorial cumulants of proton multiplicity
seen along the freeze-out curve that we explore in Section~\ref{Sec:CumulantsPlots}.
We perform our calculations using the maximum entropy freeze-out procedure along freeze-out curves parallel to the crossover curve $T_{\rm crossover}(\mu_B)$ specified by setting $T'=T_0$, namely $\Delta T'=0$, in Eq.~\eqref{Eq:quadmap}.  Specifically, we shall use three different freezeout curves specified by
\be\label{eq:FreezeoutCurve}
T_f(\mu_B)=T_{\rm crossover}(\mu_B)-\Delta T_f
\ee
with $\Delta T_f=4$, 6 and 9 MeV, where $T_{\rm crossover}(\mu_B)$ is the Ising-$r$ axis where $h=0$ and hence $\Delta T'=0$.
We show these freezeout curves in Fig.~\ref{Fig:FreeezeOutCurves}, where we also show that they are close to being curves of constant Ising-$h$.

\subsection{Contours of Cumulants of Baryon Density Fluctuations in the QCD Phase Diagram}
\label{Sec:HydroCorrelations}

\begin{figure}[t]
    \centering
    \includegraphics[width=\linewidth]{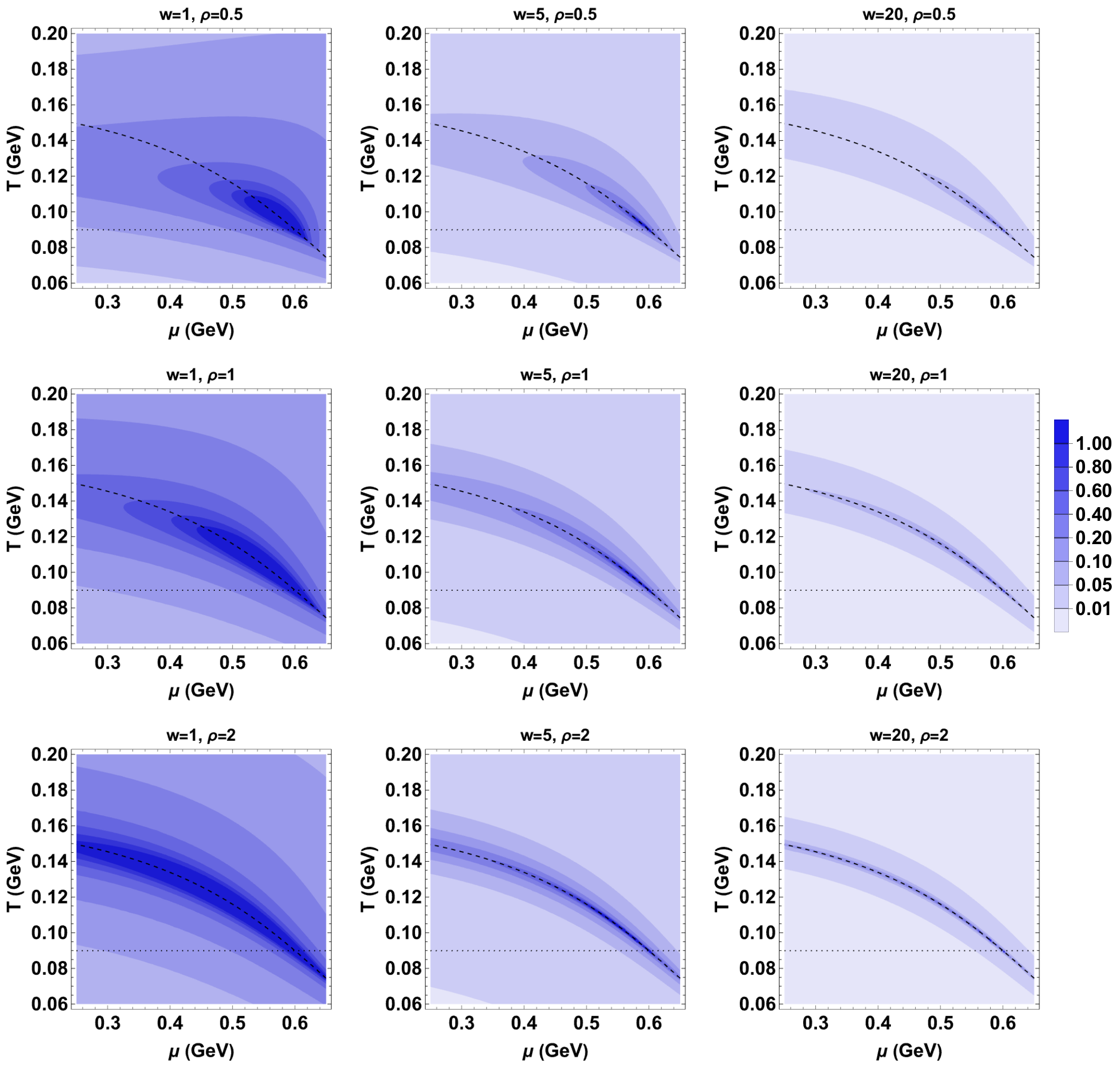}
    \caption{\justifying Contours of the critical contribution to $V T^{-3}_c \Delta H_{2n}=V T_c^{-3}\langle\delta n^2\rangle$ plotted for various values of the nonuniversal equation of state mapping parameters $w$ and $\rho$, with $\mu_c=600 \, \text{MeV}$, $T_c=90$~MeV, and  $\alpha_2=0^{\circ}$. The dotted line reflects this choice of $\alpha_2=0^{\circ}$ in that it shows the Ising $h$-axis $r=0$, while the dashed line shows the Ising $r$-axis $h=0$ which, by the construction described in Sect.~\ref{Sec:EoS}, maps onto the crossover curve on the QCD phase diagram.}
    \label{Fig:H2nAll}
\end{figure}

\begin{figure}[t]
    \centering
    \includegraphics[width=\linewidth]{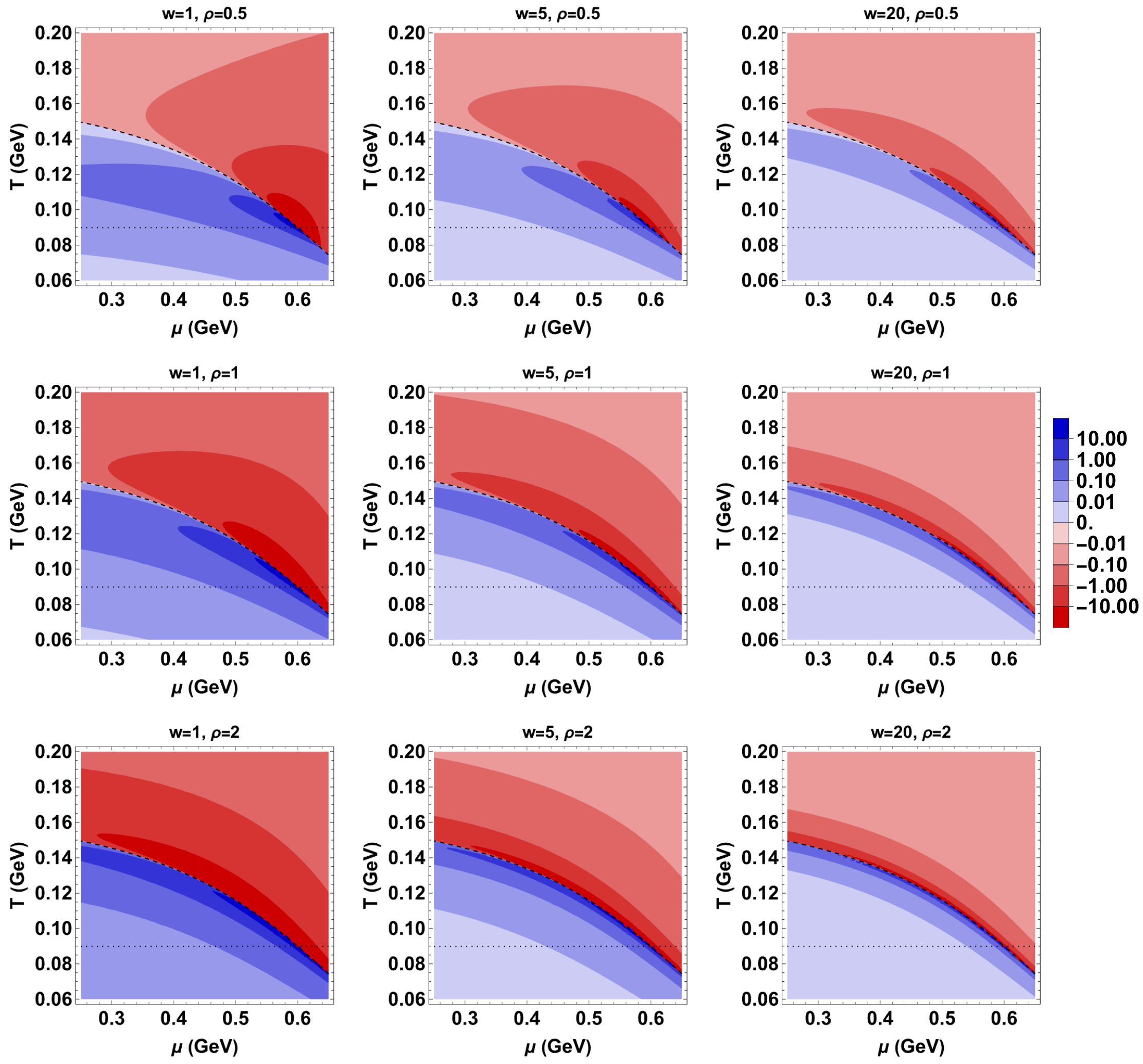}
    \caption{\justifying Contours of the critical contribution to $V^2 T^{-3}_c \Delta H_{3n}=V^2 T_c^{-3}\<\delta n^3\>$ plotted for various values of the nonuniversal mapping parameters $w$ and $\rho$ with $\mu_c=600 \, \text{MeV}$, $T_c=90$~MeV and  $\alpha_2=0^{\circ}$. The dotted line reflects this choice of $\alpha_2=0^{\circ}$ in that it shows the Ising $h$-axis, while the dashed line shows the Ising $r$-axis which is chosen to lie along the QCD crossover curve.}
    \label{Fig:H3nAll}
\end{figure}

\begin{figure}[t]
    \centering
    \includegraphics[width=\linewidth]{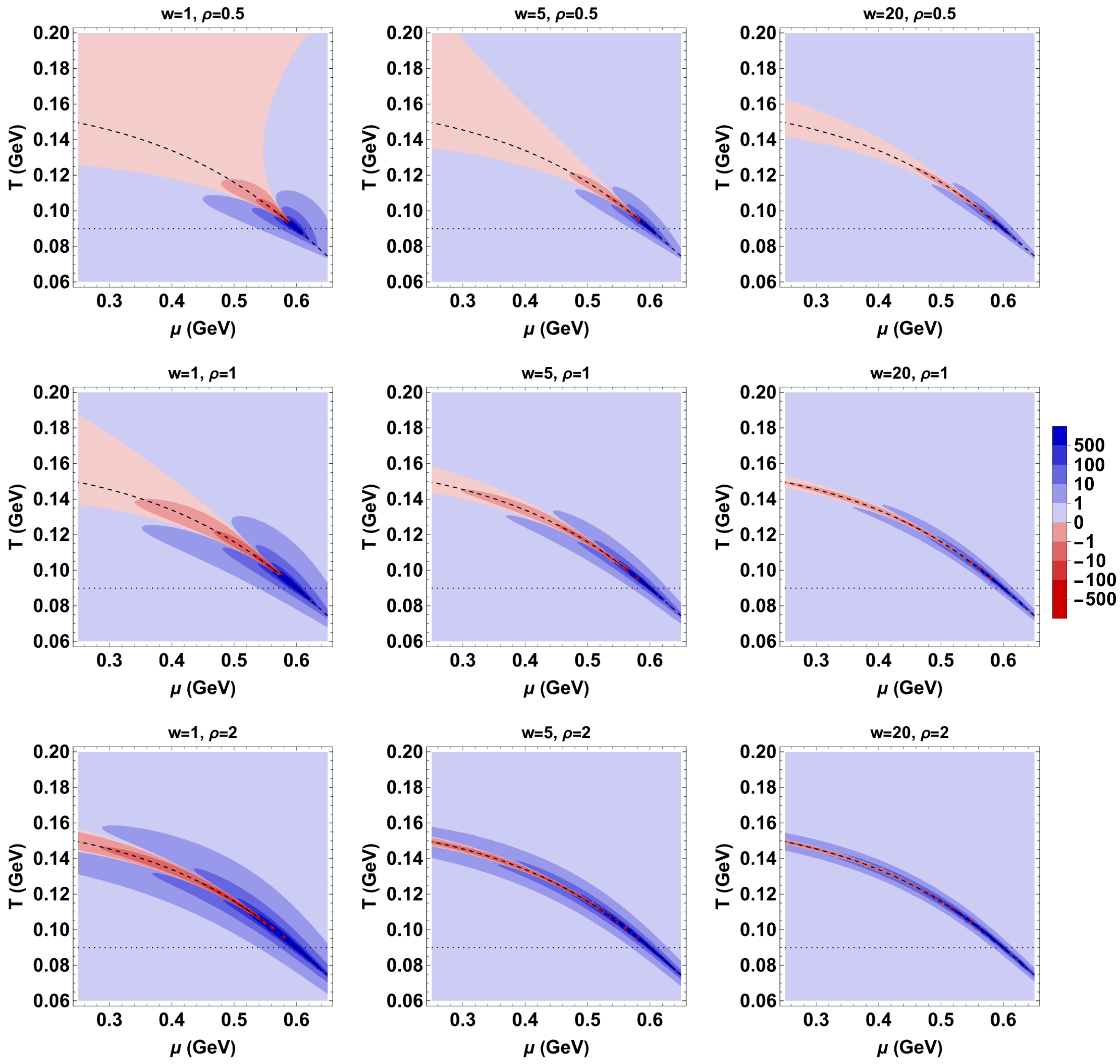}
    \caption{\justifying Contours of the critical contribution to $V^3 T^{-3}_c \Delta H_{4n}=V^{3} T_c^{-3}\<\delta n^4\>$ plotted for various values of the nonuniversal mapping parameters $w$ and $\rho$ with $\mu_c=600 \, \text{MeV}$, $T_c=90$~MeV and  $\alpha_2=0^{\circ}$. The dotted line reflects this choice of $\alpha_2=0^{\circ}$ in that it shows the Ising $h$-axis, while the dashed line shows the Ising $r$-axis which is chosen to lie along the QCD crossover curve.}
    \label{Fig:H4nAll}
\end{figure}

We present the dependence of the connected baryon density correlators $\Delta H_{a_1\dots a_k}=\Delta H_{n\dots n}
\equiv \Delta H_{kn}=\langle \delta n^k\rangle$ on the non-universal equation of state parameters $w$ and $\rho$ that govern the mapping from the Ising phase diagram to the QCD phase diagram in this section in Figs.~\ref{Fig:H2nAll}, \ref{Fig:H3nAll} and \ref{Fig:H4nAll}.
We present our results as $V^{k-1}T^{-3}_c \Delta H_{kn}$, where $V$ is the volume and $k$ is the order of the baryon density cumulant, for the choice of $\mu_B=600$ MeV and $\alpha_2=0^{\circ}$ with the same range of choices of $w$ and $\rho$ as in Sect.~\ref{Sec:CorrelationLength}. 
We have chosen to normalize our results with appropriate factors of $V$ and $T_c$ so as to ensure that the quantities whose contours we plot
in Figs.~\ref{Fig:H2nAll}, \ref{Fig:H3nAll} and \ref{Fig:H4nAll} are all dimensionless and are all independent of volume.
In each of these figures, the location of the critical point and the choice of angle $\alpha_2=0$ is evident from the dashed and dotted lines that show where the Ising model axes map onto the QCD phase diagram.
The dashed lines show the crossover curve, namely the Ising-$r$ axis $h=0$; the dotted lines show the Ising-$h$ axis $r=0$;
and, the point where the dashed and dotted lines cross
is the critical point at $(\mu,T)_c = (600~{\rm MeV},90~{\rm MeV})$.
In Figs.~\ref{Fig:H2nAll}, \ref{Fig:H3nAll} and \ref{Fig:H4nAll},
blue corresponds to positive values of $\langle\delta n^k \rangle$ while red corresponds to negative values.
In each of the three Figures, the nine panels depict our results for the choices
$w=\{1,5,20\}$ and $\rho=\{0.5,1,2\}$ of the nonuniversal mapping parameters.
Across the rows, $\rho$ remains fixed as $w$ increases, and down the columns $w$ is fixed as $\rho$ increases.
The results presented in Figs.~\ref{Fig:H2nAll}, \ref{Fig:H3nAll} and \ref{Fig:H4nAll} allow us to see how the non-universal mapping parameters $w$ and $\rho$ are related to the shape and strength of the
critical fluctuations.  One can already get an initial sense of how our results for the factorial cumulants of proton multiplicities will vary along the freezeout curves in Fig.~\ref{Fig:FreeezeOutCurves} by looking at the
results in Figs.~\ref{Fig:H2nAll}, \ref{Fig:H3nAll} and \ref{Fig:H4nAll} along these curves.

Fig. \ref{Fig:H2nAll} shows the contours of the second order fluctuations of the baryon number, $V T_c^{-3} \Delta H_{2n}$, in thermal equilibrium as a function of position on the QCD phase diagram, for various choices of $w$ and $\rho$.
This Figure can be compared directly to Fig.~\ref{Fig:CorrLengthFixxi0}, which shows the
contours of $\xi_{\rm QCD}^2/w^2$ for the same
choices of $w$ and $\rho$ -- and we see 
that they are directly comparable, as expected.

To understand the comparison between Figs.~\ref{Fig:CorrLengthFixxi0} and \ref{Fig:H2nAll} quantitatively, it is helpful to work out how $\Delta H_{2n}$ scales along the Ising-$r$ and Ising-$h$ axes, shown as the dashed and dotted curves in Fig.~\ref{Fig:H2nAll}.
We shall do this with greater generality, considering $\Delta H_{a_1\dots a_k}$, which represents an arbitrary $k^{th}$ order hydrodynamic cumulant, with $a_{i}\in\{\e, n\}$, meaning that $\Delta H_{2n}$ is given by this quantity with $k=2$ and $a_1=a_2=n$. 
To understand the behavior of $\Delta H_{a_1\dots a_k}$, we begin by observing that
the Ising Gibbs free energy scales as 
\begin{equation}\label{eq:G-Ising-x}
G_{\rm Ising}(r,h)\sim h^{1+1/\delta} g^{(0)}(x)\,,~{\rm where}
~x\equiv r h^{-1/\beta\delta} \,,
\end{equation}
and where the function $g^{(0)}(x)$ is a universal function of the scaling variable $x$ in the Ising model  that is regular at $x=0$ and that behaves as 
$g^{(0)}(x)\sim x^{\beta(1+\delta)}$ for $x\rightarrow\infty$.
For simplicity, here and throughout the following we shall write expressions for the leading scaling behavior of various quantities in the region $h\geq 0$, meaning the region where $x$ is real.  The corresponding expressions for $h<0$ can be obtained upon
noting that $G_{\rm Ising}$ is an even function of $h$ and that differentiating $G_{\rm Ising}$ an odd (even) number of times with respect to $h$ yields an odd (even) function of $h$.
In Figs.~\ref{Fig:H2nAll}, \ref{Fig:H3nAll} and \ref{Fig:H4nAll} we have plotted our results for 
$\Delta H_{kn}$ in the regions of the QCD phase diagram on both sides of the crossover curve, where $h$ takes either sign, but in the next Subsection we will focus on the freezeout curve which lies below the crossover curve. We shall choose our sign convention for $h$ such  that $h>0$ corresponds to the region of the QCD phase diagram below the crossover curve.
From Eq.~\eqref{eq:G-Ising-x}, we observe that the leading critical
contribution to the hydrodynamic correlators $\Delta H$ behaves as 
    \be
    \label{Eq:DelHlead}
    \Delta H_{a_1\dots a_k}(\mu,T)\sim h_{a_1}\dots h_{a_k}h^{1+1/\delta-k}g^{(k)}(x(\mu,T))
    \ee
where
\begin{eqnarray}
\label{Eq:HeAndHn}
h_{n}&\equiv&\(\partial h/\partial {\hmu}\)_{\text{cp}}=-\frac{\sin\alpha_1}{ w \sin(\alpha_1-\alpha_2)} \ ,\nonumber\\
h_{\e}&\equiv&-\(\partial h/\partial \beta\)_{\text{cp}}=-\frac{T_c \cos\alpha_1+\mu_c \sin \alpha_1}{ w \sin(\alpha_1-\alpha_2)}\ , 
\end{eqnarray}
and $g^{(k)}(x)$
is again universal, with $g^{(k)}(0)$ a $k$-dependent constant and
\begin{equation}\label{eq:gk}
g^{(k)}(x)\sim 
\begin{cases}
x^{\beta-(k-1)\beta\delta} & {\rm for}~x\rightarrow\infty~
\mbox{unless $k$ odd and $x>0$;}
\\
x^{\beta-k\beta\delta} & {\rm for}~x\rightarrow\infty~{\rm with}~k~{\rm odd~and}~x>0\ ,
\end{cases}
\end{equation}
meaning that along the crossover curve where $r>0$, $|h|\rightarrow 0$ and, thus, $|x|\rightarrow\infty$, the baryon number density correlators $\Delta H_{kn}$ are non-vanishing and $\sim r^{\beta-(k-1)\beta\delta}$ if $k$ is even, whereas they vanish as $r^{\beta-k\beta\delta}h$ if $k$ is odd.
The subscript $\text{cp}$ as before means that the derivative is evaluated at the critical point; 
each of the $h_{a_i}$'s contributes a factor of $1/w$. 
As we did in Sect.~\ref{Sec:CorrelationLength},
we shall analyze the scaling behavior of the expression \eqref{Eq:DelHlead} along the two Ising axes, in turn.

We begin along the Ising-$h$ axis where $r=0$ and where,  from Eq.~\eqref{eq:h-dmu2-w},  we have
\be\label{eq:Ising-h-scaling}
h^{1+1/\delta - k} \propto \left( \frac{\Delta T^{'}}{w} \right)^{1+1/\delta - k} \propto \left( \frac{\mu_c^2-\mu^2}{w} \right)^{1+1/\delta - k}
\ee
and, since $x=0$ along this axis, we have
\be\label{eq:DeltaH-scaling-along-h}
\Delta H_{a_1\dots a_k}(\mu,T)\sim
\left(\frac{1}{w}\right)^{1+1/\delta} \left(\frac{1}{\mu_c^2-\mu^2}\right)^{k-1-1/\delta}
\ee
which, upon setting $k=2$ and comparing to Eq,~\eqref{eq:correlation-scaling-h-axis}, shows that $\Delta H_{2n}$ scales in the same way that $(\xi_{\rm QCD})^{2-\eta}/w^2$ does along this axis.
We can also check that the $\rho$- and $w$-dependence of the contours in Figs.~\ref{Fig:H2nAll},~\ref{Fig:H3nAll} and \ref{Fig:H4nAll} is described 
by Eq.~\eqref{eq:DeltaH-scaling-along-h} along the Ising-$h$ axis.
From Eq.~\eqref{eq:DeltaH-scaling-along-h} we see that along the (horizontal, dotted) Ising-$h$ axis, the contours should have no dependence on $\rho$, which is immediately confirmed by
inspection of each of the columns of panels in Figs.~\ref{Fig:H2nAll},~\ref{Fig:H3nAll} and \ref{Fig:H4nAll}. 
Next, from Eq.~\eqref{eq:DeltaH-scaling-along-h} we see that along the Ising-$h$ axis, increasing $w$ should squeeze the contours toward the critical point, with the squeeze in $(\mu_c^2-\mu^2)$ proportional to $(1/w)^{(\delta+1)/(\delta k - \delta -1)}$, which can be confirmed by inspection of how the patterns of contours 
along the Ising-$h$ axis  change from left to right across each row of  panels
in Figs.~\ref{Fig:H2nAll},~\ref{Fig:H3nAll} and \ref{Fig:H4nAll}, with increasing $w$.  

Next we turn to analyzing the scaling behavior of Eq.~\eqref{Eq:DelHlead} along
the Ising-$r$ axis where $h\rightarrow 0$ and $|x|\rightarrow \infty$.   We shall focus on the $r>0$ half-axis, which
maps onto the crossover curve on the QCD phase diagram and corresponds to $x\rightarrow +\infty$. From Eq.~\eqref{eq:gk} we see that the scaling behavior of $g^{(k)}(x)$ along this axis depends on whether $k$ is even or odd, which we shall consider in turn.  For $k$ even, $\Delta H_{kn}$ is non-vanishing along the Ising-$r$ axis and is straightforward to analyze:
 $\Delta H_{kn}\sim h^{1+1/\delta-k}g^{(k)}(x)\sim r^{\beta+\beta\delta(1-k)}$. 
And, from Eq.~\eqref{Eq:quadmap} we have
\be\label{eq:r-dmu2-rho-2}
r^{\beta+\beta\delta(1-k)}\propto \left( \frac{\mu_c^2-\mu^2}{\rho w}\right)^{\beta+\beta\delta(1-k)}\ ,
\ee
and hence
\be\label{eq:DeltaH-scaling-along-r}
\Delta H_{a_1\dots a_k}(\mu,T)\sim
\left(\frac{1}{w}\right)^{k} \left( \frac{\rho w}{\mu_c^2-\mu^2}\right)^{\beta\delta(k-1)-\beta}\sim w^{k(\beta\delta-1)-\beta\delta-\beta}\left( \frac{\rho }{\mu_c^2-\mu^2}\right)^{\beta\delta(k-1)-\beta}\ ,
\ee
along the Ising-$r$ axis for even $k$.
Upon setting $k=2$, comparing to Eq.~\eqref{eq:xiscaling}, and noting that $\beta\delta-\beta=\nu(2-\eta)$, we see that $\Delta H_{2n}$  
scales in the same way that $(\xi_{\rm QCD})^{2-\eta}/w^2$ does along {\it this} axis.
Noting that $\eta\approx 0.036$ is small in the 3D Ising universality class, if we set $\eta=0$ we have demonstrated
that $\Delta H_{2n}$ and $(\xi_{\rm QCD}/w)^2$ have the same scaling as a function of $(\mu_c^2-\mu^2)$ and the same
dependence on $w$ and $\rho$ along both the Ising-$h$ and Ising-$r$ axes.  This explains the similarity between Figs.~\ref{Fig:CorrLengthFixxi0} and \ref{Fig:H2nAll}. This comparison corroborates the result from Ref.~\cite{Pradeep:2022eil} that the 
nonuniversal coupling $g_p$ between the $\sigma$ field and the proton mass employed in the analysis of Ref.~\cite{Athanasiou:2010kw} is proportional to $1/w$, as it was this result that motivated us to plot $(\xi_{\rm QCD}/w)^2$ in Fig.~\ref{Fig:CorrLengthFixxi0}.

Looking at either Fig.~\ref{Fig:CorrLengthFixxi0} or Fig.~\ref{Fig:H2nAll},
we have already analyzed how the contours behave where they cross the (horizontal, dotted) Ising-$h$
axis: increasing $w$ squeezes the contours toward the critical point and increasing $\rho$ has no effect, as in Eq.~\eqref{eq:DeltaH-scaling-along-h}.
If instead we look at how the contours behave where they cross the crossover curve (dashed line, Ising-$r$ axis), by comparing panels left-to-right across any of the three rows of panels in Fig.~\ref{Fig:CorrLengthFixxi0} or Fig.~\ref{Fig:H2nAll} we see that along the Ising-$r$ axis increasing $w$ also squeezes the contours toward the critical point, in this case as in Eq.~\eqref{eq:DeltaH-scaling-along-r}. 
It is helpful to set $\eta=0$ and $\nu=2/3$, which is a good approximation for the 3-dimensional Ising model and which yields $\beta=1/3$ and $\delta=5$. Upon making this choice, with $k=2$ the $r$-axis scaling~\eqref{eq:DeltaH-scaling-along-r}
becomes
\be\label{eq:DeltaH-scaling-along-r-k2}
\Delta H_{2n}\sim
w^{-2/3}
\left( \frac{\rho }{\mu_c^2-\mu^2}\right)^{4/3}\ .
\ee
Next, we compare panels top-to-bottom down any of the three columns of panels in
either 
Fig.~\ref{Fig:CorrLengthFixxi0} or Fig.~\ref{Fig:H2nAll}, corresponding to fixing $w$ and increasing $\rho$ by a factor of 4.
When we look at how the 
contours behave where they cross the crossover
curve (dashed line, Ising-$r$ axis) we see that
along this axis increasing $\rho$ stretches the contours away from the critical point, stretching the critical region along the crossover curve,
as in Eq.~\eqref{eq:DeltaH-scaling-along-r} and hence Eq.~\eqref{eq:DeltaH-scaling-along-r-k2}.

We can also check that the $\rho$- and $w$-dependence of the contours in Fig.~\ref{Fig:H4nAll} (in which $k$ is also even) is described 
by Eq.~\eqref{eq:DeltaH-scaling-along-r} along the Ising-$r$ axis.
With $k=4$, the $r$-axis scaling~\eqref{eq:DeltaH-scaling-along-r} takes the form
\be\label{eq:DeltaH-scaling-along-r-k4}
\Delta H_{4n}\sim
w^{2/3}
\left( \frac{\rho }{\mu_c^2-\mu^2}\right)^{14/3}\ ,
\ee
where we have set $\eta=0$ and $\nu=2/3$, meaning 
$\beta=1/3$ and $\delta=5$.
From Eq.~\eqref{eq:DeltaH-scaling-along-r-k4}, we see that along the (dashed) Ising-$r$ axis, increasing $\rho$ should stretch the pattern of contours out, pulling them away from the critical point, with the stretch in 
$(\mu_c^2-\mu^2)$ proportional to the increase in $\rho$.  Indeed, in Fig.~\ref{Fig:H4nAll} we see that
increasing $\rho$ (going from a panel in the top row to the panels below it) stretches the pattern of contours of $\Delta H_{4n}$ out along the crossover curve, as expected. 
Next, when we compare how the pattern along the direction of the crossover curve in  Fig.~\ref{Fig:H4nAll} changes from left to right across the panels in the Figure as we increase $w$, we see the opposite dependence to what we saw in 
Fig.~\ref{Fig:H2nAll}. In Fig.~\ref{Fig:H2nAll}, increasing $w$ squeezes the pattern of contours along the crossover curve toward the critical point with $(\mu_c^2-\mu^2)$ squeezed by a factor of $(1/w)^{1/2}$ as described by Eq.~\eqref{eq:DeltaH-scaling-along-r-k2}, whereas 
in Fig.~\ref{Fig:H4nAll} increasing $w$ stretches the pattern out along the crossover curve
modestly, with $(\mu_c^2-\mu^2)$ stretched by
a factor of $w^{1/7}$, 
as described by Eq.~\eqref{eq:DeltaH-scaling-along-r-k4}.
So, increasing $w$ squeezes the pattern of contours of $\Delta H_{kn}$ toward the crossover curve
for all values of $k$
as described above by the scaling along the Ising-$h$ axis from Eq.~\eqref{eq:DeltaH-scaling-along-h} and, as described by the scaling along the Ising-$r$ axis from Eq.~\eqref{eq:DeltaH-scaling-along-r}, it squeezes (stretches) the pattern along the crossover curve for $k=2$ ($k=4$).

Next we turn to analyzing the scaling behavior of Eq.~\eqref{Eq:DelHlead} along, and parallel to,
the Ising-$r$ axis when $k$ is odd, so as to understand the behavior seen near this axis in Fig.~\ref{Fig:H3nAll} with $k=3$.
From Eqs.~\eqref{Eq:DelHlead} and \eqref{eq:gk}, we see that when $k$ is odd
$\Delta H_{kn}$ vanishes along the Ising-$r$ axis with
 $\Delta H_{kn}\sim h^{1+1/\delta-k}g^{(k)}(x)\sim r^{\beta-k\beta\delta}h$. 
In the next Subsection we shall analyze the physics along freezeout curves which are parallel to and just below the $h=0$ curve that are, to a good approximation, curves of constant and small $h$. 
The features of Fig.~\ref{Fig:H3nAll} near the $h=0$ axis that are most apparent, however, 
are the tips of the ``lobes'' of each contour, and how their positions 
change with $\rho$ and $w$.  The tips of the lobes of the contours lie along a curve of constant $x$ (equivalently, along a curve of constant $\theta$ in the scaling form~\eqref{ZinnJustin}), not a curve of constant $h$. Along a curve of constant $x$, $h\propto r^{\beta\delta}$ and when $k$ is odd $\Delta H_{kn}\propto r^{\beta-k\beta\delta}h\propto r^{\beta-k\beta\delta+\beta\delta}$, which is the same as the scaling along the crossover curve for even $k$ in Eq.~\eqref{eq:DeltaH-scaling-along-r}.  With $k=3$,
the leading scaling behavior along a curve of constant $x$ therefore takes the form
\be\label{eq:DeltaH-scaling-along-r-k3}
\Delta H_{3n}\sim
w^0
\left( \frac{\rho }{\mu_c^2-\mu^2}\right)^3\ .
\ee
This tells us that (ignoring subleading effects) the position of the tips of the lobes of the contours should not change with $w$ and should stretch out away from the critical point as $\rho$ is increased,
behavior that is reasonably well confirmed by inspection of Fig.~\ref{Fig:H3nAll}.

Finally, it is straightforward to generalize the results \eqref{eq:correlation-scaling-h-axis} and \eqref{eq:xiscaling} from Sect.~\ref{Sec:CorrelationLength} to $k>2$ and then check, via comparison with Eqs.~\eqref{eq:DeltaH-scaling-along-h} and \eqref{eq:DeltaH-scaling-along-r}, that $\Delta H_{kn}$ and 
$(\xi_{\rm QCD})^{-3+k(5-\eta)/2}/w^k$ 
have the same dependence on 
$(\mu_c^2-\mu^2)$, $w$ and $\rho$ along both the Ising-$h$ and Ising-$r$ axes.
Note, though, that the baryon number cumulants $\Delta H_{kn}$ are also proportional to an additional, universal, factor 
that depends on the angle $\theta$ in Eq.~\eqref{ZinnJustin} but not on the correlation length, where this factor changes sign as a function of $\theta$ for $k>2$.
For $k=3$, this universal factor (referred to as $\tilde\lambda_3$ in Refs.~\cite{Stephanov:2008qz,Athanasiou:2010kw,Stephanov:2011pb}) 
is odd in $h$ meaning that it 
vanishes on the crossover curve ($h=0$; $|x|\rightarrow \infty$) and
changes sign across the crossover curve.
This is apparent in Fig.~\ref{Fig:H3nAll}, 
as we have already discussed.
For $k=4$, the universal factor (denoted $2\tilde\lambda_3^2-\tilde\lambda_4$ in Refs.~\cite{Stephanov:2008qz,Athanasiou:2010kw,Stephanov:2011pb})
changes sign twice -- it is negative on, and in a region around, the crossover curve and vanishes and then becomes larger
and positive farther away from $h=0$. This is apparent in Fig.~\ref{Fig:H4nAll}.

This concludes our discussion
of the results that we have obtained for the critical contributions to the second, third and fourth cumulants of the baryon number fluctuations in the QCD phase diagram, depicted in Figs.~\ref{Fig:H2nAll},
\ref{Fig:H3nAll} and \ref{Fig:H4nAll}. We have been able to use the scaling relations \eqref{eq:DeltaH-scaling-along-h} and \eqref{eq:DeltaH-scaling-along-r} to describe all the qualitative features of these plots, including their $w$- and $\rho$-dependence, and have also understood how these results are related to
the results for $\xi_{\rm QCD}$ from Section~\ref{Sec:CorrelationLength}. 
We are now ready to ``slice these plots'' along
the freezeout curves in Fig.~\ref{Fig:FreeezeOutCurves}, apply the maximum entropy freezeout procedure, and look at how the factorial cumulants of the proton multiplicity
vary along each putative freezeout curve.

\subsection{Factorial Cumulants of Proton Multiplicity in Equilibrium Along Freeze-out Curves}
\label{Sec:CumulantsPlots}

Now, we turn to our results for the critical contribution to proton factorial cumulants at second, third, and fourth order along freeze-out curves in the phase diagram.
This will allow us to study the parameter dependence of the overall magnitude of the cumulants, as well as the height, width, and location of the peak.
We shall employ the three putative freezeout curves
in Eq.~\eqref{eq:FreezeoutCurve} depicted in Fig.~\ref{Fig:FreeezeOutCurves}.
Due to the uncertainty in the determination of the chiral crossover line at finite density from lattice QCD calculations~\cite{Borsanyi:2020fev}, we utilize this parametrization of the freeze-out curve in order to study the effect of the unknown separation $\Delta T_f$ between $T_c$ at the critical point and $T_f$ at freeze-out.
Similar parametric freeze-out curves have been utilized in previous studies of fluctuations at freeze-out~\cite{Mroczek:2020rpm,Dore:2022qyz}. 
As the purpose of this work is to illustrate the consequences for 
the cumulants of the proton multiplicity of
varying various parameters including $\Delta T_f$,
we defer discussion of the freeze-out curve from experiment \cite{Cleymans:2005xv,Lysenko:2024hqp} to future work.  
We also note that all of our calculations and results assume that the hydrodynamic fluctuations are in
equilibrium at freezeout, whereas in reality they will be out of equilibrium, and various authors have noted that one important non-equilibrium effect 
is that as the hydrodynamic fluid cools the fluctuations will lag, meaning that they will ``remember'' their state at a prior higher 
temperature~\cite{Berdnikov:1999ph,Mukherjee:2015swa,Mukherjee:2016kyu,Stephanov:2017ghc,Akamatsu:2018vjr,Rajagopal:2019xwg,Du:2020bxp,Pradeep:2022mkf}.
It may therefore be a reasonable approximation 
to use our results with a value of $\Delta T_f$ that is smaller than the actual difference between the critical temperature and the freezeout temperature. This possibility, and the current uncertainties related to estimating non-equilibrium ``memory effects'', argues for treating $\Delta T_f$ at present as a free parameter to be varied, as we shall do.

At each point on a freezeout curve, from Fig.~\ref{Fig:FreeezeOutCurves}, we 
use the maximum entropy freezeout procedure
given by Eq.~\eqref{Eq:omegaP} to
compute
the factorial cumulants of the particle multiplicity
$\hat\Delta\omega_{2p}$, $\hat\Delta\omega_{3p}$ and $\hat\Delta\omega_{4p}$ from the cumulants of the energy density and the 
baryon number density, where we have plotted the latter in Figs.~\ref{Fig:H2nAll},
\ref{Fig:H3nAll} and \ref{Fig:H4nAll}.
(In Appendix~\ref{app:subleading} we 
compute the relative contribution of 
the energy density fluctuations to the factorial cumulants of the particle multiplicity and compare to those originating from
baryon number density fluctuations
and show that they are typically small.)
Recall from
Eq.~\eqref{eq:hatH} that the hydrodynamic irreducible
relative cumulants $\hat\Delta H_{kp}$ are related
to the $\Delta H_{kp}$ via various subtractions that involve the regular, noncritical, contributions to the 
fluctuations, with the irreducible relative cumulants of the particle multiplicity $\hat\Delta G_{kp}$ related
to $\Delta G_{kp}$ analogously, see Eq.~\eqref{Eq:IRCG}. 
In our calculations, we have obtained the
regular contribution to the correlations of particle multiplicity, denoted by $\bar{G}_{AB\dots}$, and the corresponding regular contribution to cumulants of thermodynamic densities, denoted by $\bar{H}_{ab\dots}$, 
determined from  the Hadron Resonance Gas (HRG) EoS by Eq.~\eqref{Eq: BarH}.
The HRG model we employ includes all the hadron species included in a recent development of the SMASH hadronic transport framework that incorporates more resonances in order to agree with the lattice QCD EoS with masses up to 3.2 GeV~\cite{SanMartin:2023zhv}.
That said, we have checked that varying the hadronic list used as input for HRG calculations does not have a large impact on the results for the proton factorial cumulants.
The proton factorial cumulants that we calculate and plot in this Section do not include the contribution from daughter protons that result from decay of resonances. In Appendix~\ref{app:resonances}, we have incorporated the contribution of resonances to the observed factorial cumulants of proton multiplicity using Eq.~(\ref{Eq:IRCOmegaGen}). We see in that Appendix that doing so does not result in any qualitative change to our results. 

\begin{figure}[t]
    \centering
    \includegraphics[width=\linewidth]{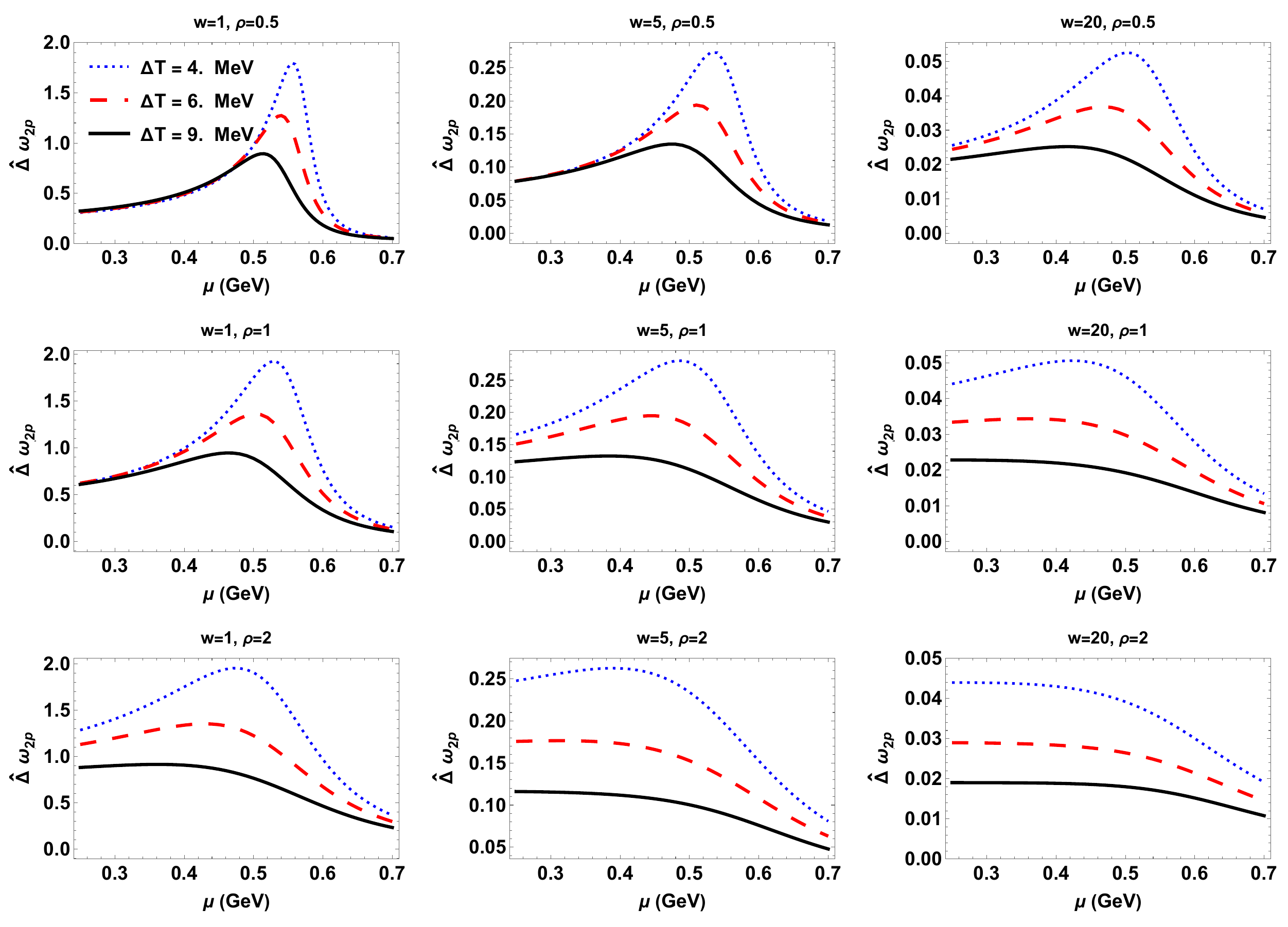}
    \caption{\justifying Second factorial cumulant of the proton multiplicity distribution, $\hat{\Delta}\omega_{2p}$, along the three freezeout curves from Fig.~\ref{Fig:FreeezeOutCurves} characterized by Eq.~\eqref{eq:FreezeoutCurve} with $\Delta T_f=4$, 6 and 9 MeV. The different panels show $\hat\Delta\omega_{2p}$ 
    for various values of the nonuniversal mapping parameters $w$ and $\rho$, with $\mu_c=600 \, \text{MeV}$, $T_c=90$~MeV and  $\alpha_2=0^{\circ}$ in all panels.}
    \label{Fig:w2pAll}
\end{figure}

\begin{figure}[t]
    \centering
    \includegraphics[width=\linewidth]{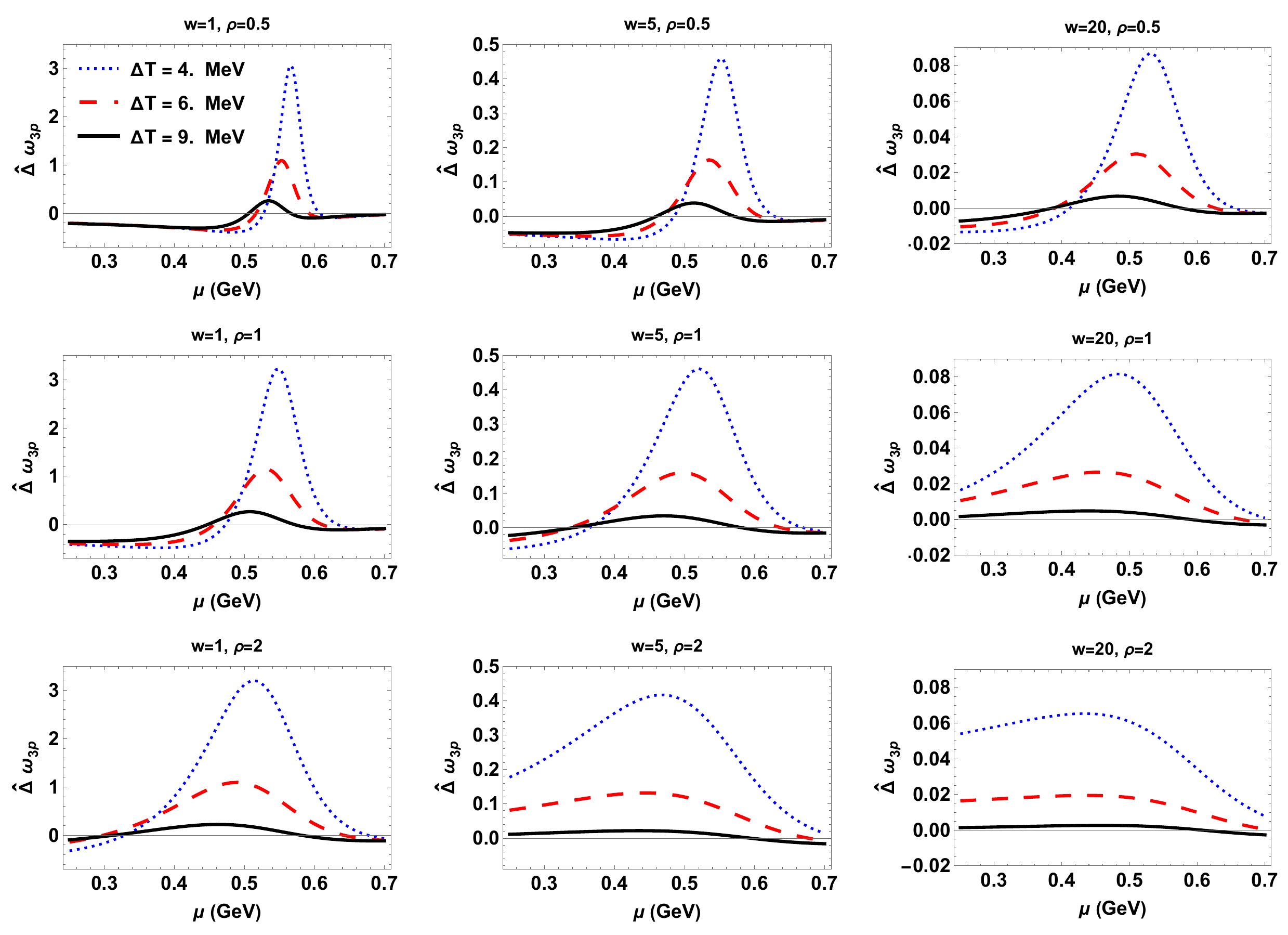}
    \caption{\justifying Third factorial cumulant of the proton multiplicity distribution, $\hat{\Delta}\omega_{3p}$, along the three freezeout curves from Fig.~\ref{Fig:FreeezeOutCurves}. The different panels show $\hat\Delta\omega_{3p}$ 
    for various values of the nonuniversal mapping parameters $w$ and $\rho$, with $\mu_c=600 \, \text{MeV}$, $T_c=90$~MeV and  $\alpha_2=0^{\circ}$ in all panels. Although $\Delta H_{3n}$ is positive all along the freezeout curve (see Fig.~\ref{Fig:H3nAll}), the third order factorial cumulant $\hat\Delta\omega_{3p}$ can be negative in some regions of the freezeout curve because it includes lower order cumulants, see Eq.~\eqref{Eq:IRCG} or Eq.~\eqref{eq:factorial-vs-ordinary-cumulants}. Specifically, where the third order ordinary cumulant is not large, the third order factorial cumulant can go negative because it includes the subtraction of a term proportional to the second order cumulant, which is positive.
        }
    \label{Fig:w3pAll}
\end{figure}

\begin{figure}[t]
    \centering
    \includegraphics[width=\linewidth]{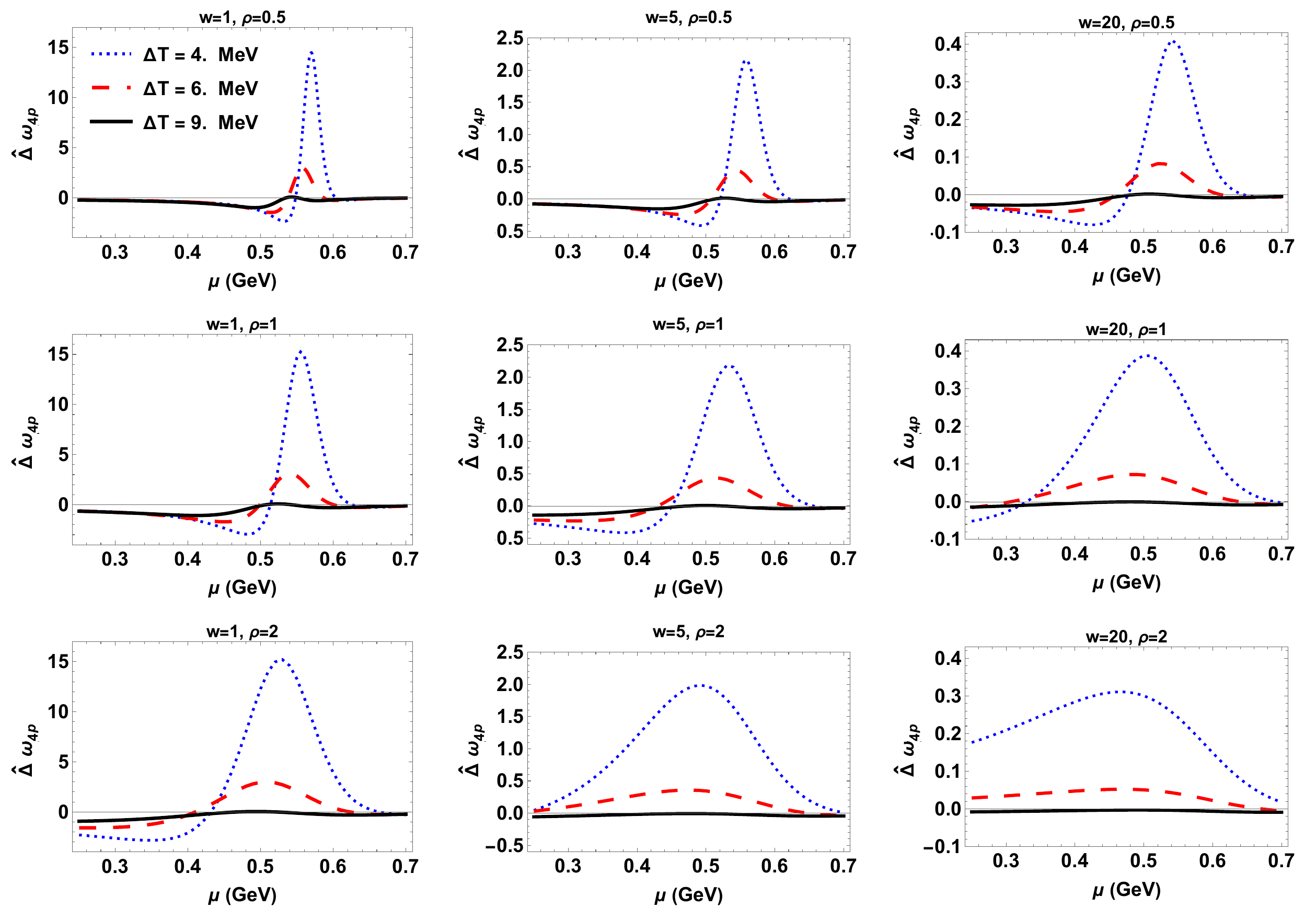}
    \caption{\justifying Fourth factorial cumulant of the proton multiplicity distribution, $\hat{\Delta}\omega_{4p}$, along the three freezeout curves from Fig.~\ref{Fig:FreeezeOutCurves}. The different panels show $\hat\Delta\omega_{4p}$ 
    for various values of the nonuniversal mapping parameters $w$ and $\rho$, with $\mu_c=600 \, \text{MeV}$, $T_c=90$~MeV and  $\alpha_2=0^{\circ}$ in all panels.
    }
    \label{Fig:w4pAll}
\end{figure}

The factorial cumulants of the proton multiplicity $\hat{\Delta}\omega_{2p}$, 
$\hat{\Delta}\omega_{3p}$ and
$\hat{\Delta}\omega_{4p}$ 
along our three possible freezeout curves are shown in Figs. \ref{Fig:w2pAll},
\ref{Fig:w3pAll},  and \ref{Fig:w4pAll}, respectively. 
In each Figure, the nine panels show results for 
the same choices of the non-universal parameters $w$ and $\rho$ as for the correlation length in Fig.~\ref{Fig:CorrLengthFixxi0} and the cumulants of the baryon number density in  
Figs.~\ref{Fig:H2nAll}, \ref{Fig:H3nAll} and \ref{Fig:H4nAll}.
In each panel in Figs. \ref{Fig:w2pAll},
\ref{Fig:w3pAll},  and \ref{Fig:w4pAll},
the blue, red and black curves show our results along the three freeze-out curves from Fig.~\ref{Fig:FreeezeOutCurves}.

Because the freezeout curves in Fig.~\ref{Fig:FreeezeOutCurves} are to a good approximation curves of constant $h$, the dependence of the heights of the peaks in all the curves 
in Figs. \ref{Fig:w2pAll},
\ref{Fig:w3pAll},  and \ref{Fig:w4pAll}
on $\Delta T_f$, $w$ and $\rho$ 
is determined by how $\Delta H_{kp}$ and, consequently, $\hat\Delta\omega_{kp}$ change with $h$.  This dependence
can be inferred from Eq.~\eqref{eq:DeltaH-scaling-along-h}, which describes how $\Delta H_{kp}$ and, consequently, $\hat\Delta\omega_{kp}$ 
scales with these parameters along the Ising-$h$ axis or upon approaching the critical point along any direction that is not parallel to lines of constant $h$.
We first observe that there is no dependence on $\rho$ in Eq.~\eqref{eq:DeltaH-scaling-along-h}, and indeed it is immediately apparent 
in Figs.~\ref{Fig:w2pAll},
\ref{Fig:w3pAll},  and \ref{Fig:w4pAll} that 
the peak heights do not change significantly as you 
vary $\rho$,  down a column of panels.  
(The small changes in the peak heights upon varying $\rho$ originate from subleading critical contributions that involve fluctuations in the energy density which are discussed in Appendix~\ref{app:subleading}.)
From Eq.~\eqref{eq:DeltaH-scaling-along-h}, we expect that as you vary $w$, going across a row of panels in any of the three Figures, the peak heights should vary like $(1/w)^{6/5}$, where we have taken $\delta=5$.
This scaling is 
reasonably well satisfied in all three figures. 

Next, let us consider the dependence of the peak heights on $\Delta T_f$.
Varying $\Delta T_f$ from 4 to 6 or from 4 to 9 MeV corresponds to increasing the $h$ where the freezeout curve crosses the $h$-axis (and consequently to a good approximation the $h$ along the freezeout curve) by the same factor
of $6/4$ or $9/4$.  
(As long as $\Delta T_f\ll T_c$ as we are assuming,
a freezeout curve displaced downward from the crossover curve by $\Delta T_f$ crosses the 
Ising-$h$ axis at $\mu \simeq \mu_c-\Delta T_f/\tan\alpha_1$, as can be seen from Fig.~\ref{Fig:FreeezeOutCurves}.
Here and throughout we are assuming for simplicity that $\alpha_2=0$, when all that we know is that it must be small. If $\alpha_2$ were nonzero, the freezeout curve displaced downward from the crossover curve by $\Delta T_f$ would cross the Ising-$h$ axis at
$\mu\simeq\mu_c-\Delta T_f\cos\alpha_1\cos\alpha_2/\sin(\alpha_1-\alpha_2)$, see Eq.~\eqref{Eq:quadmap}.)
From Eq.~\eqref{eq:DeltaH-scaling-along-h}
we see that increasing $\Delta T_f$ by a factor of (say) $6/4$ should reduce the heights of the
peaks in $\hat\Delta\omega_{kp}$ by $(4/6)^{k-6/5}$, where we have taken $\delta=5$. 
This argument would suggest that in 
Figs. \ref{Fig:w2pAll},
\ref{Fig:w3pAll},  and \ref{Fig:w4pAll}, with $k=2$, 3 and 4, respectively,
the peaks in the red curves should be lower than the peaks in the blue curves by factors that are  close to $(4/6)^{4/5}$, $(4/6)^{9/5}$ and
$(4/6)^{14/5}$, respectively.
This argument 
gets the trends correct, but inspection of the Figures shows that it is not satisfied quantitatively.
This argument
would be correct if what we were plotting here was $\Delta H_{kn}$ along the freezeout curve. However, according to the maximum entropy procedure $\hat\Delta\omega_{kp}$ 
is determined by $\hat\Delta H_{kn}$
not $\Delta H_{kn}$, see Eq.~\eqref{Eq:omegaP},
and we can then see from Eq.~\eqref{eq:hatH} 
that the calculation of $\hat\Delta\omega_{kp}$
involves subtractions of
terms composed from products of lower order cumulants and noncritical contributions.  
These subtractions
modify the peak heights more for larger $k$, and for any $k$ the subtraction of the lower order cumulant contributions 
has a larger fractional effect when the peaks are lower.  

We turn now to the widths and positions of the peaks.
It is immediately apparent in Figs.~\ref{Fig:w2pAll},
\ref{Fig:w3pAll},  and \ref{Fig:w4pAll} that increasing $\rho$ makes all the peaks in all the columns of each of the Figures wider, and shifts the location of the peaks to the left, away from the critical point. Both of these effects 
are linear in $\rho$
since, as we have seen from Eq.~\eqref{eq:DeltaH-scaling-along-r}, increasing $\rho$ stretches the patterns of contours in Figs.~\ref{Fig:H2nAll}, \ref{Fig:H3nAll} and \ref{Fig:H4nAll} out along the crossover curve, which corresponds to stretching features --- including peaks --- out along the freezeout curves.
And indeed, what we see in all three Figures is that as $\rho$ is increased from top panels to bottom panels from 0.5 to 1 to 2, the widths of the peaks and the distance in $\mu_B$ between the peaks and the critical point all stretch by a factor that is comparable  to -- but less than -- the factor by which $\rho$ is increased. 

To understand this fully, we need to take note of the second term in the mapping \eqref{Eq:quadmap} between Ising-$r$ and 
$\mu_c^2-\mu^2$ that we did not write in Eqs.~\eqref{eq:r-dmu2-rho-2} and \eqref{eq:DeltaH-scaling-along-r} because it vanishes at $h=0$.
Along a freezeout curve shifted downward from the $h=0$ curve by $\Delta T_f$ as in Fig.~\ref{Fig:FreeezeOutCurves}, 
upon noting from that Figure that our chosen 
freezeout curves are well-approximated as curves of constant $h$ meaning constant $\Delta T'$, we see from Eq.~\eqref{Eq:quadmap} that
\be\label{eq:r-dmu2-rho-w}
r \propto \frac{ \mu_c^2-\mu^2 - \frac{2\mu_c \Delta T_f}{\tan \alpha_1}  }{\rho w} 
\propto 
\frac{ \mu_c - \frac{\Delta T_f}{\tan\alpha_1} -\mu  }{\rho w} \ .
\ee
We can then state our conclusion with greater precision: when $\rho$ is increased with $w$ held fixed it is the distance in $\mu$ between $\mu$ along a freezeout curve where 
each $\hat\Delta\omega_{kp}$ has its peak
and the $\mu$ at which that freezeout curve crosses the $r=0$ Ising-$h$ axis, see Fig.~\ref{Fig:FreeezeOutCurves}, that stretches proportional to $\rho$.
Careful inspection shows that, indeed, this scaling provides a good description of how the positions, and also the widths, of the peaks in Figs.~\ref{Fig:w2pAll},
\ref{Fig:w3pAll},  and \ref{Fig:w4pAll} all stretch to the left when $\rho$ is increased.
This scaling analysis is as straightforward as it is because varying $\rho$ does not change the leading critical contribution to the heights of the peaks.

\begin{figure}[t]
  \centering
\includegraphics[width=0.43\textwidth]{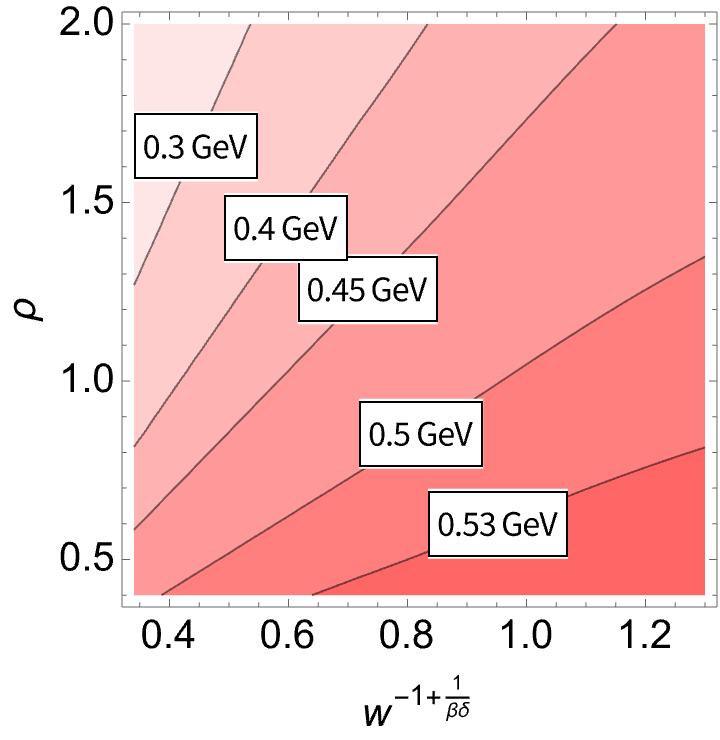}
  \caption{\justifying Contours of the baryon chemical potential $\mu^{(2)}_{\rm peak}$
  at which the second cumulant of the baryon number multiplicity distribution $\hat{\Delta}\omega_{2p}$ along the freezeout curve with $\Delta T_f=6$~MeV, namely the red dashed curves in Fig.~\ref{Fig:w2pAll}, reaches its peak value. The 
  (approximate) linearity of these contours 
  in the $w^{-1+1/\beta\delta}$-$\rho$ plane confirms that $\mu^{(2)}_{\rm peak}$  depends (approximately) on the combination $\rho w^{1-1/\beta\delta}$,  a prediction obtained in the text from the leading critical scaling Eqs.~\eqref{eq:DeltaH-scaling-along-r} and \eqref{eq:DeltaH-scaling-along-h} via Eq.~\eqref{eq:xpeak-scaling}. 
 }
   \label{Fig:mupeak-contours}
\end{figure}

The dependence of the width and position of the peaks in $\hat\Delta\omega_{kp}$ on $w$ is more subtle, because increasing $w$ squeezes the pattern of contours of $\Delta H_{kn}$ towards the crossover curve while at the same time, depending on the value of $k$, it either stretches or squeezes the pattern along the crossover curve, and it is these two effects in concert that determine how the width and position of the peaks in $\hat\Delta\omega_{kp}$ change with $w$.
The consequence of these two effects in sum is most easily understood by going back to Eq.~\eqref{Eq:DelHlead} and
noting that, since according to Fig.~\ref{Fig:FreeezeOutCurves} our chosen freezeout curves are well-approximated as curves of constant $h$, the variation of $\Delta H$ along a freezeout curve originates entirely from the 
factor $g^{(k)}(x)$ in Eq.~\eqref{Eq:DelHlead}, where we recall that $x\equiv rh^{-1/\beta\delta}$.
Along a freezeout curve, the peak in $\hat\Delta\omega_{kp}$ will occur at 
the $\mu$ where $x$ takes the value $x_{\rm peak}^{(k)}$ at which $g^{(k)}(x)$
has its peak.   Although $x_{\rm peak}^{(k)}$ does not depend on the mapping parameters $\rho$ or $w$, the value of $\mu$ along the freezeout curve 
where $x=x_{\rm peak}^{(k)}$, that we shall denote by $\mu^{(k)}_{\rm peak}$, does. The dependence of $\mu^{(k)}_{\rm peak}$ on the mapping parameters must be such that as $w$ (and $\rho$) vary 
\be\label{eq:xpeak-scaling}
x^{(k)}_{\rm peak} \propto 
\left(\frac{ \mu_c - \frac{\Delta T_f}{\tan\alpha_1} -\mu^{(k)}_{\rm peak}  }{\rho w}\right)
\left(\frac{\Delta T_f}{w}\right)^{-1/\beta\delta}
\ee
stays fixed. The first factor on the RHS of Eq.~\eqref{eq:xpeak-scaling} describes scaling along a freezeout curve parallel to the Ising-$r$ axis, as in Eq.~\eqref{eq:DeltaH-scaling-along-r}, together 
with Eq.~\eqref{eq:r-dmu2-rho-w}.  
$h$ is of course constant along such a curve, but what the second factor in Eq.~\eqref{eq:xpeak-scaling} describes is how the value of $h$ on a freezeout curve (shifted downward from the crossover curve by  $\Delta T_f$) scales when $w$ changes.
This scaling, from Eq.~\eqref{eq:Ising-h-scaling}, is the Ising-$h$-axis scaling that is responsible for understanding how the heights of the peaks in $\Delta H_{kn}$ and $\hat\Delta\omega_{kp}$ change with $w$, see Eq.~\eqref{eq:DeltaH-scaling-along-h}.  In this way, Eq.~\eqref{eq:xpeak-scaling}
combines the scaling \eqref{eq:DeltaH-scaling-along-r} along the Ising-$r$ axis with the scaling
\eqref{eq:DeltaH-scaling-along-h} along the Ising-$h$ axis.  From Eq.~\eqref{eq:xpeak-scaling}, we conclude that 
when $w$ is increased with $\rho$ held fixed the distance between the $\mu$ along a freezeout curve where 
each $\hat\Delta\omega_{kp}$ has its peak
and the $\mu=\mu_c-\Delta T_f/\tan\alpha_1$ at which that freezeout curve crosses the $r=0$ Ising-$h$ axis, see Fig.~\ref{Fig:FreeezeOutCurves}, stretches proportional to $w^{1-1/\beta\delta}\approx w^{2/5}$.
And indeed, this scaling provides a good description of how the positions, and also the widths, of the peaks in Figs.~\ref{Fig:w2pAll},
\ref{Fig:w3pAll},  and \ref{Fig:w4pAll} all stretch modestly to the left when $w$ is increased.
We can further illustrate how the scaling relation \eqref{eq:xpeak-scaling} describes
the variation of the positions of the peaks $\mu^{(k)}_{\rm peak}$ with $\rho$ and $w$ via Fig.~\ref{Fig:mupeak-contours}, where we confirm that $\mu^{(2)}_{\rm peak}$ depends on $\rho w^{1-1/\beta\delta}$ by plotting contours of $\mu^{(2)}_{\rm peak}$ in a plane with axes $\rho$ and $w^{-1+1/\beta\delta}$.
The scaling~\eqref{eq:xpeak-scaling} also predicts that the positions and widths of the peaks should stretch to the left 
as $\Delta T_f$ is increased, both on account of the shift proportional to $\Delta T_f$ in the first factor and because of the $(\Delta T_f)^{-1/\beta\delta}$-dependence of the second factor. The $\Delta T_f$-dependence of the positions of the peaks predicted by Eq.~\eqref{eq:xpeak-scaling}
is again confirmed
in Figs.~\ref{Fig:w2pAll},
\ref{Fig:w3pAll},  and \ref{Fig:w4pAll}.

The bottom line is that, via Eq.~\eqref{eq:xpeak-scaling}, Eqs.~\eqref{eq:DeltaH-scaling-along-r} and \eqref{eq:DeltaH-scaling-along-h} in concert explain the $w$-dependence, $\rho$-dependence and $\Delta T_f$-dependence of the heights, widths and positions of the peaks in all three Figs.~\ref{Fig:w2pAll},
\ref{Fig:w3pAll},  and \ref{Fig:w4pAll}.
The magnitude of the peaks in $\hat\Delta\omega_{2p}$, $\hat\Delta\omega_{3p}$ and $\hat\Delta\omega_{4p}$ are controlled by 
$w$ and $\Delta T_f$, with increasing $\Delta T_f$ and increasing $w$ resulting in lower peaks. 
The main role of $\rho$ is to stretch (increasing $\rho$) or compress (decreasing $\rho$) 
the widths of the peaks and their position in $\mu$ relative to that of the critical point. $w$ and $\Delta T_f$ play a role in this also, as the ``horizontal'' stretching or compression of the curves in 
Figs.~\ref{Fig:w2pAll},
\ref{Fig:w3pAll},  and \ref{Fig:w4pAll}
is governed by $\rho w^{1-1/\beta\delta} (\Delta T_f)^{1/\beta\delta} \approx \rho w^{2/5} (\Delta T_f)^{3/5}$.

Before concluding this Section, it is important to highlight some of the qualitative features of our results for the fourth factorial cumulant of the proton multiplicity distribution, $\hat\Delta\omega_{4p}$, plotted in Fig.~\ref{Fig:w4pAll}.  
We see from this Figure that as long as neither $w$ nor $\rho$ are too large, $\hat\Delta\omega_{4p}$ dips negative as a function of increasing $\mu$, before then rising to a positive peak at large $\mu$.  
Since Ref.~\cite{Stephanov:2011pb}, a negative dip in 
$\hat\Delta\omega_{4p}$ has been  understood as a possible harbinger of 
the presence of a critical point,
to be confirmed via the discovery of a positive peak
in $\hat\Delta\omega_{4p}$ at larger $\mu_B$.
We see from the panels toward the bottom-right of Fig.~\ref{Fig:w4pAll}, however, that if both the nonuniversal mapping parameters $w$ and $\rho$ are large, the critical contributions to $\hat\Delta\omega_{4p}$ can yield a broad peak with no preceding negative dip.  This can happen in circumstances in which subleading critical contributions, contributions that we have indeed included in all of our calculations, become important~\cite{Mroczek:2020rpm}.
We can conclude that if a 
positive peak in $\hat\Delta\omega_{4p}$ is discovered at some $\mu_B$ larger than 400 MeV in future experimental measurements, if that peak is preceded at lower $\mu_B$ by a negative dip this negative dip would not only have served its role as a harbinger of the presence of a critical point it would also allow us to rule out the possibility that both $\rho$ and $w$ are large. 

The preceding paragraph provides one hypothetical example of how future experimental measurements of $\hat\Delta\omega_{2p}$, $\hat\Delta\omega_{3p}$
and $\hat\Delta\omega_{4p}$ could be used together with the results of our analysis as plotted in Figs.~\ref{Fig:w2pAll},
\ref{Fig:w3pAll},  and \ref{Fig:w4pAll} not only to provide evidence for the presence of a QCD critical point and constrain its location on the QCD phase diagram but also to constrain the values of the nonuniversal mapping parameters $w$ and $\rho$ that are so important to understanding
how a critical point manifests itself in the QCD equation of state and consequently in hydrodynamic fluctuations and factorial cumulants of the proton multiplicity distribution.
The central remaining uncertainty is that throughout our analysis we have assumed that 
the critical hydrodynamic fluctuations are in equilibrium at freezeout, along a freezeout curve that is $\Delta T_f$ below the crossover.
In reality, critical slowing down guarantees that
the critical fluctuations cannot stay in equilibrium as the matter produced in a heavy ion collision cools and freezes out.
A quantitative analysis of the predictions of the maximum entropy procedure including non-equilibrium dynamics preceding freezeout, as for example can be described using the Hydro+ formalism, remains for future work.  Since one of the important nonequilibrium effects is that the evolution of the critical fluctuations lags behind how they would evolve in equilibrium, a crude operational way of taking this into consideration at the present time is to view $\Delta T_f$ as another unknown parameter, to be determined via comparison with experimental data, expecting that the $\Delta T_f$ so obtained should correspond to a larger temperature than the actual freezeout temperature.

\section{Concluding Remarks and a Look Ahead}
\label{Sec:Conclusion}

   We have presented a first-of-its-kind application of the maximum entropy freeze-out method~\cite{Pradeep:2022eil} to obtain estimates of experimentally measured fluctuation measures, specifically the factorial cumulants of the proton multiplicity,
   corresponding to a parametrically specified trial QCD equation of state. The equation of state from Refs.~\cite{Parotto:2018pwx,Kahangirwe:2024cny}
   employs results from lattice QCD calculations at $\mu=0$ together with universality at a critical point. The equation of state that we have used incorporates key elements of this construction, including the lattice QCD calculation of the crossover curve as well as the universal properties of the QCD critical point, by mapping the universal equation of state for a critical point in the 3-dimensional Ising model universality class onto the QCD phase diagram~\cite{Parotto:2018pwx,Kahangirwe:2024cny}. Nonuniversal aspects of this equation-of-state-mapping determine key properties of the QCD critical point including its location in $\mu$ and $T$, its strength, and the shape of its critical region, all of which we specify in terms of four non-universal mapping parameters that must ultimately be determined by experimental measurements. This is a central goal of the beam-energy scan experiments at RHIC and, in future, at FAIR.  The most important mapping parameter is $\mu_c$, the chemical potential at which the QCD critical point is located. Together with an estimate of the curvature of the crossover curve obtained from lattice QCD calculations, this specifies $T_c$ and $\alpha_1$, the temperature at the QCD critical point and the angle of the crossover curve at that point, see Fig.~\ref{Fig:FreeezeOutCurves}. The other three non-universal mapping parameters are $w$, $\rho$ and $\alpha_2$ which parametrize the strength of the critical fluctuations, the shape of the critical region, and the angle of the Ising-$h$ axis mapped onto the QCD phase diagram. The advantage of using a family of EoS parametrized by these four non-universal mapping parameters, as we do in this work, is that it allows us to cover all possibilities for the EoS near the critical point by allowing these parameters to be varied. 
   Pending investigation of important caveats that we list below and that constitute an agenda for future work, the idea of a study like ours is that predictions for the factorial cumulants of the proton multiplicity fluctuations will make it possible for a future Bayesian analysis of experimental measurements of these factorial cumulants at freezeout to constrain the values of the nonuniversal mapping parameters -- including in particular the location of the critical point, $\mu_c$.

   If we assume that fluctuations stay in thermal equilibrium (more on this assumption below) the equation of state, including the contributions to it arising from the presence of a critical point, directly determines the fluctuations of  hydrodynamic variables such as energy density and baryon density until just before freezeout.  At freezeout, these hydrodynamic fluctuations are translated into particle multiplicity fluctuations measured in experiments. The main feature of the maximum entropy freezeout procedure~\cite{Pradeep:2022eil}  is the matching of the fluctuating conserved quantities (energy; baryon number) between the hydrodynamic fluid just before freezeout and the particle multiplicities just after freezeout. Matching conserved quantities alone is not sufficient to determine the fluctuations on the particle side from those in the hydrodynamic fluid, however. Maximum entropy freezeout is the solution to this problem which maximizes relative entropy --- the entropy of the fluctuations relative to the entropy of a hadron resonance gas. 
  Such an approach is a nontrivial, but natural, generalization of the standard Cooper-Frye freezeout for averaged quantities. Maximizing the entropy without regard for
  fluctuations yields a thermal ensemble of particles with the same energy density and baryon density after freezeout as before, which is the Cooper-Frye prescription. The maximum entropy procedure applies the same logic to fluctuations. 
   
In Sect.~\ref{Sec:CorrelationLength} we have also mapped the contours of the QCD correlation length $\xi$ onto the QCD phase diagram. As can be seen by comparing Figs.~\ref{Fig:CorrLengthFixxi0} and \ref{Fig:H2nAll}, 
the ratio $\xi_{\rm QCD}^2/w^2$ (with $w$ the nonuniversal EoS parameter on which the strength of critical fluctuations in the QCD phase diagram depends) behaves very similarly to the second cumulant of the baryon density fluctuations that we have obtained directly from the EoS.
More generally, 
the scaling behavior of the results for particle multiplicity cumulants in the presence of critical fluctuations that we have obtained via the maximum entropy freezeout procedure are in agreement 
with results obtained earlier~\cite{Stephanov:2008qz,Athanasiou:2010kw,Stephanov:2011pb} 
via coupling the critical mode $\sigma$ to the masses of observed hadrons, including in particular protons, and using universal scaling relations to relate the correlation length $\xi_{\rm QCD}$ of the critical fluctuations to the factorial cumulants of proton multiplicity.
The great advantage of the maximum entropy procedure is that it yields these results, including universal scalings and nonuniversal features, directly from the EoS and the consequent hydrodynamic fluctuations.
Two additional nonuniversal parameters in the older approach (the coupling $g_p$ between the $\sigma$ mode and the proton mass, and a proportionality constant between the Ising and QCD correlation lengths) are determined implicitly by the EoS when the maximum entropy procedure is employed, and need not be determined explicitly.

In Figs.~\ref{Fig:H2nAll}, \ref{Fig:H3nAll} and \ref{Fig:H4nAll} we have plotted the results of our calculations of the second, third and fourth cumulants of baryon number density fluctuations on the QCD phase diagram for a wide range of choices of the nonuniversal EoS parameters $w$ and $\rho$ that are particularly important to specifying the strength of the critical fluctuations 
and the shape of the critical region in any equation of state.  Then in Figs.~\ref{Fig:w2pAll}, \ref{Fig:w3pAll} and \ref{Fig:w4pAll} we have employed the maximum entropy procedure to translate the hydrodynamic fluctuations (assumed to be in equilibrium) along a freezeout curve into predictions for the second, third and fourth factorial cumulants of the
proton multiplicity, that are measured in experiments.
We find that the heights of the peaks in the 
factorial cumulants as a function of $\mu_B$
are insensitive to $\rho$ and sensitive to both $\Delta T_f$ (with a dependence that is approximately proportional to $\Delta T_f^{6/5-k}$, $k$ being the order of the factorial cumulant) and $w$ (with a dependence that is approximately proportional to $1/w^{6/5}$).   The positions and widths of the peaks (in $\mu$) stretch with increasing $\rho$, $w$ or $\Delta T_f$ in a way that depends on 
$\rho w^{2/5}$ and $\Delta T_f^{3/5}$ that we have described in full in Sect.~\ref{Sec:CumulantsPlots}.  
We have understood all of the ways in which the qualitative features of the proton multiplicity factorial cumulants depend on these key nonuniversal mapping parameters in the equation of state
in terms of universal scaling behavior around an Ising critical point, in particular from Eqs.~\eqref{eq:DeltaH-scaling-along-h} and \eqref{eq:DeltaH-scaling-along-r} that describe
how the scaling behavior along the two Ising model axes maps onto scaling behavior of the cumulants of hydrodynamic fluctuations in the QCD phase diagram.

It is important to list several important simplifications made in this study. Each of these constitutes an important caveat vis a vis using our predictions for the factorial cumulants of proton multiplicity fluctuations in a future Bayesian analysis of experimental data, and as such each of these represents an important direction for further investigation.  We stress, though, that none of these simplifications are simplifications to the maximum entropy procedure itself; our study exercises the full power of this approach, applied in a context that we have simplified as follows:
    \begin{itemize}

    \item Simplification of the QCD equation of state: In the maximum entropy freezeout the particle correlations responsible for the factorial cumulants of hadron multiplicity distributions
    are benchmarked against those in the HRG, an ideal gas of noninteracting hadrons and hadronic resonances. These factorial cumulants originate from hydrodynamic fluctuations just before freezeout and, if those fluctuations are in thermal equilibrium, are fully determined by the QCD equation of state.  
    To focus on the possible effects of critical fluctuations, for the present study we have simplified the equation of state by assuming that the only deviations from the equation of state of the ideal HRG are those due to the presence of a critical point in the 3-dimensional Ising universality class. Of course, these are the dominant contributions sufficiently close to  a critical point, but in a more complete calculation 
    (and certainly in the full QCD equation of state) the equation of state deviates from that of the HRG in ways that are unrelated to a critical point. Our study could be improved by improving the equation of state employed in it, for example by including information from lattice QCD calculations of the equation of state at $\mu=0$ as done in Ref.~\cite{Parotto:2018pwx}.
    
\item Nonequilibrium dynamics and memory: We have assumed, throughout, that hydrodynamic fluctuations of the energy density and baryon number density are in equilibrium just before freezeout. This cannot be so near a critical point~\cite{Berdnikov:1999ph}, because critical slowing down means that the fluctuations cannot stay in equilibrium as the
droplet of QGP produced in a heavy ion collision 
expands and cools. The nonequilibrium dynamics of the hydrodynamic fluctuations in an expanding cooling droplet of QGP that passes near a critical point, for example as
described by 
Hydro+~\cite{Stephanov:2017ghc},
includes several dynamical effects and
can be complex~\cite{Mukherjee:2015swa,Mukherjee:2016kyu,Rajagopal:2019xwg,Du:2020bxp,Pradeep:2022mkf}, 
but the most important effect seen in all of these studies is that the dynamics of the fluctuations lags what it would have been if the fluctuations were able to stay in equilibrium.  In this sense, the fluctuations at freezeout ``remember'' conditions at times earlier than the freezeout time.
A full analysis of this effect, and other consequences of the non-equilibrium dynamics, requires applying the maximum entropy procedure to the out-of-equilibrium fluctuations at freezeout as described by Hydro+.
A crude way to use the results of our present study in a comparison with experimental data that would take at least some account of memory effects is to treat $\Delta T_f$, the difference between $T_c$ and the freezeout temperature encoded in the multiplicity factorial cumulants, as a free parameter, allowing for the possibility that the data could tell us that the multiplicity factorial cumulants remember a higher temperature than the actual freezeout temperature.

\item Other dynamical effects: In addition to memory effects, there can also be other dynamical effects which influence the magnitude of the hydrodynamic correlation functions at freezeout and consequently, via maximum entropy freezeout, the observed particle multiplicity factorial cumulants. For example, spatial inhomogeneities in the medium smear the effect of the critical point~\cite{Du:2021zqz}. The presence of a critical point can also deform the hydrodynamic trajectories in the vicinity of a critical point 
in a way that deforms the freezeout curve on the QCD phase diagram~\cite{Stephanov:1998dy,Nonaka:2004pg,Dore:2022qyz,Pradeep:2024cca}. Our simplified freezeout trajectory parallel to the crossover line does not account for such effects.

\item Resonance feed-down: Throughout most of this paper, we have focused on the fluctuations in the multiplicities of hadrons, specifically protons, produced directly at freezeout. However, resonances produced at freezeout eventually decay, and some of those decays produce protons.  The measured protons include these ``daughter protons'' as well as those produced directly. We have provided the expression~\eqref{Eq:IRCOmegaGen} for calculating the 
cumulants of particle multiplicities including the contribution of daughter particles, and have shown in Appendix~\ref{app:resonances} that doing so does not modify our results in any qualitative way.  This investigation will need to be repeated in any future calculation with a more sophisticated equation of state and/or with non-equilibrium dynamics.

\item Hadronic afterburner: We have assumed that after particlization at a freezeout hypersurface, the hadrons produced do not interact further in any way that change hadron multiplicities. That is, we have assumed that chemical freezeout and particlization are simultaneous.
If one is interested in describing measured spectra as well as multiplicity fluctuations, or if one is interested in investigating the possibility of particlization before chemical freezeout, 
one would need to extend our study to include 
a hadronic afterburner (like SMASH~\cite{SMASH:2016zqf}) after particlization. In this case,
the maximum entropy procedure can be used to turn the hydrodynamic fluctuations just before particlization into initial conditions for the hadronic afterburner. Important first steps in exactly this direction have been reported in Ref.~\cite{Hammelmann:2023aza}, and it is promising that these authors find that observable consequences of critical fluctuations in proton factorial cumulants propagate through the hadronic evolution all the way to kinetic freezeout, when particle spectra are fixed.

    \end{itemize}

We look forward to Bayesian analyses of experimental data on the factorial cumulants of proton multiplicities in heavy-ion collisions at a series of collision energies, with a series of values of the baryon chemical potential at freezeout, extending 
to as large values of $\mu$ as can be achieved.
Our investigation shows that the maximum entropy approach enables us to obtain quantitative estimates 
for the factorial cumulants measured in experiments, with the expected hierarchy of increasing $\hat\Delta\omega_{kp}$ with increasing $k$,
and with the full results being directly sensitive to the QCD equation of state including its features arising from the presence of a critical point -- including, in particular, the nonuniversal mapping parameters that can only be determined via comparison with experimental data.
We look forward to the day when confronting experimental data with such predictions 
are constraining nonuniversal parameters including 
$w$ and $\rho$ (which control the strength of the critical fluctuations and the shape of the critical region, and which we have varied over wide ranges in plotting our results) as well as $\Delta T_f$, $\mu_c$ (the location of the critical point, which we have fixed to 600 MeV in our plots for simplicity) and $\alpha_2$ (the 
angle of the Ising-$h$ axis on the QCD phase diagram, which must be small and which we have fixed to $0^\circ$ in our plots for simplicity).
As each of the simplifications that we have enumerated above is alleviated in future studies, the power of a comparison between experimental data and theoretical predictions of the proton multiplicity factorial cumulants from a calculation done via the maximum entropy freezeout procedure as we have done here will increase.

\acknowledgments

We are grateful to G.~Basar, P.~Bedaque, T.~Cohen, V.~Koch, B.~Mohanty, B.~Mueller, J.~Noronha-Hostler, P.~Parotto, C.~Ratti, F.~Renneke, R.~Steinhorst,  V.~Vovchenko and N.~Xu for helpful conversations.
This research was supported in part by the U.S.~Department of Energy, Office of Science, Office of Nuclear Physics under grant Contract Numbers DE-SC0011090, DE-FG02-93ER40762, and DE-FG02-01ER41195.  
JMK is supported by an Ascending Postdoctoral Scholar Fellowship from the National Science Foundation under Award No. 2138063.
YY acknowledges the support from NSFC under Grant No.12175282 and from CUHK-Shenzhen University Development Fund under the Grant No. UDF01003791. KR acknowledges the hospitality of the CERN Theory Department and the Aspen Center for Physics, which is supported by National Science Foundation grant PHY-2210452. MS acknowledges the hospitality of the European Center for Theoretical Studies in Nuclear Physics and Related Areas (ECT*).

\appendix

\section{Universal Scaling Equation of State}
\label{app:ScalingEoS}
The scaling equation of state which determines the behavior of the Ising model 
near its critical point can be written in a parametric form with two parameters determining the distance $R$ and the direction  $\theta$ of a displacement from the critical point~\cite{ZinnJustin}.  
The dimensionless Ising variables $r$ and $h$ can be expressed in terms of these parametric variables, as can the magnetization, $M\equiv-\partial G_{\rm Ising}/\partial h$, and the Gibbs free energy, $G_{\rm{Ising}}$:
\begin{equation} \label{IsingEoS}
    \begin{split}
    r &= R(1- \theta^2) \\
    M &= M_0 R^{\beta} \theta
    \end{split}
    \qquad \quad
    \begin{split}
    h &= h_0 R^{\beta \delta} \tilde{h}(\theta) \\
    G_{\rm{Ising}} &=  h_0M_0R^{\beta(1+\delta)}g(\theta)-Mh\,,  
    \end{split}
    \end{equation}
	where
	\begin{align*}
	\centering
    \tilde{h}(\theta) &=\theta (1+a\theta^2+b\theta^4), \\
    g(\theta) &= c_0 +  c_1(1-\theta^2) + c_2(1-\theta^2)^2 + c_3(1-\theta^2)^3, \\
    c_0 &= \frac{\beta}{2-\alpha}(1+a+b), \\
	c_1 &= -\frac{1}{2} \frac{1}{\alpha -1}((1-2\beta)(1+a+b)-2\beta(a+2b)), \\
	c_2 &= - \frac{1}{2\alpha}(2\beta b - (1-2\beta)(a+2b)), \\
	c_3 &= - \frac{1}{2(\alpha+1)}b(1-2\beta),
    \end{align*}
and the 3D Ising critical exponents are given by $\beta=0.326$  and $\delta=4.8$
and the scaling relation $2-\alpha=\beta(1+\delta)$,
with the coefficients of $\tilde{h}(\theta)$ determined from universality in the $\epsilon$-expansion as $a=-0.76201$,  $b=0.00804$ \cite{Guida:1996ep}.
The normalization constants are determined via the conventional constraints of $M(r=-1,h=0^+)=1$ and $M(r=0,h) = {\rm sgn}(h) |h|^{1/\delta}$, 
which yields $h_0=0.364$, and $M_0=0.605$ \cite{Nonaka:2004pg, Parotto:2018pwx}.  
In Section~\ref{Subsec:Mapping} we describe how to map the Ising EoS $G_{\rm Ising}$ as a function of $r$ and $h$ onto the QCD phase diagram with axes $T$ and $\mu_B$. 

\section{Contribution of Resonances to Factorial Cumulants of Observed Proton Multiplicity}
\label{app:resonances}
 When experimentalists measure the final state hadron (specifically, proton) multiplicities in heavy ion collisions, they cannot distinguish between ``direct protons'' (protons formed at freezeout) and ``daughter protons'' which come from the decay of resonance particles that were formed at freezeout and decay into protons and other hadrons later, after freezeout. In order to account for this, the  
 correlations between all the produced protons (regardless of whether they are direct protons or the products of resonance decays) 
 need to be computed. The total $\widehat{\Delta}\omega_{kp}$ would then be given by the expression~\eqref{Eq:IRCOmegaGen}. Notice that the difference between the expression~\eqref{Eq:omegaP} for $\widehat{\Delta}\omega_{kp}$ computed only using direct protons
 differs from Eq.~(\ref{Eq:IRCOmegaGen}) only via the coefficients that multiply the hydrodynamic correlation functions. This suggests that the qualitative features of $\widehat{\Delta}\omega_{kp}$ that arise from the critical contributions to hydrodynamic correlations  that we have seen in Figs.~\ref{Fig:w2pAll}, \ref{Fig:w3pAll} and \ref{Fig:w4pAll}, that we computed using Eq.~\eqref{Eq:omegaP}, should not be substantially or qualitatively modified 
 after correlations between daughter protons are included via Eq.~\eqref{Eq:IRCOmegaGen}, although there will certainly be quantitative modifications. We confirm this expectation in this Appendix.

 \begin{figure}[t]
    \centering
    \includegraphics[width=\linewidth]{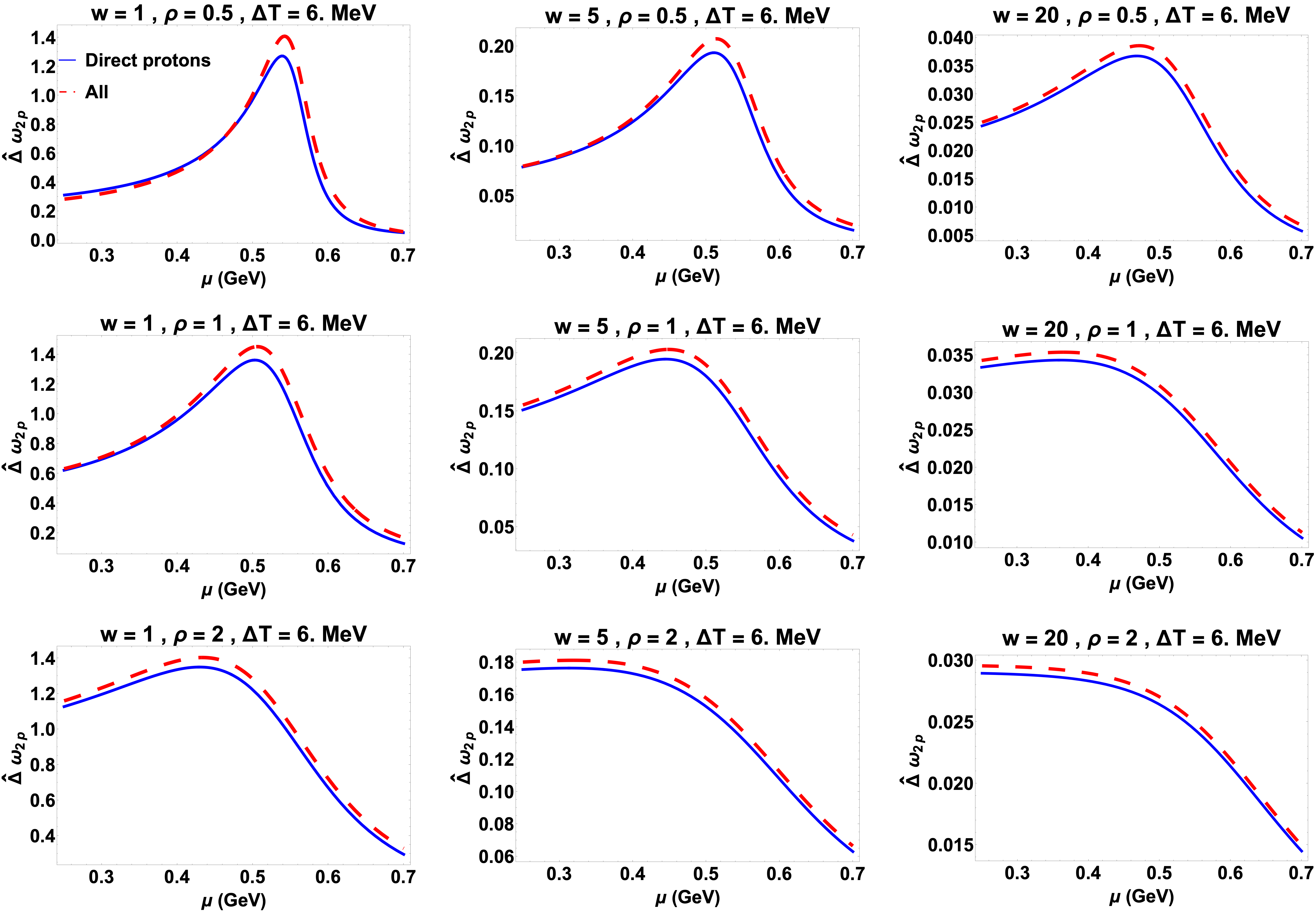}
    \caption{\justifying Second factorial cumulant of the proton multiplicity distribution, $\hat{\Delta}\omega_{2p}$, along the freezeout curve of Eq.~(\ref{eq:FreezeoutCurve}) with $\Delta T_f =6$~MeV. Different panels correspond to different choices of the nonuniversal mapping parameters $w$ and $\rho$, with $\mu_c=600$~MeV, $T_c=90$~MeV and  $\alpha_2=0^{\circ}$ in all panels. The red dashed curves show $\hat\Delta\omega_{2p}$ calculated using Eq.~(\ref{Eq:IRCOmegaGen}), which includes contributions of direct and daughter protons, whereas the blue solid curves include 
    only the direct protons, as in Fig.~\ref{Fig:w2pAll}.
    }
    \label{Fig:w2pappresonaces}
\end{figure}

 \begin{figure}[t]
    \centering
    \includegraphics[width=\linewidth]{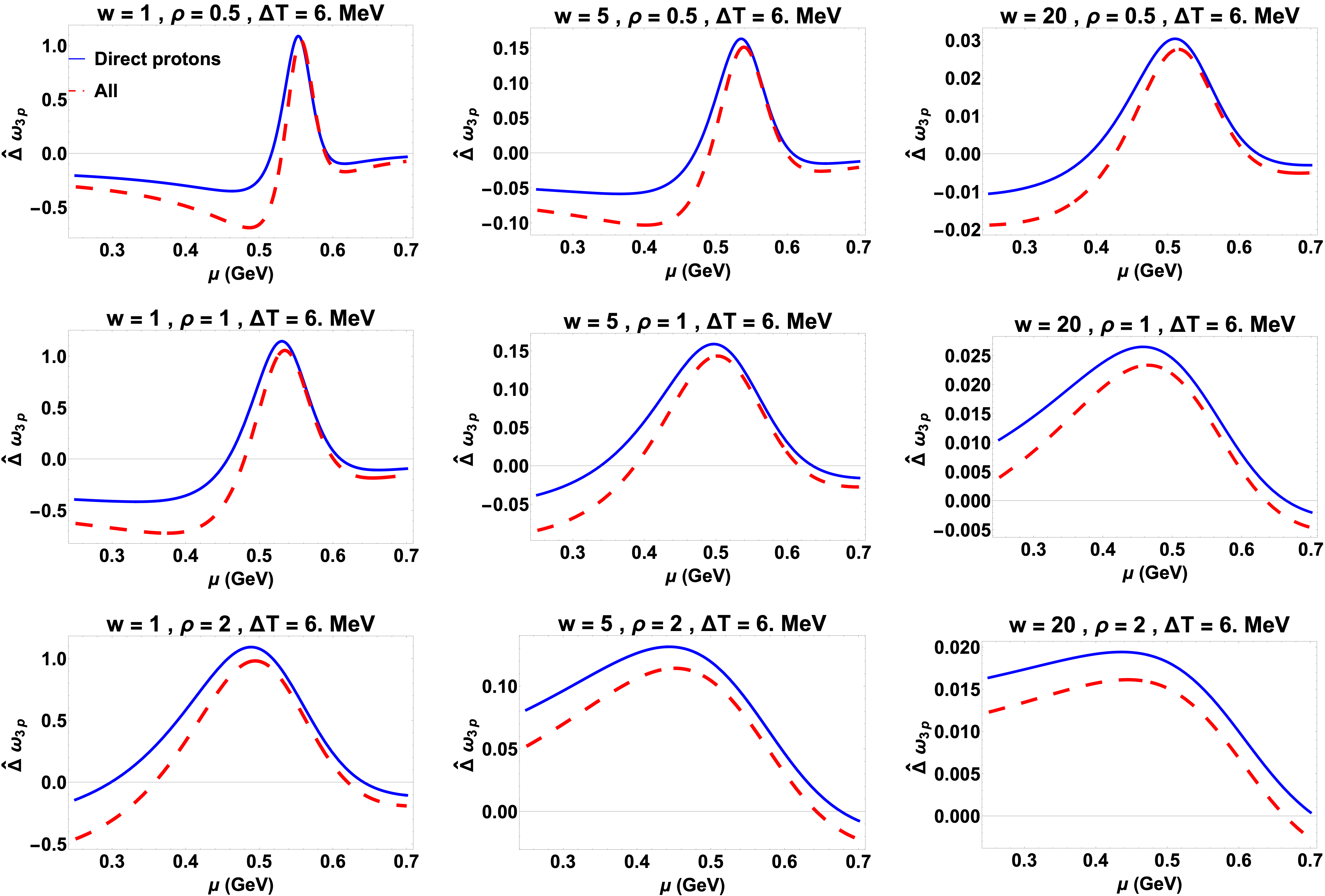}
    \caption{\justifying Third factorial cumulant of the proton multiplicity distribution, $\hat{\Delta}\omega_{3p}$, along the freezeout curve of Eq.~(\ref{eq:FreezeoutCurve}) with $\Delta T_f =6$~MeV. Different panels correspond to different choices of the nonuniversal mapping parameters $w$ and $\rho$, with $\mu_c=600$~MeV, $T_c=90$~MeV and  $\alpha_2=0^{\circ}$ in all panels. The red dashed curves show $\hat\Delta\omega_{3p}$ calculated using Eq.~(\ref{Eq:IRCOmegaGen}), which includes contributions of direct and daughter protons, whereas the blue solid curves include 
    only the direct protons, as in Fig.~\ref{Fig:w3pAll}.
    }
    \label{Fig:w3pappresonaces}
\end{figure}

 \begin{figure}[t]
    \centering
    \includegraphics[width=\linewidth]{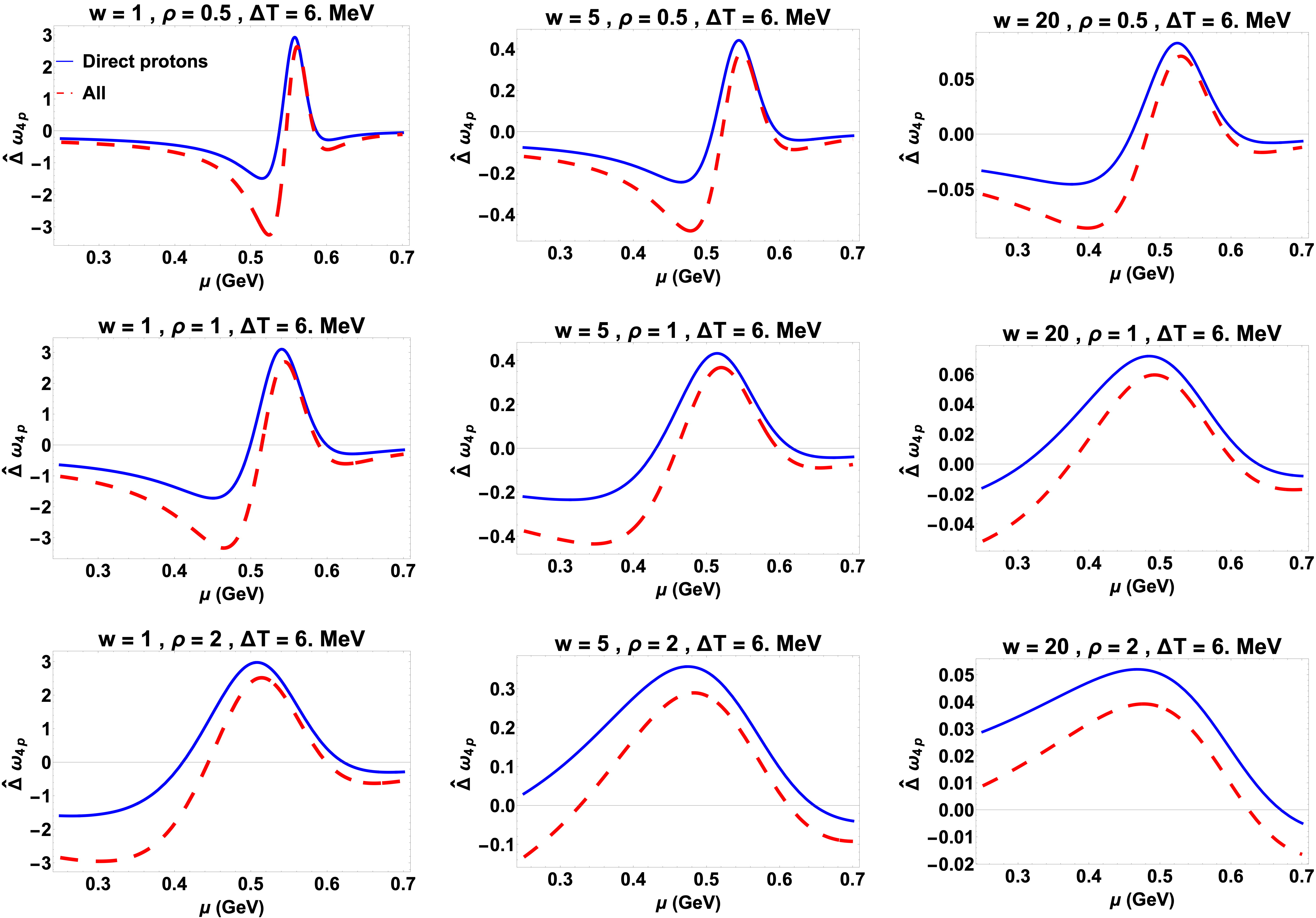}
    \caption{\justifying Fourth factorial cumulant of the proton multiplicity distribution, $\hat{\Delta}\omega_{4p}$, along the freezeout curve of Eq.~(\ref{eq:FreezeoutCurve}) with $\Delta T_f =6$~MeV. Different panels correspond to different choices of the nonuniversal mapping parameters $w$ and $\rho$, with $\mu_c=600$~MeV, $T_c=90$~MeV and  $\alpha_2=0^{\circ}$ in all panels. The red dashed curves show $\hat\Delta\omega_{4p}$ calculated using Eq.~(\ref{Eq:IRCOmegaGen}), which includes contributions of direct and daughter protons, whereas the blue solid curves include 
    only the direct protons, as in Fig.~\ref{Fig:w4pAll}.
    }
    \label{Fig:w4pappresonaces}
\end{figure}

 The blue curves in Figs.~\ref{Fig:w2pappresonaces}, \ref{Fig:w3pappresonaces} and \ref{Fig:w4pappresonaces}
 are the same as the blue curves in Figs.~\ref{Fig:w2pAll}, \ref{Fig:w3pAll} and \ref{Fig:w4pAll}: they are the second, third, and fourth factorial cumulants of the proton multiplicity computed along the freezeout curve with $\Delta T_f=6$~MeV using direct protons only.  In the  Figures in this Appendix, though, we have added the dashed red curves which show the full results for 
$\widehat{\Delta}\omega_{kp}$ when the correlations among daughter protons are included in the calculation.
The different panels correspond to the same  choices of the nonuniversal mapping parameters $w$ and $\rho$ that we have used in previous Figures.
From Fig.~\ref{Fig:w2pappresonaces} we see that including the daughter protons in the calculation makes almost no difference to $\hat\Delta\omega_{2p}$.
In Figs.~\ref{Fig:w3pappresonaces} and \ref{Fig:w4pappresonaces} we see that the correlations among daughter protons introduces quantitative changes to $\hat\Delta\omega_{3p}$ and $\hat\Delta\omega_{4p}$ as expected, but all qualitative features remain as in the blue curves, which is to say as when computed using direct protons only as in Figs.~\ref{Fig:w3pAll}
and \ref{Fig:w4pAll}.

\section{Contribution of Energy Density Fluctuations Compared to Baryon Number Fluctuations}
\label{app:subleading}

In the main text, we calculated the  factorial cumulants of the proton multiplicity 
distribution 
via the maximum entropy method for the first time, upon assuming that the critical contributions to hydrodynamic fluctuations are in equilibrium at freezeout. 
Specifically, in Section~\ref{Sec:main} we calculated the factorial cumulants $\hat{\Delta}\omega_{2p}$, $\hat\Delta\omega_{3p}$ and $\hat\Delta\omega_{4p}$ just after freezeout by first determining the ordinary  cumulants of the fluctuations of the two hydrodynamic densities, namely the energy density cumulants 
$\Delta H_{k\epsilon}$ and the baryon number density cumulants $\Delta H_{kn}$ as well as their mixed 
cumulant counterparts, just before freezeout
using the maximum entropy prescription \eqref{Eq:omegaP}. 
The singular contribution to the $\Delta H$'s 
coming from critical contributions was obtained 
using the 3D Ising EoS
mapped to the QCD phase diagram (see Eq.~(\ref{eq:ExplicitCorrelations})).
In order to obtain the IRCs (generalized factorial cumulants) of particle multiplicities shown in Eq.~\eqref{Eq:omegaP}, we utilized the expressions for the hydrodynamic IRCs in terms of ordinary cumulants given in Eq.~(\ref{eq:hatH}).
The calculation of all of the $\Delta H$'s is described in  Section~\ref{Sec:ME},
but in Section~\ref{Sec:main} we have only plotted
our results for $\Delta H_{2n}$, $\Delta H_{3n}$ and $\Delta H_{4n}$.
As described in detail in the main text, one of the ways in which the maximum entropy freeze-out prescription differs from earlier treatments of freeze-out is that it includes the contribution of {\em energy\/} density correlators to the factorial cumulants of proton multiplicity. 
In this Appendix, we evaluate the relative
importance of the contribution of energy density
cumulants to the factorial cumulants of proton multiplicity, relative to the contribution of 
the cumulants of baryon number density.

We begin by rewriting the expression~\eqref{Eq:IRCOmegaGen} 
for the factorial cumulants of the proton multiplicity distribution including contributions from resonance decays as
\begin{equation}
    \hat\Delta\omega_{kp}
    =
    \frac{\hat\Delta H_{kn}}{\langle N_p\rangle}
    \(
    \sum_{A}
    \Gamma_{A\rightarrow p}P^{n}_{A}
    \)^k
    \left(1+\sum_{j=1}^k R_{j\epsilon(k-j)n}\right)
\end{equation}
where 
$R_{j\epsilon(k-j)n}$ 
denotes the ratio of the contribution of the correlator $\hat\Delta H_{j\epsilon(k-j)n}$
to the factorial cumulant
$\hat\Delta\omega_{kp}$ to the contribution
of the correlator $\hat\Delta H_{kn}$ involving only baryon number density fluctuations.
Here, $\hat\Delta H_{j\epsilon(k-j)n}$ involves $j$ energy density variables and $j$ fewer baryon  
density variables. 
We shall focus on $R_{1\epsilon(k-1)n}$ which quantifies the relative importance of correlators with a single energy density variable to those with only baryon density variables that we plotted in Section~\ref{Sec:HydroCorrelations}.
According to Eq.~\eqref{Eq:IRCOmegaGen}, this ratio is given by
\begin{equation}
    R_{1\epsilon(k-1)n}
    =
    \frac{\hat\Delta H_{1\epsilon(k-1)n}}
    {\hat\Delta H_{kn}}\,
    \frac{\sum_{A}
    \Gamma_{A\rightarrow p}P^{\epsilon}_{A}}{\sum_{A}
    \Gamma_{A\rightarrow p}P^{n}_{A}}
    \equiv 
    \frac{\hat\Delta H_{1\epsilon(k-1)n}}
    {\mu\hat\Delta H_{kn}} \bar R_{\epsilon n}\,
\end{equation}
where $P_A^\epsilon$ and $P_A^n$ were defined in Eq.~\eqref{eq:P-lowerA-uppera-defn} and where we have introduced the dimensionless ratio 
\begin{equation}\label{eq:Rbar-defn}
\bar R_{\epsilon n}\equiv \mu\, \frac{\sum_{A}
    \Gamma_{A\rightarrow p}P^{\epsilon}_{A}}{\sum_{A}
    \Gamma_{A\rightarrow p}P^{n}_{A}}\,.
\end{equation}
For arbitrary $j\leq k$, we can write
\begin{equation}\label{Eq:RJeKmj}
    R_{j\epsilon(k-j)n}
    =
    \frac{\hat\Delta H_{j\epsilon(k-j)n}}
    {\hat\Delta H_{kn}}\,
    \(\frac{\sum_{A}
    \Gamma_{A\rightarrow p}P^{\epsilon}_{A}}{\sum_{A}
    \Gamma_{A\rightarrow p}P^{n}_{A}}\)^{j}
    \equiv 
    \frac{\hat\Delta H_{j\epsilon(k-j)n}}
    {\mu^{j}\hat\Delta H_{kn}}  \(\bar{R}_{\epsilon n}\)^{j}\,,
\end{equation}
indicating that the contribution of a correlator with $j$ energy density variables is suppressed relative to that with only baryon density variables by {a factor of order $(\bar R_{\epsilon n})^j$.  In the left panel of Fig.~\ref{FigRNew}, we plot $\bar R_{\epsilon n}$ along the crossover curve as a function of $\mu_B$. At the critical point where $\mu=\mu_c=600$~MeV, $\bar R_{\epsilon n}\approx -0.12$.

\begin{figure}[t]
 \begin{minipage}{0.45\textwidth}
  \includegraphics[width=\textwidth]{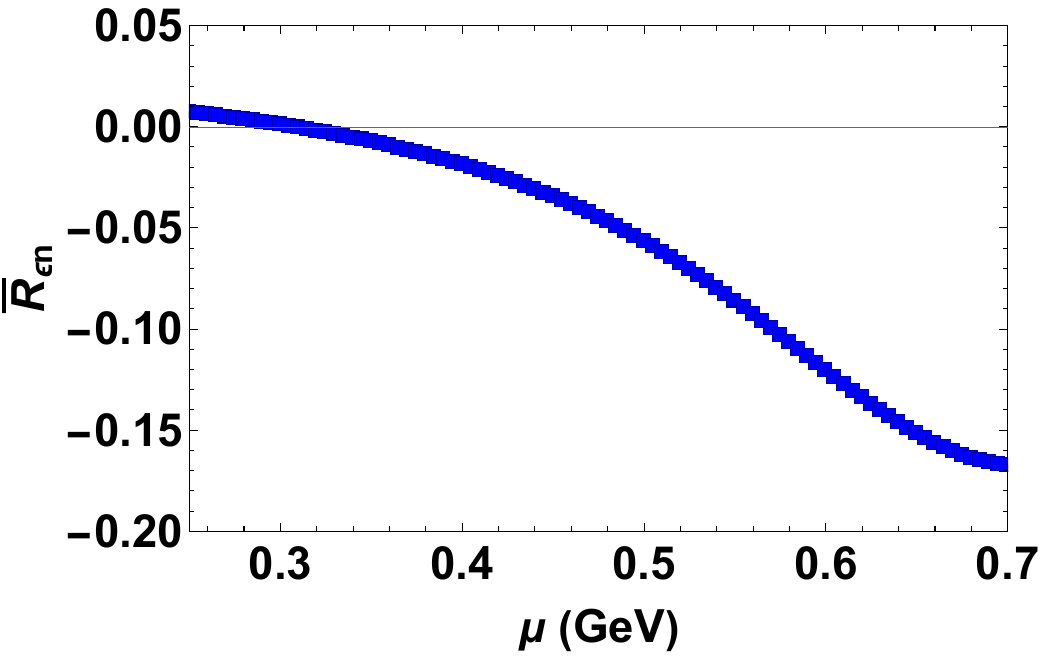}
  \end{minipage}
   \begin{minipage}{0.45\textwidth}
  \includegraphics[width=\textwidth]{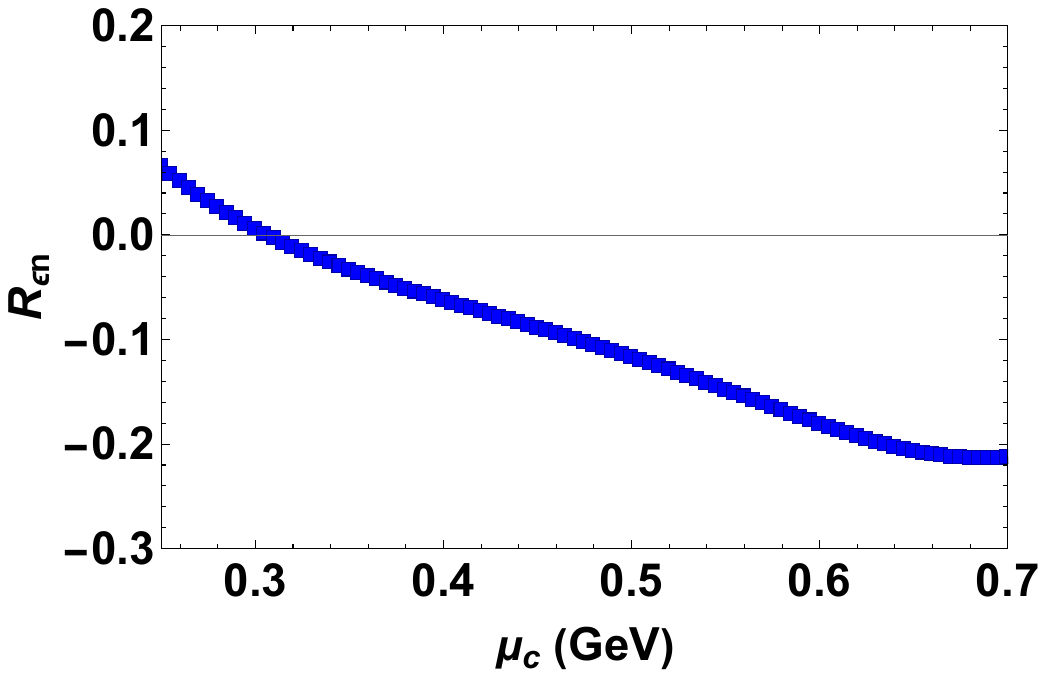}
   \end{minipage}
   \caption{\justifying Left: The quantity $\bar{R}_{\e n}$, defined in Eq.~\eqref{eq:Rbar-defn} and appearing Eq.~\eqref{Eq:RJeKmj}, along the crossover curve as a function of the baryon chemical potential $\mu_B$ in GeV.  Right:  The ratio $R_{\e n}$ at the critical point, as estimated in Eq.~\eqref{Eq:Ren} upon assuming that the hydrodynamic fluctuations are dominated by the leading critical contribution, as a function of $\mu_c$, the baryon chemical potential at which the critical point is located.
   One can read off from this plot that for a critical point at $\mu_c= 600\, \text{MeV}$, $R_{\e n}\approx -0.18$. The upturn in the curve below $\mu_c=300$~MeV reflects the fact that $R_{\epsilon n}$ diverges as $\mu_c\rightarrow 0$, as discussed in the text.} 
 \label{FigRNew}
 \end{figure}

We can make our estimate of the suppression $R_{j\epsilon (k-j)n}$ in the vicinity of the critical point more concrete by setting $\mu=\mu_c$ in Eq.\eqref{Eq:RJeKmj}
and noting that
if we ignore the subleading contributions to $\widehat{\Delta}H_{j\e(k-j)n}$ at the critical point, we can make the approximation $\widehat{\Delta}H_{j\e(k-j)n}\approx \Delta H_{j\e(k-j)n}$. From Eq.~\eqref{Eq:DelHlead} and the discussion around it, we can then see that the leading critical contributions to the hydrodynamic IRCs that arise in Eq.~(\ref{Eq:RJeKmj}) are
$\widehat{\Delta}H_{j\e(k-j)n}\propto h^{k-j}_{n}h_{\e}^{j}\partial^{k} G/\partial h^{k}$, where $h_n$ and $h_\e$ are defined in Eq.~(\ref{Eq:HeAndHn}).  
Thus, approximating the hydrodynamic IRCs by their leading critical contributions yields the estimate
\be
R_{j\epsilon(k-j)n}
    \approx\left( 1+\frac{T_c}{\mu_c\tan\alpha_1} \right)^j \left(\bar{R}_{\e n}\right)^{j}\ 
    \label{Eq:Ren}
\ee
at (and in the vicinity of) the critical point.
In the right panel of Fig.~\ref{FigRNew} we set $k=2$ and $j=1$ and plot $R_{\epsilon n}$ 
from Eq.~\eqref{Eq:Ren} as a function of $\mu_c$.
At $\mu_c=600$~MeV, $R_{\epsilon n}\approx 1.5 \bar R_{\epsilon n} \approx -0.18$, meaning that in the vicinity of a critical point at $\mu_c=600$~MeV we find $R_{j\epsilon(k-j)n}\approx (-0.18)^j$.

The EoS that we have used throughout this paper is not designed to describe physics at $\mu=0$, since it does not incorporate what we know from lattice QCD at $\mu=0$. We can nevertheless 
glean from Eq.~\eqref{Eq:Ren} an important qualitative point about the difference between the physics of fluctuations near a critical point at (for example) $\mu_c=600$~MeV and the physics of critical fluctuations if there were a critical point near $\mu=0$.  If we take $\mu_c\rightarrow 0$ in Eq.~\eqref{Eq:Ren}, at such a critical point $\bar R_{\epsilon n}\propto \mu_c$ (see Eq.~\eqref{eq:Rbar-defn}) and 
$\bar R_{j\epsilon(k-j)n}\propto \mu_c^j$, meaning that the RHS of Eq.~\eqref{Eq:Ren},
$R_{j\epsilon(k-j)n}\propto (\mu_c+T_c/\tan\alpha_1)^j$, diverges since $\alpha_1\rightarrow 0$ as $\mu_c\rightarrow 0$.
This confirms that if there were a critical point at very small $\mu$, the contribution of critical fluctuations to the factorial cumulants of the proton multiplicity would be dominated by the effects of the cumulants of the hydrodynamic fluctuations of the energy density. Fluctuations of the baryon density would be relatively unimportant in this case, with their contribution to the factorial cumulants of the proton multiplicity suppressed as the angle $\alpha_1$ vanishes and the crossover curve becomes horizontal.

Of course, we have long known from lattice QCD calculations and from experimental data that there is no critical point with $\mu_c\rightarrow 0$.
If a critical point were to be found at $\mu_c=600$~MeV, the curvature of the crossover line as determined by lattice QCD calculations suggests that at such a critical point $T_c\simeq 90$~MeV and $\alpha_1\simeq 16.6^\circ$. As already noted, at such a critical point $R_{\epsilon n}\approx 1.5 \bar R_{\epsilon n} \approx -0.18$ and $R_{j\epsilon(k-j)n}\approx (-0.18)^j$.
This means that the dominant contribution to the factorial cumulants of the proton multiplicity comes from the critical fluctuations of the baryon density, but the contributions coming from fluctuations of the energy density and from mixed correlators of the energy and baryon density are not negligible.  
Thus, it is important to take into account these additional contributions in order to make quantitative estimates of the proton factorial cumulants, as we have done throughout Section.~\ref{Sec:main}.

\section{Glossary/Index of Notation}
This Appendix summarizes the notations utilized in this work.
\label{app:notations}
\begin{list}{}{}


\item
$ G_{A_1...A_k}$ ---  $k^{\rm{th}}$ order particle multiplicity correlation function after freezeout,  Eq. \eqref{Eq:G};

\item
$ \Delta G_{A_1...A_k}$
--- 
the difference of $G_{A_1...A_k}$ from the HRG value $\bar{G}_{A_1...A_k}$,
Eq. \eqref{Eq:DelG};

\item
$ \hat{\Delta}G_{A_1\dots A_k},~ \hat\Delta H_{a_1\dots a_k}$  --- $k^{\rm{th}}$ order irreducible relative cumulants (IRCs), Eqs.~\eqref{Eq:IRCG},~\eqref{eq:hatH};

\item
$ H_{a_1...a_k}$ --- $k^{\rm{th}}$ order hydrodynamic correlation function before freezeout, Eq. \eqref{Eq:Hdef}; 

\item
$ \Delta H_{a_1\dots a_{k}}$
--- the difference of $ H_{a_1...a_k}$ from its value in HRG, $ \bar{H}_{a_1...a_k}$,
Eq. \eqref{Eq:DelH};

\item
$ H_{kn}$ --- 
shorthand for $H_{a_1\dots a_{k}}$ with $a_1=\dots=a_k=n$;

\item
$\<N_{\hA}\>$ --- mean multiplicity of particle species $\hA$;

\item 
$P_a^A$ --- contribution of the occupied particle state $A$ to conserved density $a$, Eqs.~\eqref{Eq:conEN},~\eqref{Eq:P}.

\item
$P_A^a$ --- the component of the fluctuation of the multiplicity of particle $A$ matching the fluctuation of hydrodynamic density $a$, Eqs.~\eqref{Eq:IRCGAsToIRCH},~\eqref{eq:P-lowerA-uppera-defn}.

\item
$ T_{\rm crossover}(\mu_B) \equiv T_{h=0}(\mu_B)$ --- the temperature along the crossover curve 
as a function of
$\mu_B$;

\item
$ T_{f}(\mu_B)$ --- the temperature along the freezeout curve as a function of baryon chemical potential, Eq.~\eqref{eq:FreezeoutCurve};

\item
$ \Delta T'$ --- the distance from the transition line $h=0$ in terms of $T'(\mu,T)$,
Eq. \eqref{eq:DT'-def};

\item
$ \Delta T_f$ --- the temperature difference between the crossover curve and the freezeout curve, which, in this work, is taken as a constant, independent of chemical potential, Eq.~\eqref{eq:FreezeoutCurve};


\item
$ f_{\xi}$ --- 
ratio of $\xi_{\rm QCD}$ to the correlation length $\xi$ in the Ising model matching QCD at the critical point, Eq.~\eqref{eq:fxi};


\item
$ \xi_{\rm{QCD}}$ --- correlation length in QCD at given $T$ and $\mu$;

\item
$ \omega_{\hA_1\dots \hA_{k}}$  --- normalized mixed cumulant of the multiplicities of species  $\{\hA_1\dots \hA_k\}$,
 Eq. \eqref{Eq:Omegak};

\item
$ \Delta \omega_{\hA_1\dots \hA_{k}}$  --- 
the difference of $\omega_{\hA_1\dots \hA_{k}}$
from its value in HRG, $\bar\omega_{\hA_1\dots \hA_{k}}$,
Eq. \eqref{eq:Delta-omega-Ahats};

\item
$ \widehat{\Delta} \omega_{\hA_1\dots \hA_{k}}$ --- integrated
IRC of particle multiplicities, 
Eq.~\eqref{Eq:IRCOmega};

\item 
$\hat\Delta\omega_{kp}$ --- 
shorthand for
$\widehat{\Delta}\omega_{\hat A_1\dots\hat A_k}$ with $\hat A_1=\dots=\hat A_k=p$, i.e., $k^{\rm{th}}$ order integrated 
IRC for protons;

\item
$ \[\dots\]_{\overline{ABCD}}$ --- the average over the permutations of listed indices.

\end{list}

\bibliography{refs}

\end{document}